\documentclass[
	superscriptaddress,
	aps,
	pra,
	nofootinbib,
	notitlepage,
	10pt
]{revtex4-1}
\pdfoutput=1
\usepackage{changepage}
\usepackage{graphicx}
\usepackage{amsmath}
\usepackage{amssymb}
\usepackage{comment}
\usepackage{placeins}
\usepackage[caption=false]{subfig}
\usepackage[colorlinks]{hyperref}
\usepackage{hypcap}
\usepackage{color}
\usepackage{subfig}

\newtheorem{theorem}{Theorem}
\newtheorem{lemma}{Lemma}
\newcommand{\mlfac}{\zeta}

\newcommand{\eq}[1]{Eq.~\hyperref[eq:#1]{(\ref*{eq:#1})}}
\renewcommand{\sec}[1]{\hyperref[sec:#1]{Section~\ref*{sec:#1}}}
\newcommand{\app}[1]{\hyperref[app:#1]{Appendix~\ref*{app:#1}}}
\newcommand{\tab}[1]{\hyperref[tab:#1]{Table~\ref*{tab:#1}}}
\newcommand{\fig}[1]{\hyperref[fig:#1]{Figure~\ref*{fig:#1}}}
\newcommand{\figa}[2]{\hyperref[fig:#1]{Figure~\ref*{fig:#1}#2}}
\newcommand{\figx}[2]{\hyperref[fig:#1]{Figure~\ref*{fig:#1}(#2)}}
\newcommand{\thm}[1]{\hyperref[thm:#1]{Theorem~\ref*{thm:#1}}}
\newcommand{\lem}[1]{\hyperref[lem:#1]{Lemma~\ref*{lem:#1}}}
\newcommand{\cor}[1]{\hyperref[cor:#1]{Corollary~\ref*{cor:#1}}}
\newcommand{\defn}[1]{\hyperref[def:#1]{Definition~\ref*{def:#1}}}
\newcommand{\alg}[1]{\hyperref[alg:#1]{Algorithm~\ref*{alg:#1}}}

\def\avg#1{\mathinner{\langle{#1}\rangle}}
\def\bra#1{\mathinner{\langle{#1}|}}
\def\ket#1{\mathinner{|{#1}\rangle}}
\newcommand{\braket}[2]{\langle #1|#2\rangle}

\newcommand{\abs}[1]{\ensuremath{\left| #1 \right|}}
\newcommand{\norm}[1]{\ensuremath{\left\| #1 \right\|}}

\newcommand\R{{\mathrm {I\!R}}}

\newcommand{\dd}[1]{\ensuremath{\mathrm{d} #1}}

\newcommand{\ignore}[1]{}

\newcommand{\be}{\begin{equation}}
\newcommand{\ee}{\end{equation}}
\newcommand{\ba}{\begin{eqnarray}}
\newcommand{\ea}{\end{eqnarray}}

\newcommand{\nn}{\nonumber \\}
\newcommand{\kets}[1]{ |{#1} \rangle}

\newcommand{\bs}[1]{\boldsymbol{#1}}
\newcommand{\cH}{\aleph}
\newcommand{\tcH}{{\widetilde{\aleph}}}

\input{Qcircuit}

\newcommand{\inter}{\delta}
\begin{document}

\title{Exponentially More Precise Quantum Simulation of Fermions
	in the Configuration Interaction Representation}

\date{\today}
\author{Ryan Babbush}
\email[Corresponding author: ]{babbush@google.com}
\affiliation{Google, Venice, CA 90291, USA}
\affiliation{Department of Chemistry and Chemical Biology, Harvard University, Cambridge, MA 02138}
\author{Dominic W. Berry}
\email[Corresponding author: ]{dominic.berry@mq.edu.au}
\affiliation{Department of Physics and Astronomy, Macquarie University, Sydney, NSW 2109, Australia}
\author{Yuval R. Sanders}
\affiliation{Department of Physics and Astronomy, Macquarie University, Sydney, NSW 2109, Australia}
\author{Ian D. Kivlichan}
\affiliation{Department of Chemistry and Chemical Biology, Harvard University, Cambridge, MA 02138}
\affiliation{Department of Physics, Harvard University, Cambridge, MA 02138, USA}
\author{Artur Scherer}
\affiliation{Department of Physics and Astronomy, Macquarie University, Sydney, NSW 2109, Australia}
\author{Annie Y. Wei}
\affiliation{Department of Chemistry and Chemical Biology, Harvard University, Cambridge, MA 02138}
\author{Peter J. Love}
\affiliation{Department of Physics and Astronomy, Tufts University, Medford, MA 02155}
\author{Al\'{a}n Aspuru-Guzik}
\affiliation{Department of Chemistry and Chemical Biology, Harvard University, Cambridge, MA 02138}

\begin{abstract}
We present a quantum algorithm for the simulation of molecular systems that is
asymptotically more efficient than all previous algorithms in the literature
in terms of the main problem parameters.  As in previous work
[Babbush \textit{et al.},  \textit{New Journal of Physics} \textbf{18},
033032 (2016)], we employ a recently developed technique for simulating
Hamiltonian evolution, using a truncated Taylor series to obtain logarithmic
scaling with the inverse of the desired precision.  The algorithm of this paper
involves simulation under an oracle for the sparse, first-quantized
representation of the molecular Hamiltonian known as the configuration
interaction (CI) matrix.  We construct and query the CI matrix oracle to allow
for on-the-fly computation of molecular integrals in a way that is exponentially
more efficient than classical numerical methods.  Whereas second-quantized
representations of the wavefunction require $\widetilde{\cal O}(N)$ qubits,
where $N$ is the number of single-particle spin-orbitals, the CI matrix
representation requires $\widetilde{\cal O}(\eta)$ qubits where $\eta \ll N$
is the number of electrons in the molecule of interest.  We show that the gate
count of our algorithm scales at most as $\widetilde{\cal O}(\eta^2 N^3 t)$.
\end{abstract}

\maketitle

\section{Introduction}
\label{sec:intro}

The first quantum algorithm for quantum chemistry was introduced nearly a
decade ago \cite{Aspuru-Guzik2005}.  That algorithm was based on the
Trotter-Suzuki decomposition, which Lloyd and Abrams first applied to quantum
simulation in \cite{Lloyd1996,Abrams1997}.  The Trotter-Suzuki decomposition has
been used in almost all quantum algorithms for quantum chemistry since then
\cite{Jones2012,Veis2010,Wang2014,Li2011,Yung2013,Kassal2008,
      Toloui2013,Whitfield2013b,Whitfield2015,Wecker2014,Hastings2015,
      Poulin2014,McClean2014,BabbushTrotter},
with the exception of the adiabatic algorithm detailed in \cite{BabbushAQChem},
the variational quantum eigensolver approach described in
\cite{Peruzzo2013,McClean2015,OMalley2016}, and in our prior papers using
the Taylor series technique \cite{BabbushSparse1,Kivlichan2016}.  Recently, there has been
substantial renewed interest in these algorithms due to the low qubit
requirement compared with other algorithms such as factoring, together with the
scientific importance of the electronic structure problem.  This led to a series
of papers establishing formal bounds on the cost of simulating various molecules
\cite{Wecker2014,Hastings2015,Poulin2014,McClean2014,BabbushTrotter}.

Whereas qubit requirements for the quantum chemistry problem seem modest,
using arbitrarily high-order Trotter formulas, the tightest-known upper bound on
the gate count of the second-quantized, Trotter-based quantum simulation of
chemistry is $\widetilde{\cal O}(N^{8+o(1)} t / \epsilon^{o(1)})$
\cite{Berry2006,Wiebe2011}\footnote{We use the typical computer science
convention that $f\in \Theta(g)$, for any functions $f$ and $g$, if $f$ is
asymptotically upper and lower bounded by multiples of $g$, ${\cal O}$ indicates
an asymptotic upper bound, $\widetilde{{\cal O}}$ indicates an asymptotic upper
bound suppressing any polylogarithmic factors in the problem parameters,
$\Omega$ indicates the asymptotic lower bound and $f \in o(g)$ implies
$f / g \rightarrow 0$ in the asymptotic limit.}, where $N$ is the number of
spin-orbitals and $\epsilon$ is the required accuracy. However, using
significantly more practical Trotter decompositions, the best known gate
complexity for this quantum algorithm is
$\widetilde{\cal O}(N^9 \sqrt{t^3 / \epsilon})$ \cite{Hastings2015}.
Fortunately, recent numerics suggest that the scaling for real molecules is
closer to $\widetilde{\cal O}(N^6 \sqrt{t^3 / \epsilon})$ \cite{Poulin2014} or
$\widetilde{\cal O}(Z^3 N^4 \sqrt{t^3 / \epsilon})$ \cite{BabbushTrotter}, where
$Z$ is the largest nuclear charge in the molecule.  Still, the Trotter-based
quantum simulation of many molecular systems remains a costly proposition \cite{Gibney2014,Mueck2015}.

In Ref.~\cite{BabbushSparse1}, we introduced two novel quantum algorithms for
chemistry based on the truncated Taylor series simulation method of
\cite{Berry2015}, which are exponentially more precise than algorithms using the
Trotter-Suzuki decomposition.  Our first algorithm, referred to as the
``database'' algorithm, was shown to have gate count scaling as
$\widetilde{\cal O}(N^4 \| H\| t)$. Our second algorithm, referred to as the
``on-the-fly'' algorithm, was shown to have the lowest scaling of any approach
to quantum simulation previously in the literature, $\widetilde{\cal O}(N^5 t)$.
Both of these algorithms use a second-quantized representation of the
Hamiltonian; in this paper we employ a more compressed, first-quantized
representation of the Hamiltonian known as the configuration interaction (CI)
matrix. 
We also analyze the on-the-fly integration strategy far more rigorously, by making the
assumptions explicit and rigorously deriving error bounds.
Our approach combines a number of improvements:
\begin{itemize}
\item a novel 1-sparse decomposition of the CI matrix (improving over that in \cite{Toloui2013}),
\item a self-inverse decomposition of 1-sparse matrices as introduced in \cite{Berry2013},
\item the exponentially more precise simulation techniques of \cite{Berry2015},
\item and the on-the-fly integration strategy of \cite{BabbushSparse1}.
\end{itemize}

The paper is outlined as follows.
In \sec{summary},
we summarize the key results of this paper,
and note the improvements presented here over previous approaches.
In \sec{encoding},
we introduce the configuration basis encoding of the wavefunction.
In \sec{decomp1},
we show how to decompose the Hamiltonian into 1-sparse unitary matrices.
In \sec{oracle},
we use the decomposition of \sec{decomp1} to construct a circuit which provides
oracular access to the Hamiltonian matrix entries,
assuming access to $\textsc{sample}(w)$ from \cite{BabbushSparse1}.
In \sec{simulation},
we review the procedures in \cite{Berry2015} and \cite{BabbushSparse1} to
demonstrate that this oracle circuit can be used to effect a quantum simulation
which is exponentially more precise than using a Trotter-Suzuki decomposition
approach.
In \sec{conclusion},
we discuss applications of this algorithm and future research directions.

\section{Summary of Results}
\label{sec:summary}

In our previous work \cite{BabbushSparse1}, simulation of the molecular
Hamiltonian was performed in second quantization using Taylor series simulation
methods to give a gate count scaling as $\widetilde{\cal O}(N^5 t)$.
In this work, we use the configuration interaction representation of the
Hamiltonian to provide an improved scaling of
$\widetilde{\cal O}(\eta^2 N^3 t)$.
This result is summarized by the following Theorem.

\begin{theorem}
\label{thm:maintheorem}
Using atomic units in which $\hbar$, Coulomb's constant, and the charge and mass
of the electron are unity, we can write the molecular Hamiltonian as
\begin{align}
\label{eq:electronic}
 H = - \sum_i \frac{\nabla_{\vec r_i}^2}{2} -
       \sum_{i,j} \frac{Z_i}{\|\vec R_i - \vec r_j \|} +
       \sum_{i, j>i} \frac{1}{\|\vec r_i - \vec r_j \|}
\end{align}
where $\vec R_i$ are the nuclear coordinates,
      $\vec r_j$ are the electron coordinates, and
      $Z_i$ are the nuclear atomic numbers.
Consider a basis set of $N$ spin-orbitals satisfying the following conditions:
\begin{enumerate}
\item each orbital takes significant values up to a distance at most logarithmic
      in $N$,
\item beyond that distance the orbital decays exponentially,
\item the maximum value of each orbital, and its first and second derivatives,
      scale at most logarithmically in $N$,
\item and the value of each orbital can be evaluated with complexity
      $\widetilde {\cal O}(1)$.
\end{enumerate}
Evolution under the Hamiltonian of \eq{electronic} can be simulated in this
basis for time $t$ within error $\epsilon>0$ with a gate count scaling as
$\widetilde{\cal O}(\eta^2 N^3 t)$, where $\eta$ is the number of electrons in
the molecule.
\end{theorem}

We note that these conditions will be satisfied for most, but not all, quantum
chemistry simulations.  To understand the limitations of these conditions, we
briefly discuss the concept of a model chemistry (i.e.\ standard basis set
specifications) and how model chemistries are typically selected for electronic
structure calculations.  There are thousands of papers which study the effectiveness of various basis sets
developed for the purpose of representing molecules \cite{Huzinaga85}.  These
model chemistries associate specific orbital basis functions with each atom in a
molecule.  For example, wherever Nitrogen appears in a molecule a model
chemistry would mandate that one add to the system certain basis functions which
are centered on Nitrogen and have been pre-optimized for Nitrogen chemistry;
different basis functions would be associated with each Phosphorus, and so on.
In addition to convenience, the use of standardized model chemistries helps
chemists to compare different calculations and reproduce results.

Within a standard model chemistry, orbital basis functions are almost always
represented as linear combinations of pre-fitted Gaussians which are centered on
each atom.  Examples of such model chemistries include Slater Type Orbitals
(e.g.\ STO-3G), Pople Basis Sets (e.g.\ 6-31G*) and correlation consistent basis
sets (e.g.\ cc-DVTZ).  We note that all previous studies on quantum algorithms
for quantum chemistry in an orbital basis have advocated the use of one of these
models.  Simulation within any of these model chemistries would satisfy the
conditions of our theorem because the basis functions associated with each atom
have maximum values, derivatives and distances beyond which each orbital decays
exponentially.

Similarly, when molecular instances grow because more atoms are added to the
system it is standard practice to perform these progressively larger
calculations using the same model chemistry and the conditions of Theorem 1 are
satisfied.  For instance, in a chemical series such as progressively larger
Hydrogen rings or progressively longer alkane chains or protein sequences, these
conditions would be satisfied.  We note that periodic systems such as conducting
metals might require basis sets (e.g.\ plane waves) violating the conditions of
Theorem 1.  When systems grow because atoms in the molecule are replaced with
heavier atoms, the orbitals do tend to grow in volume and their maximum values
might increase (even within a model chemistry).  However, there are only a
finite number of elements on the periodic table so this is irrelevant for
considerations of asymptotic complexity. Finally, we point out that these conditions do not hold if the simulation is
performed in the canonical molecular orbital basis, but this is not a problem
for our approach since the Hartree-Fock state can easily be prepared in the
atomic orbital basis at cost that is quadratic in the number of spin-orbitals.
We discuss this procedure further in \sec{encoding}.

The simulation procedure of Ref.~\cite{Berry2015} requires a decomposition of
the Hamiltonian into a weighted sum of unitary matrices.
In \cite{BabbushSparse1}, we decomposed the molecular Hamiltonian in such a way
that all the coefficients were integrals, i.e.
\begin{align}
\label{eq:unit_sum}
H = \sum_\ell W_\ell H_\ell  \quad \quad \quad \quad 
W_\ell = \int \! w_\ell \!\left(\vec z\right) \, \dd\vec z,
\end{align}
where the $H_\ell$ are unitary operators, and the $w_\ell \!\left(\vec z\right)$
are determined by the procedure.  We then showed how one could evolve under $H$
while simultaneously computing these integrals. In this paper, we investigate a different representation of the molecular
Hamiltonian with the related property that the Hamiltonian matrix elements
$H^{\alpha\beta}$ can be expressed as integrals,
\begin{equation}
\label{eq:int_H2}
H^{\alpha\beta} = \int \! \aleph^{\alpha\beta}(\vec z) \, \dd\vec z,
\end{equation}
or a sum of a limited number of integrals.  We decompose the Hamiltonian into a
sum of one-sparse Hamiltonians, each of which has only a single integral in its
matrix entries.  We then decompose the Hamiltonian by discretizing the
integrals and then further decompose the Hamiltonian into a sum of self-inverse
operators, ${\cal H}_{\ell,\rho}$. Using this decomposition, we construct a circuit called
$\textsc{select}({\cal H})$ which selects and applies the self-inverse operators so that
\begin{equation}
\label{eq:selectH}
\textsc{select}\left({\cal H}\right) \ket{\ell} \ket{\rho} \ket{\psi} =
\ket{\ell} \ket{\rho} {\cal H}_{\ell,\rho} \ket{\psi}.
\end{equation}
By repeatedly calling $\textsc{select}({\cal H})$, we are able to evolve under
$H$ with an exponential improvement in precision over Trotter-based algorithms.

The CI matrix is a compressed representation of the molecular Hamiltonian that
requires asymptotically fewer qubits than all second-quantized algorithms for
chemistry.  Though the CI matrix cannot be expressed as a sum of polynomially
many local Hamiltonians, a paper by Toloui and Love \cite{Toloui2013}
demonstrated that the CI matrix can be decomposed into a sum of ${\cal O}(N^4)$
1-sparse Hermitian operators, where $N$ is the number of spin-orbitals.
We provide in this paper a new decomposition of the CI matrix into a sum of
${\cal O}(\eta^2 N^2)$ 1-sparse Hermitian operators, where $\eta \ll N$ is the
number of electrons in the molecule.  This new decomposition enables our improved scaling.  Using techniques introduced in \cite{Berry2013}, we further decompose these 1-sparse operators into unitary operators which are
also self-inverse.  As a consequence of the self-inverse decomposition, the
Hamiltonian is an equally weighted sum of unitaries.
$\textsc{select}({\cal H})$ requires the ability to compute the entries of the
CI matrix; accordingly, we can use the same strategy for computing integrals
on-the-fly that was introduced in \cite{BabbushSparse1}, but this time our
Hamiltonian is of the form in \eq{int_H2}. 

Using this approach, the simulation of evolution over time $t$ then requires
$\widetilde{\cal O}(\eta^2 N^2 t)$ calls to $\textsc{select}({\cal H})$.  To implement
$\textsc{select}({\cal H})$, we make calls to the CI matrix
oracle as described in \sec{oracle}, which requires $\widetilde{\cal O}(N)$
gates.  This scaling is due to using a database approach to computing the
orbitals, where a sequence of $N$ controlled operations is performed.  This
causes our overall approach to require $\widetilde{\cal O}(\eta^2 N^3 t)$ gates.
As in \cite{Toloui2013}, the number of qubits is $\widetilde{\cal O}(\eta)$
rather than $\widetilde{\cal O}(N)$, because the compressed representation
stores only the indices of occupied orbitals, rather than occupation numbers of
all orbitals. To summarize, our algorithm with improved gate count scaling of
$\widetilde{\cal O}(\eta^2 N^3 t)$ proceeds as follows:
\begin{enumerate}
\item Represent the molecular Hamiltonian in \eq{electronic} in first
      quantization using the CI matrix formalism.  This requires selection of a
      spin-orbital basis set, chosen such that the conditions in
      \thm{maintheorem} are satisfied.
\item Decompose the Hamiltonian into sums of self-inverse matrices approximating
      the required molecular integrals \textit{via} the method of \sec{decomp1}.
\item Query the CI matrix oracle to evaluate the above self-inverse matrices,
      which we describe in \sec{oracle}.
\item Simulate the evolution of the system over time $t$ using the method of
      \cite{Berry2015}, which is summarized in \sec{simulation}.
\end{enumerate}

\section{The CI Matrix Encoding}
\label{sec:encoding}

The molecular electronic structure Hamiltonian describes electrons interacting in a nuclear potential that is fixed under the Born-Oppenheimer approximation. Except for the proposals in \cite{Kassal2008,Toloui2013,Whitfield2013b,Whitfield2015,Kivlichan2016}, all prior quantum algorithms for chemistry use second quantization. While in second quantization antisymmetry is enforced by the fermionic anti-commutation relations, in first quantization the wavefunction itself is explicitly antisymmetric.
The representation of \eq{electronic} in second quantization is
\begin{equation}
H = \sum_{ij} h_{ij} a_i^{\dagger}a_j + \frac{1}{2} \sum_{ijk\ell} h_{ijk\ell} a_i^{\dagger} a_j^{\dagger} a_k a_\ell
\label{eq:2nd}
\end{equation}
where the operators $a_i^\dagger$ and $a_j$ in \eq{2nd} obey antisymmetry due to the fermionic anti-commutation relations,
\begin{align}
\label{eq:anticomm}
 \{a_i^\dagger, a_j\} = \delta_{ij} \quad \quad \quad \quad \{a_i^\dagger, a_j^\dagger\} = \{a_i, a_j\} = 0.
\end{align}
The one-electron and two-electron integrals in \eq{2nd} are
\begin{align}
 &h_{ij} = \int \varphi_i^*(\vec r) \left(-\frac{\nabla^2}{2} - \sum_{q} \frac{Z_q}{ \|\vec R_q - \vec r \|} \right)\varphi_j(\vec r)  \,\dd\vec r  ,  \label{eq:single_int1}\\
 &h_{ijk\ell} = \int  \frac{ \varphi_i^*(\vec r_1) \, \varphi_j^*(\vec r_2) \, \varphi_\ell(\vec r_1) \, \varphi_k(\vec r_2) }{\|\vec r_1 - \vec r_2\|} \, \dd\vec r_1\, \dd\vec r_2. \label{eq:double_int}
\end{align}
where (throughout this paper), $\vec r_j$ represents the position of the $j^\textrm{th}$ electron, and $\varphi_i(\vec r_j)$ represents the $i^\textrm{th}$ spin-orbital when occupied by that electron. To ensure that the integrand in \eq{single_int1} is symmetric, we can alternatively write the integral for $h_{ij}$ as
\begin{equation}\label{eq:single_int}
 h_{ij} = \frac 12\int \nabla \varphi_i^*(\vec r) \cdot \nabla\varphi_j(\vec r)  \,\dd\vec r 
  - \int \sum_{q} \varphi^*_i(\vec r) \frac{Z_q}{ \|\vec R_q - \vec r \|}\varphi_j(\vec r) \,\dd\vec r.
\end{equation}
The second-quantized Hamiltonian in \eq{2nd} is straightforward to simulate because one can explicitly represent the fermionic operators as tensor products of Pauli operators, using either the Jordan-Wigner transformation \cite{Jordan1928,Somma2002} or the Bravyi-Kitaev transformation \cite{Bravyi2002,Seeley2012,Tranter2015}.

With the exception of real-space algorithms described in \cite{Kassal2008,Kivlichan2016}, all quantum algorithms for chemistry represent the system in a basis of $N$ single-particle spin-orbital functions, usually obtained as the solution to a classical mean-field treatment such as Hartree-Fock \cite{Helgaker2002}. However, the conditions of \thm{maintheorem} only hold when actually performing the simulation in the atomic orbital basis\footnote{The basis of atomic orbitals is not necessarily orthogonal. However, this can be fixed using the efficient Lowdin symmetric orthogonalization procedure which seeks the closest orthogonal basis \cite{Helgaker2002, McClean2014}.} (i.e.\ the basis prescribed by the model chemistry). The canonical Hartree-Fock orbitals are preferred over the atomic orbitals because initial states are easier to represent in the basis of Hartree-Fock orbitals. These orbitals are actually a unitary rotation of the orthogonalized atomic orbitals prescribed by the model chemistry. This unitary basis transformation takes the form
\begin{align}
\tilde{\varphi}_i & = \sum_{j=1}^N \varphi_j U_{ij}\\
U = e^{- \kappa}, \quad & \quad \kappa = - \kappa^\dagger = \sum_{ij} \kappa_{ij} a^\dagger_i a_j,
\end{align}
and $\kappa$ is anti-Hermitian.
For $\kappa$ and $U$, the quantities $\kappa_{ij}$ and $U_{ij}$ respectively correspond to the matrix elements of these operators in the basis of spin orbitals.
It is a consequence of the Thouless theorem that this unitary transformation is efficient to apply.

The canonical Hartree-Fock orbitals and $\kappa$ are obtained by performing a self-consistent field procedure to diagonalize a mean-field Hamiltonian for the system which is known as the Fock matrix. Because the Fock matrix describes a system of non-interacting electrons it can be expressed as the following $N$ by $N$ matrix:
\begin{equation}
\label{eq:fock}
f_{ij} = h_{ij} + \frac{1}{2} \sum_{k} h_{i k k j} - h_{i k j k}.
\end{equation}
The integrals which appear in the Fock matrix are defined by \eq{single_int1} and \eq{double_int}. Importantly, the canonical orbitals are defined to be the orbitals which diagonalize the Fock matrix. Thus, the integrals in the definition of the Fock matrix are defined in terms of the eigenvectors of the Fock matrix so \eq{fock} is a recursive definition. The canonical orbitals are obtained by repeatedly diagonalizing this matrix until convergence with its own eigenvectors. The Hartree-Fock procedure is important because the Hartree-Fock state (which is a product state in the canonical basis with the lowest $\eta$ eigenvectors of the Fock matrix occupied and the rest unoccupied) has particularly high overlap with the ground state of $H$.

As stated before, the conditions of \thm{maintheorem} do not apply if we represent the Hamiltonian in the basis of canonical orbitals. But this is not a problem for us because we can still prepare the Hartree-Fock state in the basis of orthgonalized atomic orbitals (which do satisfy the conditions) and then apply the operator $U = e^{- \kappa}$ to our initial state at cost $\widetilde{\cal O}(N^2)$. Note that the use of a local basis has other advantages, as pointed out in  \cite{McClean2014}. In particular, in the limit of certain large molecules, use of a local basis allows one to truncate terms from the Hamiltonian so that there are $\widetilde{\cal O}(N^2)$  terms instead of ${\cal O}(N^4)$ terms. However, \thm{maintheorem} exploits an entirely different property of basis locality which does not require any approximation from truncating terms.

The spatial encoding of \eq{2nd} requires $\Theta(N)$ qubits, one for each spin-orbital; under the Jordan-Wigner transformation, the state of each qubit indicates the occupation of a corresponding spin-orbital. Many states representable in second quantization are inaccessible to molecular systems due to symmetries in the Hamiltonian. For instance, molecular wavefunctions are eigenstates of the total spin operator so the total angular momentum is a good quantum number, and this insight can be used to find a more efficient spatial encoding \cite{Whitfield2013b,Whitfield2015}. Similarly, the Hamiltonian in \eq{2nd} commutes with the number operator, $\nu$, whose expectation value gives the number of electrons, $\eta$,
\begin{equation}
\nu = \sum_{i=1}^N a_i^{\dagger} a_i, \quad \quad \left[H, \nu\right] = 0, \quad \quad \eta = \avg{\nu}.
\end{equation}
Following the procedure in \cite{Toloui2013}, our algorithm makes use of an encoding which reduces the number of qubits required by recognizing $\eta$ as a good quantum number.

Conservation of particle number implies there are only $\xi = \binom{N}{\eta}$ valid configurations of these electrons, but the second-quantized Hilbert space has dimension $2^N$, which is exponentially larger than $\xi$ for fixed $\eta$.
We work in the basis of Slater determinants, which are explicitly antisymmetric functions of both space and spin associated with a particular $\eta$-electron configuration.
We denote these states as $\ket{\alpha} =  \ket{\alpha_0, \alpha_1, \cdots, \alpha_{\eta-1}}$, where $\alpha_i \in \{1,\ldots, N\}$ and $\alpha \in \{1,\ldots, N^{\eta}\}$.
We emphasize that $ \alpha_i$ is merely an integer which indexes a particular spin-orbital function $ \varphi_{\alpha_i}(\vec r)$. While each configuration requires a specification of $\eta$ occupied spin-orbitals, there is no sense in which $\alpha_i$ is associated with ``electron $i$'' since fermions are indistinguishable. Specifically,
\begin{equation}
\label{eq:det}
 \braket{\vec r_0,\ldots,\vec r_{\eta-1}}{\alpha} = \braket{\vec r_0,\ldots,\vec r_{\eta-1}}{\alpha_0,  \alpha_1, \cdots, \alpha_{\eta-1}}
= \frac{1}{\sqrt{\eta!}}
\begin{vmatrix}
\varphi_{ \alpha_0}\!\left(\vec r_0\right) & \varphi_{ \alpha_1}\!\left(\vec r_0\right) & \cdots & \varphi_{\alpha_{\eta-1}} \!\left(\vec r_0\right) \\
\varphi_{\alpha_0}\!\left(\vec r_1\right) & \varphi_{ \alpha_1}\!\left(\vec r_1\right) & \cdots & \varphi_{ \alpha_{\eta-1}} \!\left(\vec r_1\right) \\
\vdots & \vdots & \ddots & \vdots\\
\varphi_{\alpha_0}\!\left(\vec r_{\eta-1}\right) & \varphi_{ \alpha_1}\!\left(\vec r_{\eta-1}\right) & \cdots & \varphi_{ \alpha_{\eta-1}} \!\left(\vec r_{\eta-1}\right) \end{vmatrix}
\end{equation}
where the bars enclosing the matrix in \eq{det} denote a determinant. Because determinants have the property that they are antisymmetric under exchange of any two rows, this construction ensures that our wavefunction obeys the Pauli exclusion principle. We note that although this determinant can be written equivalently in different orders (e.g.~by swapping any two pairs of orbital indices), we avoid this ambiguity by requiring the Slater determinants to only be written in ascending order of spin-orbital indices.

The representation of the wavefunction introduced in \cite{Toloui2013} uses $\eta$ distinct registers to encode the occupied set of spin-orbitals, thus requiring $\Theta(\eta \log N) = \widetilde{\cal O}(\eta)$ qubits. However, it would be possible to use a further-compressed representation of the wavefunction based on the direct enumeration of all Slater determinants, requiring only $\Theta(\log \xi)$ qubits. When using very small basis sets (such as the minimal basis), it will occasionally be the case that the spatial overhead of $\Theta(N)$ for the second-quantized algorithm is actually less than the spatial complexity of our algorithm. However, for a fixed $\eta$, the CI matrix encoding requires exponentially fewer qubits.

\section{The CI Matrix Decomposition}
\label{sec:decomp1}

The molecular Hamiltonian expressed in the basis of Slater determinants is known to chemists as the CI matrix. Elements of the CI matrix are computed according to the Slater-Condon rules \cite{Helgaker2002}, which we will express in terms of the one-electron and two-electron integrals in \eq{single_int1} and \eq{double_int}. In order to motivate our 1-sparse decomposition, we state the Slater-Condon rules for computing the matrix element
\begin{equation}
H^{\alpha\beta} = \bra{\alpha} H \ket{\beta}
\end{equation}
by considering the spin-orbitals which differ between the determinants $\ket{\alpha}$ and $\ket{\beta}$ \cite{Helgaker2002}:
\begin{enumerate}
\item If $\ket{\alpha}$ and $\ket{\beta}$ contain the same spin-orbitals $\{\chi_i\}_{i=1}^{\eta}$ then we have a diagonal element
\begin{equation}
\label{eq:slate1}
H^{\alpha\beta} = \sum_{i=1}^{\eta} h_{\chi_i\chi_i} + \sum_{i = 1}^{\eta - 1}\sum_{j = i + 1}^{\eta} \left(h_{\chi_i\chi_j \chi_i \chi_j} - h_{\chi_i \chi_j \chi_j \chi_i}\right).
\end{equation}
\item If $\ket{\alpha}$ and $\ket{\beta}$ differ by exactly one spin-orbital such that $\ket{\alpha}$ contains spin-orbital $k$ where $\ket{\beta}$ contains spin-orbital $\ell$, but otherwise contain the same spin-orbitals $\{\chi_i\}_{i=1}^{\eta - 1}$, then
\begin{equation}
\label{eq:slate2}
H^{\alpha\beta} = h_{k\ell} + \sum_{i=1}^{\eta-1} \left(h_{k\chi_i \ell \chi_i} - h_{k \chi_i \chi_i \ell}\right).
\end{equation}
\item If $\ket{\alpha}$ and $\ket{\beta}$ differ by exactly two spin-orbitals such that occupied spin-orbital $i$ in $\ket{\alpha}$ is replaced with spin-orbital $k$ in $\ket{\beta}$, and occupied spin-orbital $j$ in $\ket{\alpha}$ is replaced with spin-orbital $\ell$ in $\ket{\beta}$, then
\begin{equation}
\label{eq:slate3}
H^{\alpha\beta} = h_{ijk\ell} - h_{ij\ell k}.
\end{equation}
\item If $\ket{\alpha}$ and $\ket{\beta}$ differ by more than two spin-orbitals,
\begin{equation}
\label{eq:slate4}
H^{\alpha\beta} = 0.
\end{equation}
\end{enumerate}

These rules assume that $\alpha$ and $\beta$ have the list of occupied orbitals given in a corresponding order, so all corresponding occupied orbitals are listed in the same positions.
In contrast, we will be giving the lists of occupied orbitals in ascending order.
In order to use the rules, we therefore need to change the order of the list of occupied orbitals.
In changing the order of the occupied orbitals, there is a sign flip on the state for an odd permutation.
This sign flip needs to be included when using the above rules.

In general, there is no efficient way to decompose the CI matrix into a polynomial number of tensor products of Pauli operators. It is thus inefficient to directly simulate this Hamiltonian in the same fashion with which we simulate local Hamiltonians. However, the CI matrix is sparse and there exist techniques for simulating arbitrary sparse Hamiltonians. A $d$-sparse matrix is one which contains at most $d$ nonzero elements in each row and column. As discussed in \cite{Toloui2013,Wecker2014}, the Slater-Condon rules imply that the sparsity of the CI matrix is
\begin{equation}
d  = \binom{\eta}{2}\binom{N - \eta}{2} + \binom{\eta}{1}\binom{N - \eta}{1} + 1
 = \frac{\eta^4}{4} - \frac{\eta^3 N}{2} + \frac{\eta^2 N^2}{2} + {\cal O}\left(\eta^2N+\eta N^2\right).
\end{equation}
Because $N$ is always greater than $\eta$, we find that the CI matrix is $d$-sparse where $d \in {\cal O}(\eta^2 N^2)$. This should be compared with the second-quantized Hamiltonian which is also $d$-sparse, but where $d\in{\cal O}(N^4)$. Our strategy here parallels the second-quantized decomposition, but works with the first-quantized wavefunction. This decomposition is explained in four steps, as follows.
\begin{enumerate}
\item[A.] Decompose the molecular Hamiltonian into ${\cal O}(\eta^2 N^2)$ 1-sparse matrices.
\item[B.] Further decompose each of these 1-sparse matrices into 1-sparse matrices with entries proportional to a  sum of a  constant number of molecular integrals.
\item[C.] Decompose those 1-sparse matrices into sums approximating the integrals in Eqs.~\eqref{eq:double_int} and \eqref{eq:single_int}.
\item[D.] Decompose the integrands from those integrals into sums of self-inverse matrices.
\end{enumerate}

\subsection{Decomposition into 1-sparse matrices}
\label{sec:dec1}

In order to decompose the molecular Hamiltonian into 1-sparse matrices, we require a unique and reversible graph coloring between nodes (Slater determinants). We introduce such a graph coloring here, with the details of its construction and proof of its properties given in \app{1sparseproof}. The graph coloring can be summarized as follows.

\begin{enumerate}
	\item Perform the simulation under $\sigma_x \otimes H$, where $\sigma_x$ is the Pauli $x$ matrix, in order to create a bipartite Hamiltonian of the same sparsity as $H$.
	\item Label the left nodes $\alpha$ and the right nodes $\beta$. We seek a procedure to take $\alpha$ to $\beta$, or vice versa, with as little additional information as possible, and without redundancy or ambiguity.
	\item Provide an 8-tuple $\gamma = (a_1,b_1,i,p,a_2,b_2,j,q)$ which determines the coloring. The coloring must uniquely determine $\alpha$ given $\beta$ or vice versa. Using the 8-tuples, proceed via either Case 1, 2, 3, or 4 in \app{1sparseproof} to determine the other set of spin-orbitals, using an intermediate list of orbitals $\chi$. The 4-tuples $(a_1,b_1,i,p)$ and $(a_2,b_2,j,q)$ each define a differing orbital. For a single difference, we can set $p = 0$, and for no differences, we can set $p=q=0$.
\end{enumerate}

The basic idea is that we give the positions $i$ and $j$ of those orbitals which differ in $\alpha$, as well as by how much the occupied orbital indices shift, which we denote by $p$ and $q$. This allows us to determine $\beta$ from $\alpha$. However, it does not allow us to unambiguously determine $\alpha$ from $\beta$. To explain how to resolve this ambiguity, we consider the case of a single differing orbital. We will denote by $i$ the position of the differing orbital in $\alpha$, and by $k$ the position of the differing orbital in $\beta$.

\begin{figure}
\centering
\subfloat[][]{\includegraphics[width=0.4\textwidth,trim={0in 4.75in 6in -0.5in},clip]{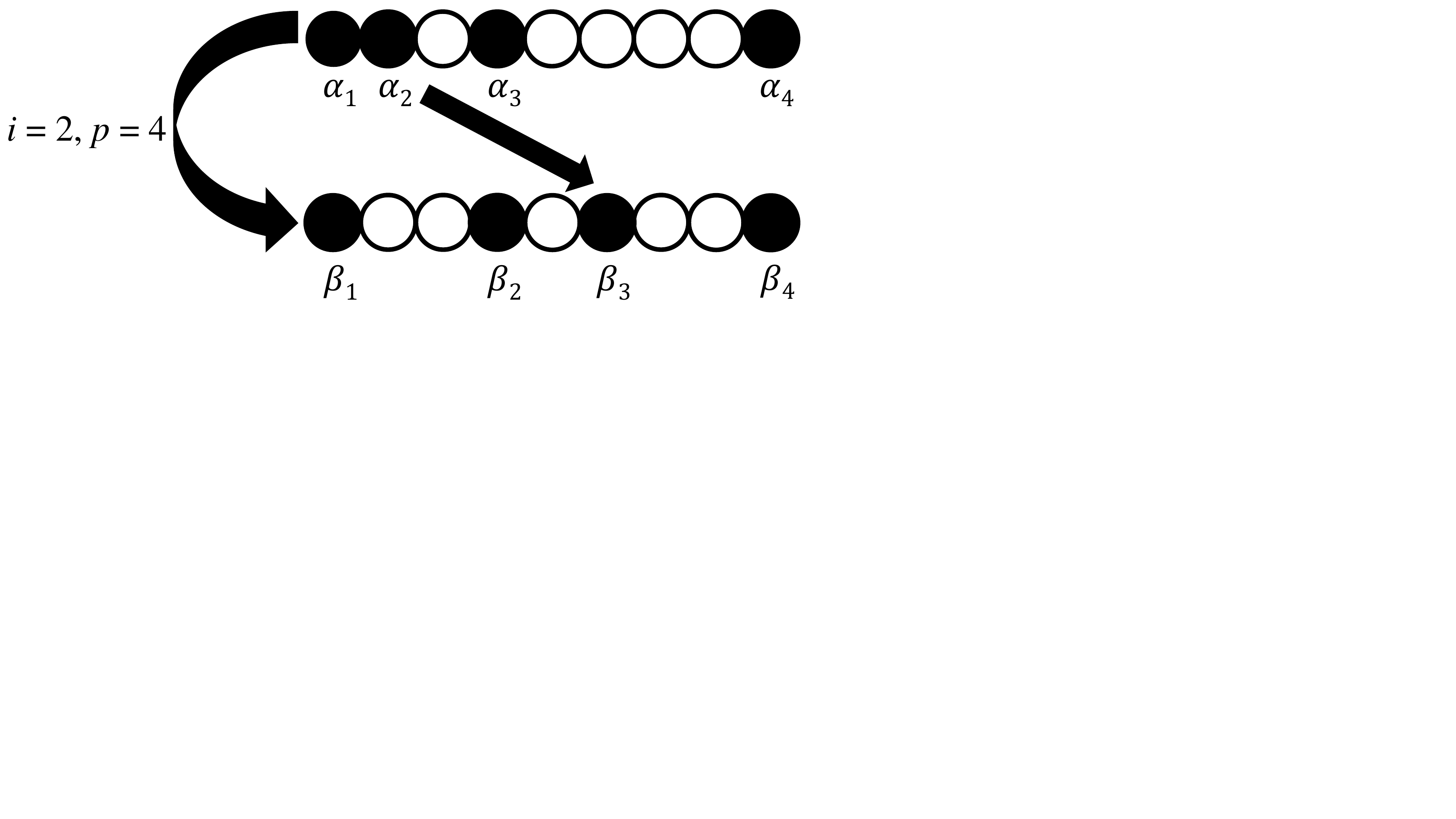} \label{fig:subfigA}} \\
\subfloat[][]{\includegraphics[width=0.4\textwidth,trim={0in 4.75in 6in 0in},clip]{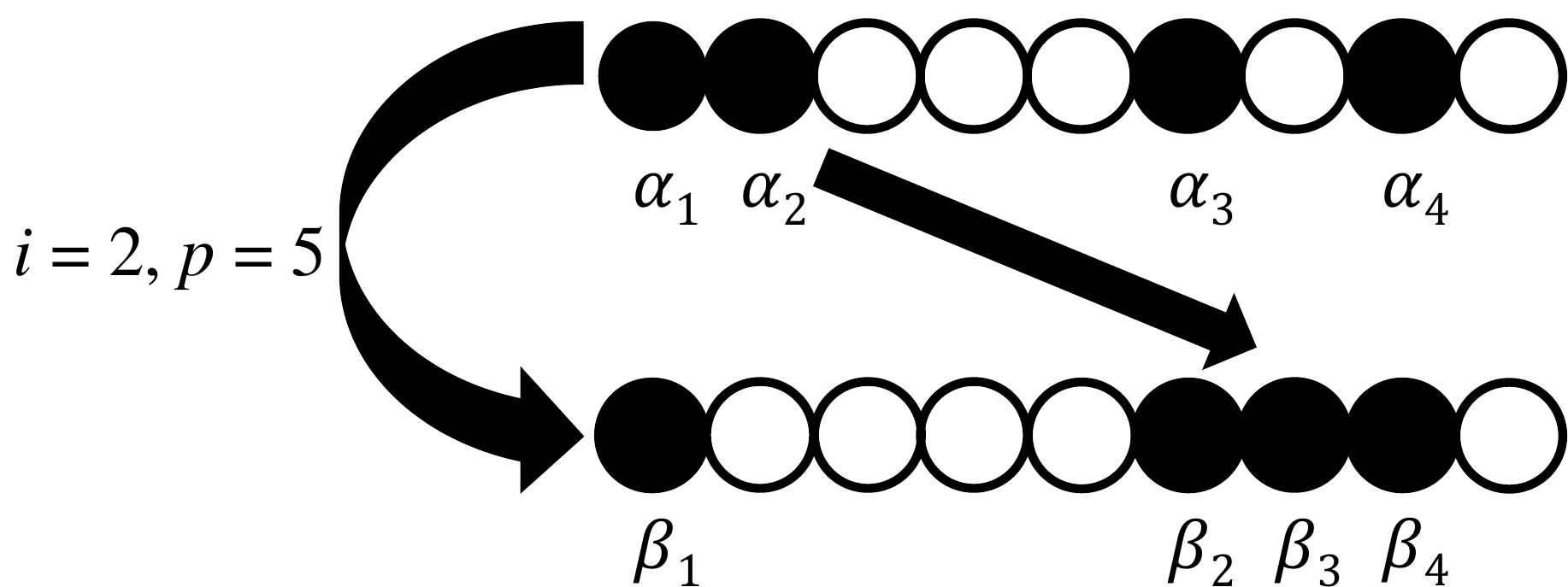} \label{fig:subfigB}}
\caption{Example of the 1-sparse coloring, where $i$ is the position of the occupied orbital in $\alpha$ that must be moved. \protect\subref{fig:subfigA} $i = 2$, $p = 4$ is sufficient to determine $\beta$ from $\alpha$, as well as to determine $\alpha$ from $\beta$. \protect\subref{fig:subfigB} $i = 2$, $p = 5$ is sufficient to determine $\beta$ from $\alpha$, but not the reverse: subtracting $p=5$ from $\beta_2$, $\beta_3$, or $\beta_4$ all give different valid values for $\alpha_i = \alpha_2$. The spacing condition means that we would need to give the position of the occupied orbital for $\beta$ instead.}
\end{figure}

Consider the example in \hyperref[fig:subfigA]{Figure~1(\protect\subref*{fig:subfigA})}: given $i$ which is the position in $\alpha$, the position $k$ in $\beta$ can be immediately determined.
But given $\beta$, multiple potential positions of occupied orbitals would need to be tested to see if they put the occupied orbital in position $i=2$ in $\alpha$.
In this case, given $\beta$ there is only one orbital which can be shifted to position 2 for $\alpha$ so the position in $\beta$ is unambiguous.
Now consider \hyperref[fig:subfigB]{Figure~1(\protect\subref*{fig:subfigB})}: multiple positions in $\beta$ could lead to position 2 in $\alpha$.

The difference between the two cases is that in \hyperref[fig:subfigA]{Figure~1(\protect\subref*{fig:subfigA})} there is a larger spacing between orbitals for $\beta$, whereas in \hyperref[fig:subfigB]{Figure~1(\protect\subref*{fig:subfigB})} there is a larger spacing for $\alpha$.
More specifically, for \hyperref[fig:subfigA]{Figure~1(\protect\subref*{fig:subfigA})} the spacing between $\alpha_1$ and $\alpha_3$ is $3$, whereas the spacing between $\beta_2$ and $\beta_4$ is larger at $5$.
For \hyperref[fig:subfigB]{Figure~1(\protect\subref*{fig:subfigB})} the spacing between $\alpha_1$ and $\alpha_3$ is $5$, whereas the spacing between $\beta_2$ and $\beta_4$ is smaller at $2$.
It is the spacing between the occupied orbitals adjacent to the one that is moved that should be compared.

For the situation in \hyperref[fig:subfigB]{Figure~1(\protect\subref*{fig:subfigB})},
rather than specifying the position in $\alpha$ we should specify the position in $\beta$ to resolve the ambiguity.
The bit $a$ determines whether we are specifying the position in $\alpha$ or in $\beta$; this is done depending on the relative spacing of the adjacent occupied orbitals in the two.
However, this spacing condition does not completely resolve the ambiguity: there are potentially two different choices for the occupied orbital.
The choice is made by the bit $b$. The coloring for the two differing orbitals is done by doing this twice with an intermediate list of occupied orbitals $\chi$.

There are ${\cal O}(\eta^2 N^2)$ possible colors: there are two possible choices of each of the bits $a_1$, $a_2$, $b_1$, and $b_2$, $\eta$ choices each of $i$ and $j$, and $N$ choices each of $p$ and $q$.

\subsection{Decomposition into $h_{ij}$ and $h_{ijk\ell}$}
\label{sec:dec2}

Each 1-sparse matrix from \sec{dec1} is associated with some 8-tuple $\gamma = (a_1,b_1,i,p,a_2,b_2,j,q)$.
However, without further modification, some of these 1-sparse matrices have entries given by a sum over a number of molecular integrals that grows with $\eta$, namely the diagonal terms as in \eq{slate1}, and the single-orbital terms as in \eq{slate2}. Here, we further decompose those matrices into a sum of 1-sparse matrices $H_{\gamma}$, which have entries proportional to the sum of a constant number of molecular integrals, in order to remove this changing upper bound.

We want to have a new set of 1-sparse matrices, each with entries corresponding to a single term in the sum over molecular integrals. To be more specific, the combinations of $\gamma$ correspond to terms in \eq{slate1} to \eq{slate3} as follows.
\begin{enumerate}
\item
If $p=q=0$, this indicates that we have a diagonal 1-sparse matrix.
In \eq{slate1}, the entries on the diagonal would be a sum of ${\cal O}(\eta^2)$ terms.
As we have freedom in how to use $i$ and $j$, we use these to give terms in the sum.
When $i=j$ for $p=q=0$, we take the 1-sparse matrix to have diagonal elements given by $h_{i i}$.
If $i < j$ for $p = q = 0$ we take the 1-sparse matrix to have diagonal entries $h_{\chi_i\chi_j \chi_i \chi_j} - h_{\chi_i \chi_j \chi_j \chi_i}$.
We do not allow tuples $\gamma$ such that $i>j$ for $p=q=0$ (alternatively we could just give zero in this case).
The overall result is that the sum over $i$ and $j$ for the 1-sparse matrices for $\gamma$ with $p=q=0$ yields the desired sum in \eq{slate1}.

\item
Next, if $p = 0$ and $q \neq 0$, then this indicates that we have a 1-sparse matrix with entries where $\alpha$ and $\beta$ differ by only one spin-orbital.
According to \eq{slate2}, each entry would normally be a sum of ${\cal O}(\eta)$ terms.
Instead, when $p = 0$ and $q \neq 0$, we use the value of $i$ to index terms in the sum in \eq{slate2}, though we only yield a nonzero result when $i$ is in the Slater determinant.
In particular, the 1-sparse matrix has entries $h_{k \chi_i \ell \chi_i} - h_{k \chi_i \chi_i \ell}$.
We allow an additional value of $i$ to indicate a 1-sparse matrix with entries $h_{k \ell}$.
Then the sum over 1-sparse matrices for different values of $i$ gives the desired sum \eq{slate2}.
We do not allow $\gamma$ such that $q=0$ but $p\neq 0$.

\item
Finally, if both $p$ and $q$ are nonzero, then we have a 1-sparse matrix with entries where $\alpha$ and $\beta$ differ by two orbitals.
In this case, there is no sum in \eq{slate3}, so there is no additional decomposition needed.
\end{enumerate}

Combining these three steps we find that the decomposition into 1-sparse matrices $H_\gamma$ can be achieved with the indices $(a_1,b_1,i,p,a_2,b_2,j,q)$. Thus, there are ${\cal O}(\eta^2 N^2)$ terms without any redundancies. Note that sorting of the spin-orbital indices requires only $\widetilde{\cal O}(\eta)$ gates, which is less than the number of complexity of evaluating the spin-orbitals. In the following sections, we denote the total number of terms given by the above decomposition by $\Gamma$, and the sum over $H_\gamma$ yields the complete CI matrix,
\begin{equation}
\label{eq:gamma_decomp}
H = \sum_{\gamma=1}^{\Gamma} H_\gamma.
\end{equation}

\subsection{Discretizing the integrals}
\label{sec:Riemann}
Next we consider discretization of the integrals for $h_{ij}$ and $h_{ijk\ell}$.
In \cite{Berry2015} it is shown how to simulate Hamiltonian evolution with an exponential improvement in the scaling with $1/\epsilon$, as compared to methods based on Trotter formulas.
In this approach, the time-ordered exponential for the evolution operator is approximated by a Taylor series up to an order $K$.
The time $t$ is broken into $r$ segments, and the integrals are discretized in the following way on each segment:
\begin{align}
{\cal T} \exp\left[-i \int_{0}^{t/r} \!\! H(t) \, d t \right] 
&\approx \sum_{k=0}^K \frac{(-i)^k}{k!}\int_{0}^{t/r} {\cal T} H(t_k)\dots H(t_1)\,  \dd\bs t \nn
&\approx \sum_{k=0}^K \frac{(-it/r)^k}{\mu^k k!}\sum_{j_1,\ldots,j_k=0}^{\mu-1} H(t_{j_k})\dots H(t_{j_1}),
\end{align}
where ${\cal T}$ is the time-ordering operator.
In our case the Hamiltonian does not change in time, so the time-ordering is unimportant.

The Hamiltonian is expanded as a sum of $H_\gamma$ as in \eq{gamma_decomp}, and each of those terms has matrix entries that
can be given in the form of an integral as
\begin{align}
H^{\alpha\beta}_\gamma = \int \cH^{\alpha\beta}_\gamma(\vec z) \, \dd \vec z \, .
\end{align}
In cases where $H^{\alpha\beta}_\gamma$ corresponds to $h_{ij}$, the integral is over a three-dimensional region, and where $H^{\alpha\beta}_\gamma$ corresponds to $h_{ijk\ell}$ the integral is over a six-dimensional region, so $\vec z$ represents six parameters.

Ideally, each integral can be truncated to a finite domain $D$ with volume ${\cal V}$.
Using a set of grid points $\vec z_{\rho}$, we can approximate the integral by
\begin{align}
\label{eq:double}
H^{\alpha\beta}_\gamma & \approx \int_D \cH^{\alpha\beta}_\gamma(\vec z) \, \dd \vec z \approx \frac{\cal V}{\mu} \sum_{\rho=1}^{\mu} \aleph^{\alpha\beta}_\gamma(\vec z_{\rho}) \, .
\end{align}
The complexity will then be logarithmic in the number of points in the sum, $\mu$, and linear in the volume times the maximum value of the integrand.

In practice the situation is more complicated than this.
That is because the integrals are all different.
As well as the dimensionality of the integrals (three for $h_{ij}$ and six for $h_{ijk\ell}$), there will be differences in the regions that the integrals will be over, as well as some integrals being in spherical polar coordinates.
To account for these differences, it is better to write the discretized integral in the form
\begin{equation}
\label{eq:double2}
H^{\alpha\beta}_\gamma  \approx \sum_{\rho=1}^{\mu} \aleph^{\alpha\beta}_{\gamma,\rho} \, .
\end{equation}
The Hamiltonian $H_\gamma$ can then be written as the sum
\begin{equation}
\label{eq:double3}
H_\gamma \approx \sum_{\rho=1}^{\mu} \cH_{\gamma,\rho} \, .
\end{equation}

As discussed in \cite{BabbushSparse1}, the discretization is possible because the integrands can be chosen to decay exponentially \cite{Helgaker2002}.
The required properties of the orbitals are given in \thm{maintheorem}.
Here we present a more precise formulation of the required properties, and provide specific results on the number of terms needed.
We make the following three assumptions about the spin-orbitals $\varphi_\ell$.
\begin{enumerate}
	\item		There exists a positive real number $\varphi_\text{max}$ such that,
  				for all spin-orbital indices $\ell$ and for all $\vec{r} \in \R^3$,
          \begin{equation}
          \label{eq:spin-orbital_bound}
            \abs{\varphi_\ell (\vec{r})} \leq \varphi_\text{max}.
          \end{equation}
	\item		For each spin-orbital index $\ell$, there exists a vector
	  			$\vec{c}_\ell \in \R^3$ (called the center of $\varphi_\ell$)
	  			and a positive real number $x_\text{max}$ such that, whenever
	  			$\abs{\vec{r} - \vec{c}_\ell} \geq x_\text{max}$ for some
					$\vec{r} \in \R^3$,
	  			\begin{equation}
          \label{eq:spin-orbital_decay}
	  				\abs{\varphi_\ell (\vec{r})} \leq \varphi_\text{max} \exp \left(
	  					-\frac{\alpha}{x_\text{max}} \norm{\vec{r}-\vec{c}_\ell}
	  				\right),
	  			\end{equation}
	  			where $\alpha$ is some positive real constant.
	\item		For each spin-orbital index $\ell$, $\varphi_\ell$ is
					twice-differentiable and there exist positive real constants
					$\gamma_1$ and $\gamma_2$ such that
					\begin{equation}
          \label{eq:spin-orbital_first_derivative_bound}
						\norm{\nabla \varphi_\ell (\vec{r})}
						\leq  \gamma_1 \frac{\varphi_\text{max}}{x_\text{max}}
					\end{equation}
					and
					\begin{equation}
          \label{eq:spin-orbital_second_derivative_bound}
						\abs{\nabla^2 \varphi_\ell (\vec{r})}
						\leq  \gamma_2 \frac{\varphi_\text{max}}{x_\text{max}^2}
					\end{equation}
					for all $\vec{r} \in \R^3$.
\end{enumerate}
Note that $\alpha$, $\gamma_1$ and $\gamma_2$ are dimensionless constants,
whereas $x_{\max}$ has units of distance, and
$\varphi_\text{max}$ has the same units as $\varphi_\ell$.
The conditions of \thm{maintheorem} mean that $\varphi_\text{max}$ and $x_\text{max}$ grow at most
logarithmically with the number of spin-orbitals.
Note that we use $x_{\max}$ in a different way than in \cite{BabbushSparse1}, where it was the size of the cutoff on the region of integrals, satisfying $x_{\max}={\cal O}(\log(Nt/\epsilon))$.
Here we take $x_{\max}$ to be the size scale of the orbitals independent of $t$ or $\epsilon$, and the cutoff will be a multiple of $x_{\max}$.
We also assume that
$x_{\max}$ is bounded below by a constant, so
the first and second derivatives of the spin-orbitals grow no more than
logarithmically as a function of the number of spin-orbitals.

We next define notation used for the integrals for $h_{ij}$ and $h_{ijk\ell}$.
These integrals are
\begin{equation}
	S_{ij}^{(0)}\! \left( D_0 \right)
	:=  -\frac{1}{2} \int_{D_0}
			\varphi_i^* (\vec{r}) \nabla^2 \varphi_j (\vec{r}) \dd{\vec{r}},
\end{equation}
\begin{equation}
	S_{ij}^{(1,\,q)}\! \left( D_{1,q} \right)
	:=  -Z_q \int_{D_{1,q}} \frac{
				\varphi_i^* (\vec{r}) \, \varphi_j (\vec{r})
			}{
				\|\vec{R}_q - \vec{r}\|
			} \dd{\vec{r}},
\end{equation}
and
\begin{equation}
	S_{ijk\ell}^{(2)}\! \left( D_2 \right)
	:=	\int_{D_2}
			\frac{
				\varphi_i^*\!\left(\vec{r}_1\right) \varphi_j^*\!\left(\vec{r}_2\right)
				\varphi_k\!\left(\vec{r}_2\right) \varphi_\ell\!\left(\vec{r}_1\right)
			}{ \|\vec{r}_1 - \vec{r}_2\| }
			\dd{\vec{r}_1} \dd{\vec{r}_2},
\end{equation}
for any choices of $D_0, D_{1,q} \subseteq \R^3$ and $D_2 \subseteq \R^6$.
Thus
\begin{equation}
	h_{ij}		=		S_{ij}^{(0)}\! \left( \R^3 \right) +
								\sum_q S_{ij}^{(1,\,q)}\! \left( \R^3 \right)
\end{equation}
and
\begin{equation}
	h_{ijk\ell}	=		S_{ijk\ell}^{(2)}\! \left( \R^6 \right).
\end{equation}
Using the assumptions on the properties of the orbitals, we can bound the number of terms needed in a
Riemann sum that approximates each integral to within a specified accuracy, $\inter$ (which is distinct from the accuracy of the overall simulation, $\epsilon$).
These bounds are summarized in the following three lemmas.

\begin{lemma}
\label{lem:int0}
  Let $\inter$ be any real number that satisfies
  \begin{equation}
  \label{eq:sensible0}
    0 < \inter \leq e^{-\alpha/2} K_0 \varphi_\text{max}^2 x_\text{max}\, ,
  \end{equation}
  where
  \begin{equation}
    K_0
    :=		\frac{26 \gamma_1}{\alpha^2} +
          \frac{8\pi\gamma_2}{\alpha^3} +
          32\sqrt{3} \gamma_1 \gamma_2 \, .
  \end{equation}
  Then $S_{ij}^{(0)}\! \left( \R^3 \right)$ can be
	approximated to within error $\inter$ using a Riemann sum with
  \begin{equation}
  \label{eq:lem1mu}
  \mu \le  \left\lceil
      \frac{K_0 \varphi_\text{max}^2 x_\text{max}}{\inter}
      \left[
        \frac{2}{\alpha} \log \left(
          \frac{K_0 \varphi_\text{max}^2 x_\text{max}}{\inter}
        \right)
      \right]^4
    \right\rceil^3
  \end{equation}
  terms, where
  the terms in the sum have absolute value no larger than
  \begin{equation}
  \label{eq:lem1bnd}
    \frac{1}{\mu} \times 32 \frac{\gamma_1^2}{\alpha^3}
    \varphi_\text{max}^2 x_\text{max} \left[
      \log \left(
        \frac{K_0 \varphi_\text{max}^2 x_\text{max}}{\inter}
      \right)
    \right]^3.
  \end{equation}
\end{lemma}

\begin{lemma}
\label{lem:int1}
  Let $\inter$ be any real number that satisfies
  \begin{equation}
  \label{eq:sensible1}
    0 < \inter \leq e^{-\alpha/2} K_1 Z_q \varphi_\text{max}^2 x_\text{max}^2 \, ,
  \end{equation}
  where
  \begin{equation}
	  K_1
	  :=    \frac{8\pi^2}{\alpha^3}\left( \alpha+2 \right) +
	        1121 \left( 8 \gamma_1 + \sqrt{2} \right).
	\end{equation}
  Then $S_{ij}^{(1,q)}\! \left( \R^3 \right)$ can be approximated to within
  error $\inter$ using a Riemann sum with
	\begin{equation}\label{eq:lem2mu}
	\mu \le	\left\lceil
			\frac{K_1 Z_q \varphi_\text{max}^2 x_\text{max}^2}{\inter}
				\left[
					\frac{2}{\alpha} \log \left(
						\frac{K_1 Z_q \varphi_\text{max}^2 x_\text{max}^2}{\inter}
					\right)
				\right]^4
			\right\rceil^3
	\end{equation}
  terms, where the terms in the sum have absolute value no larger than
  \begin{equation}\label{eq:lem2bnd}
  \frac{1}{\mu} \times \frac{256\pi^2}{\alpha^3}
  Z_q \varphi_\text{max}^2 x_\text{max}^2 \left[
  \log \left(
  				\frac{K_1 Z_q \varphi_\text{max}^2 x_\text{max}^2}{\inter}
  			\right)
  		\right]^3.
  \end{equation}
\end{lemma}

\begin{lemma}
\label{lem:int2}
  Let $\inter$ be any real number that satisfies
  \begin{equation}
  \label{eq:sensible2}
    0 < \inter \leq e^{-\alpha} K_2 \varphi_\text{max}^4 x_\text{max}^5 \, ,
  \end{equation}
  where
  \begin{equation}
    K_2
    :=      \frac{128\pi}{\alpha^6}(\alpha+2) +
            2161 \pi^2 \left( 20 \gamma_1 + \sqrt{2} \right).
  \end{equation}
  Then $S_{ijk\ell}^{(2)} \! \left( \R^6 \right)$ can be approximated to within
  error $\inter$ using a Riemann sum with
	\begin{equation}\label{eq:lem3mu}
	\mu \le	\left\lceil
			\frac{K_2 \varphi_\text{max}^4 x_\text{max}^5}{\inter}
      \left[
        \frac{1}{\alpha} \log \left(
          \frac{K_2 \varphi_\text{max}^4 x_\text{max}^5}{\inter}
        \right)
      \right]^7
    \right\rceil^6
	\end{equation}
	terms, where the terms in the sum have absolute value no larger than
  \begin{equation}\label{eq:lem3bnd}
    \frac{1}{\mu} \times
    \frac{672\pi^2}{\alpha^6} \varphi_\text{max}^4 x_\text{max}^5 \left[
      \log \left(
        \frac{K_2 \varphi_\text{max}^4 x_\text{max}^5}{\inter}
      \right)
    \right]^6.
  \end{equation}
\end{lemma}

The conditions in Eqs.~\eqref{eq:sensible0}, \eqref{eq:sensible1} and
\eqref{eq:sensible2} are just used to ensure that we are considering a
reasonable combination of parameters, and not for example a very large
allowable error $\inter$ or a small value of $x_{\max}$.
We prove these Lemmas in \app{integral_discretization}. Specifically, we prove
\lem{int0} in \app{integral_discretization/int0_proof},
\lem{int1} in \app{integral_discretization/int1_proof} and
\lem{int2} in \app{integral_discretization/int2_proof}.
In discretizing these integrals it is important that the integrands are
Hermitian, because we need ${\cal H}_{\gamma,\rho}$ to be Hermitian.
The integrands of these integrals are not Hermitian as discretized in the way given in the proofs in \app{integral_discretization}.
This is because the regions of integration are chosen in a non-symmetric way.
For example, the region of integration for $S_{ij}^{(0)}$ is chosen centered on the orbital $\varphi_i$, so the integrand is not symmetric.
It is simple to symmetrize the integrands, however.
For example, for $S_{ij}^{(0)}$ we can add $(S_{ji}^{(0)})^*$ and divide by two.
That ensures that the integrand is symmetric, with just a factor of two overhead in the number of terms in the sum.

As a consequence of these Lemmas, we see that the terms of any Riemann sum
approximation to one of the integrals that define the Hamiltonian coefficients
$h_{ij}$ and $h_{ijk\ell}$ have absolute values bounded by
\begin{equation}
  \mathcal{O} \left(
    \frac{\varphi_\text{max}^4 x_\text{max}^5}{\mu}
    \left[
      \log \left(
        \frac{\varphi_\text{max}^4 x_\text{max}^5}{\inter}
      \right)
    \right]^6
  \right),
\end{equation}
where $\mu$ is the number of terms in the Riemann sum and
$\inter$ is the desired accuracy of the approximation.
Here we have taken $Z_q$ to be $\mathcal{O}(1)$.

\subsection{Decomposition into self-inverse operators}
\label{sec:decomp2}

The truncated Taylor series strategy introduced in \cite{Berry2015} requires that we can represent our Hamiltonian as a weighted sum of unitaries.
To do so, we follow a procedure in \cite{Berry2013} which shows how 1-sparse matrices can be decomposed into a sum of self-inverse matrices with eigenvalues $\pm 1$.
Specifically, we decompose each $\cH_{\gamma,\rho}$ into a sum of $M \in \Theta\big(\max_{\gamma,\rho}\big\|\cH_{\gamma,\rho}\big\|_\textrm{max}/\zeta\big)$ 1-sparse unitary matrices of the form
\begin{equation}
\cH_{\gamma,\rho} \approx \tcH_{\gamma,\rho} \equiv \zeta\sum_{m=1}^{M} C_{\gamma,\rho,m}
\end{equation}
where $\zeta$ is the desired precision of the decomposition.

First, we construct a new matrix $\tcH_{\gamma,\rho}$ by rounding each entry of $\cH_{\gamma,\rho}$ to the nearest multiple of $2\,\zeta$, so that $\big\|\cH_{\gamma,\rho}-\tcH_{\gamma,\rho}\big\|_{\textrm{max}} \leq \zeta$. We define $C_{\gamma,\rho}\equiv\cH_{\gamma,\rho}/\zeta$ so that $\left \| C_{\gamma,\rho} \right\|_{\textrm{max}}\leq 1+ \|\cH_{\gamma,\rho} \|_\textrm{max}/\zeta$. We decompose each $C_{\gamma,\rho}$ into $ \|C_{\gamma,\rho} \|_{\textrm{max}}$ 1-sparse matrices, indexed by $m$, with entries in $\{0, -2, 2\}$, as follows:
\begin{equation}
\label{eq:Cthing}
C_{\gamma,\rho,m}^{\alpha\beta}\equiv
\begin{cases}
+2 & C_{\gamma,\rho}^{\alpha\beta}\geq 2m\\
-2 & C_{\gamma,\rho}^{\alpha\beta} < 2m\\
0 & \text{otherwise}.
\end{cases}
\end{equation}
Finally, we remove zero eigenvalues by further dividing each $C_{\gamma,\rho,m}$ into two matrices $C_{\gamma,\rho,m,1}$ and $C_{\gamma,\rho,m,2}$ with entries in $\{0, -1, +1\}$.
For every all-zero column $\beta$ in $C_{\gamma,\rho,m}$, we choose $\alpha$ so that $(\alpha, \beta)$ is the location of the nonzero entry in column $\beta$ in the original matrix ${\cal H}_{\gamma,\rho}$.
Then the matrix $C_{\gamma,\rho,m,1}$ has $+1$ in the $(\alpha, \beta)$ position, and $C_{\gamma,\rho,m,2}$ has $-1$ in the $(\alpha, \beta)$ position.
Thus, we have decomposed each $H_\gamma$ into a sum of 1-sparse, unitary matrices with eigenvalues $\pm 1$.

We now use a simplified notation where $\ell$ corresponds to the triples $(s,m,\gamma)$, and $\aleph_{\ell,\rho} \equiv C_{\gamma,\rho,m,s}$.
We denote the number of values of $\ell$ by $L$, and can write the Hamiltonian as a sum of ${\cal O}(N^4 \mu M )$ unitary, 1-sparse matrices
\begin{equation}\label{eq:selfinvdec}
H = \zeta \sum_{\ell=1}^{L} \sum_{\rho=1}^{\mu} {\cal H}_{\ell,\rho}.
\end{equation}
That is, the decomposition is of the form in \eq{unit_sum}, but in this case $W_\ell $ is independent of $\ell$.

To summarize, we decompose the molecular Hamiltonian into a sum of self-inverse matrices in four steps:
\begin{enumerate}
	\item Decompose the molecular Hamiltonian into a sum of 1-sparse matrices using the bipartite graph coloring given in \app{1sparseproof}, summarized in \sec{dec1}.
	\item Decompose these 1-sparse matrices further, such that each entry corresponds to a single term in the sum over molecular integrals. This does not change the number of terms, but simplifies calculations.
	\item Discretize the integrals over a finite region of space, subject to the constraints and bounds given in \cite{BabbushSparse1}.
	\item Decompose into self-inverse operators by the method proposed in \cite{Berry2013}.
\end{enumerate}
This decomposition gives an overall gate count scaling contribution of ${\cal O}(\eta^2 N^2)$.

\section{The CI Matrix Oracle}
\label{sec:oracle}

In this section, we discuss the construction of the circuit referred to in our introduction as $\textsc{select}({\cal H})$, which applies the self-inverse operators in a controlled way.
As discussed in \cite{BabbushSparse1}, the truncated Taylor series technique of \cite{Berry2015} can be used with a selection oracle for an integrand which defines the molecular Hamiltonian. This method will then effect evolution under this Hamiltonian with an exponential increase in precision over Trotter-based methods. For clarity of exposition, we describe the construction of $\textsc{select}({\cal H})$ in terms of two smaller oracle circuits which can be queried to learn information about the 1-sparse unitary integrands. This information is then used to evolve an arbitrary quantum state under a specific 1-sparse unitary.

The first of the oracles described here is denoted as $Q^{\textrm{col}}$ and is used to query information about the sparsity pattern of a particular 1-sparse Hermitian matrix from \eq{gamma_decomp}. The second oracle is denoted as $Q^{\textrm{val}}$ and is used to query information about the value of integrands for elements in the CI matrix. We construct $Q^{\textrm{val}}$ by making calls to a circuit constructed in \cite{BabbushSparse1} where it is referred to as ``$\textsc{sample}(w)$''. The purpose of $\textsc{sample}(w)$ is to sample the integrands of the one-electron and two-electron integrals $h_{ij}$ and $h_{ijk\ell}$ in \eq{double_int} and \eq{single_int}. The construction of $\textsc{sample}(w)$ in \cite{BabbushSparse1} requires $\widetilde{\cal O}(N)$ gates.

The oracle $Q^{\textrm{col}}$ uses information from the index $\gamma$.
The index $\gamma$ is associated with the indices $(a_1,b_1,i,p,a_2,b_2,j,q)$ which describe the sparsity structure of the 1-sparse Hermitian matrix $H_{\gamma}$ according to the decomposition in \sec{dec2}.
$Q^{\textrm{col}}$ acts on a register specifying a color $\ket{\gamma}$ as well a register containing an arbitrary row index $\ket{\alpha}$ to reveal a column index $\ket{\beta}$ so that the ordered pair ($\alpha$, $\beta$) indexes the nonzero element in row $\alpha$ of $H_{\gamma}$,
\begin{align}
Q^\textrm{col} \ket{\gamma} \ket{\alpha}\ket{0}^{\otimes \eta \log N}
& =  \ket{\gamma}\ket{\alpha}\ket{\beta}.
\end{align}
From the description in \sec{dec2}, implementation of the unitary oracle $Q^\textrm{col}$ is straightforward.

To construct $\textsc{select}({\cal H})$ we need a second oracle that returns the value of the matrix elements in the decomposition. This selection oracle is queried with a register $\ket{\ell} =  \ket{s}\ket{m}\ket{\gamma}$ which specifies which part of the 1-sparse representation we want, as well as a register $\ket{\rho}$ which indexes the grid point $\rho$ and registers $\ket{\alpha}$ and $\ket{\beta}$ specifying the two Slater determinants. Specifically, the entries in the tuple identify the color ($\gamma$) of the 1-sparse Hermitian matrix from which the 1-sparse unitary matrix originated, which positive integer index ($m \leq M$) it corresponds to in the further decomposition of $\cH_{\gamma,\rho}$ into $C_{\gamma,\rho,m}$, and which part it corresponds to in the splitting of $C_{\gamma,\rho,m}$ into $C_{\gamma,\rho,m,s}$ (where $s\in\{1,2\}$).

As a consequence of the Slater-Condon rules shown in Eqs.~\eqref{eq:slate1}, \eqref{eq:slate2}, \eqref{eq:slate3}, and \eqref{eq:slate4}, $Q^\textrm{val}$ can be constructed given access to $\textsc{sample}(w)$, which samples the integrand of the integrals in Eqs.~\eqref{eq:double_int} and \eqref{eq:single_int} \cite{BabbushSparse1}.
Consistent with the decomposition in \sec{dec2}, the $i$ and $j$ indices in the register containing $\gamma = (i,p,j,q)$ specify the dissimilar spin-orbitals in $\ket{\alpha}$ and $\ket{\beta}$ that are needed in the integrands defined by the Slater-Condon rules; therefore, the determination of which spin-orbitals differ between $\ket{\alpha}$ and $\ket{\beta}$ can be made in ${\cal O}(\log N)$ time (only the time needed to read their values from $\gamma$). As $\textsc{sample}(w)$ is comprised of $\widetilde{\cal O}(N)$ gates, $Q^{\textrm{val}}$ has time complexity $\widetilde{\cal O}(N)$ and acts as
\begin{align}
&\quad Q^\textrm{val}\ket{\ell}\ket{\rho}\ket{\alpha}\ket{\beta}= {\cal H}_{\ell,\rho}^{\alpha \beta} \ket{\ell}\ket{\rho}\ket{\alpha}\ket{\beta},
\end{align}
where ${\cal H}_{\ell,\rho}^{\alpha \beta}$ is the value of the matrix entry at $(\alpha,\beta)$ in the self-inverse matrix ${\cal H}_{\ell,\rho}$.
When either $\ket{\alpha}$ or $\ket{\beta}$ represents an invalid Slater determinant (with more than one occupation on any spin-obital), we take ${\cal H}_{\ell,\rho}^{\alpha \beta} = 0$ for $\alpha \ne \beta$. This ensures there are no transitions into Slater determinants which violate the Pauli exclusion principle.
The choice of ${\cal H}_{\ell,\rho}^{\alpha \alpha}$ for invalid $\alpha$ will not affect the result, because the state will have no weight on the invalid Slater determinants.

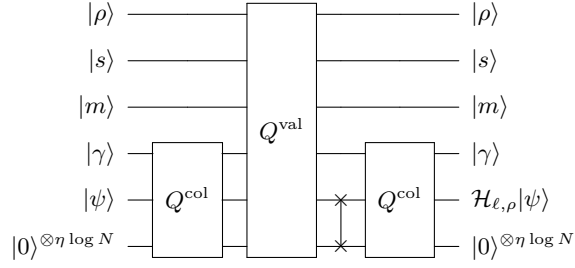
\begin{figure}[h!]
\[\Qcircuit @R 1em @C 1em {
	\lstick{\kets{\rho}} & \qw & \multigate{5}{Q^\textrm{val}} & \qw & \qw &\rstick{\kets{\rho}} \qw\\
    \lstick{\kets{s}} & \qw & \ghost{Q^\textrm{val}} & \qw & \qw & \rstick{\kets{s}} \qw\\
    \lstick{\kets{m}} & \qw & \ghost{Q^\textrm{val}} & \qw & \qw & \rstick{\kets{m}} \qw\\
    \lstick{\kets{\gamma}} & \multigate{2}{Q^\textrm{col}} &  \ghost{Q^\textrm{val}} & \qw &  \multigate{2}{Q^\textrm{col}}& \rstick{\kets{\gamma}} \qw\\
    \lstick{\kets{\psi}} & \ghost{Q^\textrm{col}} & \ghost{Q^\textrm{val}} & \qswap & \ghost{Q^\textrm{col}} &\rstick{{\cal H}_{\ell,\rho}\kets{\psi}} \qw\\
    \lstick{\kets{0}^{\otimes \eta \log N}} & \ghost{\widetilde{Q}_\gamma^{col}} & \ghost{Q^\textrm{val}} &  \qswap \qwx & \ghost{Q^\textrm{col}} & \rstick{\kets{0}^{\otimes \eta \log N}} \qw\\
 }\]
\caption{\label{fig:select_circuit} The circuit implementing $\textsc{select}({\cal H})$, which applies the term ${\cal H}_\ell(\vec z_\rho)$ labeled by $\ell=(\gamma, m, s)$ in the unitary 1-sparse decomposition to the wavefunction $\ket{\psi}$.}
\end{figure}

Having constructed the column and value oracles, we are finally ready to construct $\textsc{select} ({\cal H})$. This involves implementing 1-sparse unitary operations. The method we describe is related to the scheme presented in \cite{Aharonov2003} for evolution under 1-sparse Hamiltonians, but is simplified due to the simpler form of the operators. As in \eq{selectH}, $\textsc{select}({\cal H})$ applies the term ${\cal H}_{\ell,\rho}$ in the 1-sparse unitary decomposition to the wavefunction $\ket{\psi}$. Writing $\ket{\psi}=\sum_\alpha c_\alpha\ket{\alpha}$, we require that $\textsc{select}({\cal H})$ first call $Q^\textrm{col}$ to obtain the columns, $\beta$, corresponding to the rows, $\alpha$, for the nonzero entries of the Hamiltonian:
\begin{align}
\ket{\ell}\ket{\rho}\ket{\psi}\ket{0}^{\otimes \eta \log N} & \mapsto \sum_\alpha c_\alpha
Q^\textrm{col}\ket{\ell}\ket{\rho}\ket{\alpha}\ket{0}^{\otimes \eta \log N}\nonumber\\
& =\sum_\alpha c_\alpha \ket{\ell}\ket{\rho}\ket{\alpha}\ket{\beta}.
\end{align}
Now that we have the row and column of the matrix element, we apply $Q^\textrm{val}$ which causes each Slater determinant to accumulate the phase factor $k_\alpha = {\cal H}_{\ell,\rho}^{\alpha \beta} = \pm 1$:
\begin{align}
\sum_\alpha c_\alpha \ket{\ell}\ket{\rho}\ket{\alpha}\ket{\beta} & \mapsto \sum_\alpha c_\alpha Q^\textrm{val}\ket{\ell}\ket{\rho}\ket{\alpha}\ket{\beta}\\
& =\sum_\alpha c_\alpha k_\alpha\ket{\ell}\ket{\rho}\ket{\alpha}\ket{\beta}\nonumber.
\end{align}
Next, we swap the locations of $\alpha$ and $\beta$ in order to complete multiplication by the 1-sparse unitary,
\begin{align}
\sum_\alpha c_\alpha k_\alpha\ket{\ell}\ket{\rho}\ket{\alpha}\ket{\beta} & \mapsto \sum_\alpha c_\alpha k_\alpha\ket{\ell}\ket{\rho}\text{SWAP}\ket{\alpha}\ket{\beta}\nonumber\\
& =\sum_\alpha c_\alpha k_\alpha\ket{\ell}\ket{\rho}\ket{\beta}\ket{\alpha}.
\end{align}
Finally we apply $Q^\textrm{col}$ again but this time $\beta$ is in the first register. Since $Q^\textrm{col}$ is self-inverse and always maps $\ket{\alpha}\ket{b}$ to $\ket{\alpha}\ket{b \oplus \beta}$ and $\ket{\beta}\ket{b}$ to $\ket{\beta}\ket{b \oplus \alpha}$, this allows us to uncompute the ancilla register.
\begin{align}
\sum_\alpha c_\alpha k_\alpha\ket{\ell}\ket{\rho}\ket{\beta}\ket{\alpha} & \mapsto \sum_\alpha c_\alpha k_\alpha Q^\textrm{col}\ket{\ell}\ket{\rho}\ket{\beta}\ket{\alpha}\nonumber\\
& = \sum_\alpha c_\alpha k_\alpha\ket{\ell}\ket{\rho}\ket{\beta}\ket{0}^{\otimes \eta \log N} \nn
&= \ket{\ell}{\cal H}_{\ell,\rho}\ket{\psi}\ket{0}^{\otimes \eta \log N}.
\end{align}
Note that this approach works regardless of whether the entry is on-diagonal or off-diagonal; we do not need separate schemes for the two cases. The circuit for $\textsc{select}({\cal H})$ is depicted in \fig{select_circuit}.

\section{Simulating Hamiltonian Evolution}
\label{sec:simulation}

The simulation technique we now discuss is based on that of Ref.~\cite{Berry2015}.
We partition the total time $t$ into $r$ segments of duration $t/r$.
For each segment, we expand the evolution operator $e^{-iHt/r}$ in a Taylor series up to order $K$,
\begin{align}
\label{eq:Ur}
U_r  :=  e^{-i H t / r} \approx \sum_{k=0}^K\frac{\left(-i Ht/r\right)^k}{k!}.
\end{align}
The error due to truncating the series at order $K$ is bounded by
\begin{equation}
\label{eq:error}
{\cal O}\left(\frac{\left(\left\| H \right\| t / r\right)^{K + 1}}{\left(K+1\right)!}\right).
\end{equation}
In order to ensure the total simulation has error no greater than $\epsilon$, each segment should have error bounded by $\epsilon/r$.
Therefore, provided $r \ge \| H\| t$, the total simulation will have error no more than $\epsilon$ for
\begin{equation}
\label{eq:K}
K \in {\cal O} \left(\frac{\log\left(r/\epsilon\right)}{\log\log\left(r/\epsilon\right)}\right).
\end{equation}

Using our full decomposition of the Hamiltonian from \eq{selfinvdec} in the Taylor series formula of \eq{Ur}, we obtain
\begin{align}
\label{eq:Ur3}
U_r \approx \sum_{k=0}^K\frac{\left(-i t \zeta \right)^k}{r^k k!} \! \!  \sum_{\ell_1, \cdots, \ell_k=1}^{L} \sum_{\rho_1, \cdots, \rho_k=1}^{\mu}
\!\!\! {\cal H}_{\ell_1,\rho_1} \cdots {\cal H}_{\ell_k,\rho_k}.
\end{align}
The sum in \eq{Ur3} takes the form
\begin{align}
\label{eq:bV}
& \widetilde{U} =\sum_{j} \beta_j V_j, \quad \quad j = \left(k, \ell_1, \cdots, \ell_k,{\rho_1} \cdots {\rho_k} \right),\nn
& V_j =\left(-i\right)^k{\cal H}_{\ell_1,\rho_1} \cdots {\cal H}_{\ell_k,\rho_k}, \quad  \beta_j = \frac{t^k \zeta^k }{r^k k!},
\end{align}
where $\widetilde{U}$ is close to unitary and the $V_j$ are unitary.
Note that in contrast to \cite{BabbushSparse1}, $\beta_j\ge 0$, consistent with the convention used in \cite{Berry2015}. Our simulation will make use of an ancillary ``selection'' register $\ket{j} = \ket{k}\ket{\ell_1} \cdots \ket{\ell_K} \ket{\rho_1} \cdots \ket{\rho_K}$ for $0\leq k\leq K$, with $1\leq \ell_\upsilon\leq L$ and $1 \leq \rho_\upsilon \leq \mu$ for all $\upsilon$. It is convenient to encode $k$ in unary, as $\ket{k}=\ket{1^k0^{K-k}}$, which requires $\Theta(K)$ qubits.
Additionally, we encode each $\ket{\ell_k}$ in binary using $\Theta (\log L )$ qubits and each $\ket{\rho_k}$ in binary using $\Theta(\log \mu)$ qubits.
We denote the total number of ancilla qubits required as $J$, which scales as
\begin{equation}
\label{eq:J}
J \in \Theta\left(K\log \left(\mu L\right)\right) = {\cal O}\left(\frac{\log \left(\mu L\right)\log\left(r/\epsilon\right)}{\log\log\left(r/\epsilon\right)}\right).
\end{equation}

\begin{figure}
\[\Qcircuit @R 1em @C 0.25em {
	&&&&&\lstick{\kets{0}} &\gate{R_y\left(\theta_1\right)} & \ctrl{1} & \qw &&&& \cdots &&&& & \qw & \qw & \qw \\
	&&&&&\lstick{\kets{0}} &\qw & \gate{R_y\left(\theta_2\right)} & \qw &&&& \cdots &&&& & \qw & \qw & \qw \\
	&\vdots &&&&&&&&&&& \ddots&& &&  & \vdots & \\\\
	&&&&&\lstick{\kets{0}} &\qw & \qw & \qw &&&& \cdots& &&& & \ctrl{1} & \qw & \qw \\
	&&&&&\lstick{\kets{0}} &\qw & \qw & \qw &&&& \cdots&&& && \gate{R_y\left(\theta_K\right)} & \qw & \qw \\
&&&&&\lstick{\kets{0}^{\otimes \log L}} & {/} \qw & \gate{H\!d^{\otimes \log L}}  &\qw & \qw& \qw & \qw &\qw & \qw& \qw & \qw &\qw & \qw& \qw & \qw  \\
&&&&&\lstick{\kets{0}^{\otimes \log\mu}} & {/} \qw & \gate{H\!d^{\otimes \log \mu}}  &\qw & \qw & \qw & \qw &\qw & \qw& \qw & \qw&\qw & \qw& \qw & \qw \\
}\]
\caption{\label{fig:B_circuit} The circuit for $B$ as described in \eq{B}. An expression for $\theta_k$ is given in \eq{theta_k}.}
\end{figure}
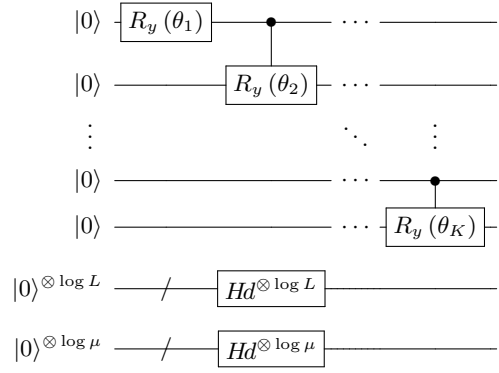

To implement the truncated evolution operator in \eq{bV}, we wish to prepare a superposition state over $j$, then apply a controlled $V_j$ operator.
Following the notation of Ref.~\cite{Berry2015}, we denote the state preparation operator as $B$, and it has the effect
\begin{equation}
\label{eq:B}
B\ket{0}^{\otimes J} = \sqrt{\frac{1}{\lambda}}\sum_{j} \sqrt{\beta_j} \ket{j},
\end{equation}
where $\lambda$ is a normalization factor.
We can implement $B$ by applying Hadamard gates (which we denote as $H\!d$) to every qubit in the $\ket{\ell_\upsilon}$ and $\ket{\rho_\upsilon}$ registers, in addition to $K$ (controlled) single-qubit rotations applied to each qubit in the $\ket{k}$ register.
We need to prepare the superposition state over $\ket k$
\begin{equation}
\left(\sum_{k=0}^K \frac{(\zeta L \mu t)^k}{r^k k!}\right)^{-1/2}\sum_{k=0}^K \sqrt{\frac{(\zeta L \mu t)^k}{r^k k!}} \ket{k}.
\end{equation}
Using the convention that $R_y(\theta_k) := \exp [-i\, \theta_k\, \sigma_y / 2 ]$, this can be prepared by applying $R_y(\theta_1)$ followed by $R_y (\theta_k)$ controlled on qubit $k-1$ for all $k \in [2, K]$, where
\begin{equation}
\label{eq:theta_k}
\theta_k := 2\arcsin{\left(\sqrt{ 1- \frac{(\zeta L \mu t)^{k-1}}{r^{k-1}\left(k-1\right)!} \left( \sum_{s = k}^K \frac{(\zeta L \mu t)^{s}}{r^s s!} \right)^{-1}} \right)}.
\end{equation}
The Hadamard gates are applied to each of the qubits in the $2 K$ remaining components of the selection register $\ket{\ell_1} \cdots \ket{\ell_K} \ket{\rho_1} \cdots \ket{\rho_K}$. This requires ${\cal O}(K\log (\mu L ))$ gates; parallelizing the Hadamard transforms leads to circuit depth ${\cal O} (K)$ for $B$. A circuit implementing $B$ is shown in \fig{B_circuit}.

We then wish to implement an operator to apply the $V_j$ which is referred to in \cite{BabbushSparse1,Berry2015} as $\textsc{select}(V)$,
\begin{equation}
\label{eq:selectV}
\textsc{select}\left(V\right)\ket{j}\ket{\psi} =\ket{j}V_j\ket{\psi}.
\end{equation}
This operation can be applied using ${\cal O}(K)$ queries to a controlled form of the oracle $\textsc{select}({\cal H})$ defined in \sec{oracle}.
One can apply $\textsc{select}({\cal H})$ $K$ times, using each of the $\ket{\ell_\upsilon}$ and $\ket{\rho_\upsilon}$ registers.
Thus, given that $\textsc{select}({\cal H})$ requires $\widetilde{\cal O}(N)$ gates, our total gate count for $\textsc{select}(V)$ is $\widetilde {\cal O}(K N)$. \tab{parameters} lists relevant parameters along with their bounds, in terms of chemically relevant variables. \tab{operators} lists relevant operators and their gate counts.

\begin{table*}[ht]
\caption{Taylor series simulation parameters and bounds}
\label{tab:parameters}
\begin{tabular}{  c   |   c   |   c }
\hline\hline
Parameter & Explanation & Bound\\
\hline
$r$ & number of time segments, \eq{r} & $ \zeta L\mu t/\ln(2)$\\
$L$ & terms in unitary decomposition, \eq{L} & ${\cal O}\left(\eta^2 N^2 \max_{\gamma,\rho}\|\aleph_{\gamma,\rho} \|_{\max}/\zeta\right)$\\
$K$ & truncation point for Taylor series, \eq{K} & ${\cal O}\left(\frac{\log(r/\epsilon)}{\log\log(r/\epsilon)}\right)$\\
$J$ & ancilla qubits in selection register, \eq{J} & $\Theta\left(K\log \left(\mu L\right)\right)$\\
\hline
\end{tabular}
\end{table*}
\begin{table*}[ht]
\caption{Taylor series simulation operators and complexities}
\label{tab:operators}
\begin{tabular}{  c   |   c   |   c }
\hline\hline
Operator & Purpose & Gate Count\\
\hline
$\textsc{select}\left({\cal H}\right)$ & applies specified terms from decomposition, \eq{selectH} & $\widetilde{\cal O}\left(N\right)$\\
$\textsc{select}\left(V\right)$ & applies specified strings of terms, \eq{selectV} & $\widetilde{\cal O}\left(N K\right) $\\
$B$ & prepares superposition state, \eq{B} & $ {\cal O}\left(K \log \left(\mu L\right) \right) $\\
${\cal W}$ & probabilistically performs simulation under $H t / r$, \eq{W} & $ \widetilde{\cal O}\left(N K\right) $\\
$P$ & projector onto $\ket{0}^{\otimes J}$ state of selection register, \eq{P} & $ {\cal O}\left(K \log \left(\mu L\right)\right) $\\
$G$ & amplification operator to implement sum of unitaries, \eq{G} & $ \widetilde{\cal O}\left(N K \right) $\\
$\left(PG\right)^r$ & entire algorithm & $ \widetilde{\cal O}\left(r N K\right) $\\
\hline
\end{tabular}
\end{table*}

A strategy for implementing the evolution operator in \eq{bV} becomes clear if we consider the combined action of $B$ followed by $\textsc{select}(V)$ on state $\ket{\psi}$,
\begin{equation}
\label{eq:BV}
\textsc{select}\left(V\right) B \ket{0}^{\otimes J} \ket{\psi} = \sqrt{\frac{1}{\lambda}}\sum_{j}  \sqrt{\beta_j} \ket{j} V_j \ket{\psi}.
\end{equation}
This state is similar to that desired for $\widetilde{U}\ket{\psi}$ except for the presence of the register $\ket{j}$.
The need to disentangle from that register motivates the operator \cite{Berry2015}
\begin{align}
\label{eq:W}
& {\cal W} := \left( B\otimes \openone \right)^\dagger \textsc{select}\left(V\right)\left(B \otimes \openone\right), \\
& {\cal W} \ket{0}^{\otimes J}\ket{\psi} = \frac{1}{\lambda}\ket{0}^{\otimes J}\widetilde{U}\ket{\psi} + \sqrt{1 - \frac{1}{\lambda^2}}\ket{\Phi},
\end{align}
where $\ket{\Phi}$ is a state for which the selection register ancilla qubits are orthogonal to $\ket{0}^{\otimes J}$. Accordingly, we define a projection operator $P$ so that
\begin{align}
\label{eq:P}
P = \left(\ket{0}\!\bra{0}\right)^{\otimes J} \otimes \openone , \\
P {\cal W} \ket{0}^{\otimes J} \ket{\psi} = \frac{1}{\lambda}\ket{0}^{\otimes J} \widetilde{U} \ket{\psi}.
\end{align}
Using these definitions, we follow the procedure for oblivious amplitude amplification of \cite{Berry2015} to deterministically execute the intended unitary. To this end, we define the oblivious amplitude amplification operator
\begin{equation}
\label{eq:G}
G \equiv - {\cal W} \left( \openone - 2 P\right) {\cal W}^\dagger \left( \openone - 2 P\right) {\cal W}.
\end{equation}
We use the amplification operator $G$ in conjunction with the projector onto the empty ancilla state $P$ so that
\begin{equation}
PG\ket{0}\ket{\psi}=\ket{0}\left(\frac{3}{\lambda}\widetilde{U}-\frac{4}{\lambda^3}\widetilde{U}\widetilde{U}^\dagger\tilde{U}\right)\ket{\psi}.
\end{equation}

The sum of the absolute values of the coefficients in the self-inverse decomposition of the Hamiltonian in \eq{selfinvdec} is $\zeta L\mu$. If we choose the number of segments as $r=\zeta L\mu t / \ln(2)$, then our choice of $K$ as in \eq{K} ensures that $\big\|\widetilde{U}-U_r\big\|_{\textrm{max}} \in {\cal O}(\epsilon/r)$, and hence \cite{Berry2015}
\begin{equation}
\left\|PG\ket{0}\ket{\psi}-\ket{0}U_r\ket{\psi}\right\|_{\textrm{max}} \in {\cal O}\left(\epsilon/r\right).
\end{equation}
Then the total error due to oblivious amplitude amplification on all segments will be ${\cal O}(\epsilon)$.
Therefore, the complexity of the total algorithm is $r$ times the complexity of implementing $\textsc{select}(V)$ and $B$.
While we implement $B$ with gate count ${\cal O}(K \log (\mu L))$, our construction of $\textsc{select}({\cal H})$ has gate count $\widetilde{\cal O}(N K)$.

The gate count of our entire algorithm depends crucially on $r$.  Above we have
taken $r \in {\cal O}(\zeta L \mu t)$ where
\begin{align}
\label{eq:L}
L & \in {\cal O}\left(M \Gamma\right),\\
M & \in \Theta\left(\max_{\gamma,\rho}\left \|\aleph_{\gamma,\rho}\right\|_{\max}/\zeta\right),\\
\Gamma & \in {\cal O}\left(\eta^2 N^2 \right).
\end{align}
As a result, we may bound $r$ as
\begin{equation}
\label{eq:r}
r\in{\cal O}\left(\eta^2 N^2 t \mu \max_{\gamma,\rho}\left \|\aleph_{\gamma,\rho}\right\|_{\max}\right).
\end{equation}
As a consequence of \lem{int0}, \lem{int1} and \lem{int2},
$\mu \max_{\gamma,\rho}\left \|\aleph_{\gamma,\rho}\right\|_{\max}$ can be replaced with
\begin{equation}
   \mathcal{O} \left(
     \varphi_{\max}^4 x_{\max}^5
     \left[
       \log \left(
         \frac{\varphi_{\max}^4 x_{\max}^5}{\inter}
       \right)
     \right]^6
   \right),
 \end{equation}
where $\varphi_{\max}$ is the maximum value taken by the orbitals, and $x_{\max}$ is the
scaling of the spatial size of the orbitals.
To relate $\epsilon$ to $\inter$, in \sec{dec2} the Hamiltonian is broken up into
a sum of ${\cal O}(\eta^2 N^2)$ terms, each of which contains one or two of the integrals.
Therefore, the error in the Hamiltonian is ${\cal O}(\inter \eta^2 N^2)$.
The Hamiltonian is simulated for time $t$, so the resulting error in the simulation will be ${\cal O}(\inter t \eta^2 N^2)$.
To ensure that the error is no greater than $\epsilon$, we should therefore choose $\inter=\Theta(\epsilon/( t \eta^2 N^2))$.
Since we are considering scaling with large $\eta$ and $N$, $\inter$ will be small and
the conditions \eq{sensible0}, \eq{sensible1} and \eq{sensible2} will be satisfied.
In addition, the conditions of \thm{maintheorem} mean that $\varphi_{\max}$ and $x_{\max}$ are logarithmic in $N$.
Hence one can take, omitting logarithmic factors,
\begin{equation}
r\in \widetilde{\cal O}(\eta^2 N^2 t).
\end{equation}

The complexity of $B$ does not affect the scaling, because it is lower order in $N$.
Therefore, our overall algorithm has gate count
\begin{equation}
\widetilde{\cal O}\left(r N K \right) = \widetilde{\cal O}\left(\eta^2 N^3 t\right),
\end{equation}
as stated in \thm{maintheorem}.
This scaling represents an exponential improvement in precision as compared to
Trotter-based methods. However, we suspect that the actual scaling of these
algorithms is much better for real molecules, just as has been observed for the
Trotter-based algorithms \cite{Poulin2014,BabbushTrotter}. Furthermore, the
approach detailed here requires fewer qubits than any other approach to quantum
simulation of chemistry in the literature.

\section{Discussion}
\label{sec:conclusion}

We have outlined a method to simulate the quantum chemistry Hamiltonian in a basis of Slater determinants using recent advances from the universal simulation literature. We find an oracular decomposition of the Hamiltonian into 1-sparse matrices based on an edge coloring routine first described in \cite{Toloui2013}. We use that oracle to simulate evolution under the Hamiltonian using the truncated Taylor series technique described in \cite{Berry2013}.
We discretize the integrals which define entries of the CI matrix, and use the sum of unitaries approach to effectively exponentially compress evaluation of these discretized integrals.

Asymptotic scalings suggest that the algorithms described in this paper series will allow for the quantum simulation of much larger molecular systems than would be possible using a Trotter-Suzuki decomposition.
Recent work \cite{Wecker2014,Hastings2015,Poulin2014,McClean2014,BabbushTrotter} has demonstrated markedly more efficient implementations of the original Trotter-Suzuki-based quantum chemistry algorithm \cite{Aspuru-Guzik2005,Whitfield2010}; similarly, we believe the implementations discussed here can still be improved upon, and that numerical simulations will be crucial to this task.

Finally, we note that the CI matrix simulation strategy discussed here opens up the possibility of an interesting approach to adiabatic state preparation. An adiabatic algorithm for quantum chemistry was suggested in second quantization in \cite{BabbushAQChem} and studied further in \cite{Veis2014}. However, those works did not suggest a compelling adiabatic path to take between an easy-to-prepare initial state (such as the Hartree-Fock state) and the ground state of the exact Hamiltonian. We note that one could start the system in the Hartree-Fock state, and use the CI matrix oracles discussed in this paper to ``turn on'' a Hamiltonian having support over a number of configuration basis states which increases smoothly with time.

\section*{Acknowledgements}
The authors thank Jhonathan Romero Fontalvo, Jarrod McClean, Borzu Toloui, and Nathan Wiebe for helpful discussions. D.W.B.\ is funded by an Australian Research Council Future Fellowship (FT100100761). P.J.L.\ acknowledges the support of the National Science Foundation under grant number PHY-0955518. A.A.G.\ and P.J.L.\ acknowledge the support of the Air Force Office of Scientific Research under award number FA9550-12-1-0046. A.A.-G.\ acknowledges the Army Research Office under Award: W911NF-15-1-0256.

\bibliographystyle{apsrev4-1}
\bibliography{library,library_ryan}

\begin{thebibliography}{40}%
\makeatletter
\providecommand \@ifxundefined [1]{%
 \@ifx{#1\undefined}
}%
\providecommand \@ifnum [1]{%
 \ifnum #1\expandafter \@firstoftwo
 \else \expandafter \@secondoftwo
 \fi
}%
\providecommand \@ifx [1]{%
 \ifx #1\expandafter \@firstoftwo
 \else \expandafter \@secondoftwo
 \fi
}%
\providecommand \natexlab [1]{#1}%
\providecommand \enquote  [1]{``#1''}%
\providecommand \bibnamefont  [1]{#1}%
\providecommand \bibfnamefont [1]{#1}%
\providecommand \citenamefont [1]{#1}%
\providecommand \href@noop [0]{\@secondoftwo}%
\providecommand \href [0]{\begingroup \@sanitize@url \@href}%
\providecommand \@href[1]{\@@startlink{#1}\@@href}%
\providecommand \@@href[1]{\endgroup#1\@@endlink}%
\providecommand \@sanitize@url [0]{\catcode `\\12\catcode `\$12\catcode
  `\&12\catcode `\#12\catcode `\^12\catcode `\_12\catcode `\%12\relax}%
\providecommand \@@startlink[1]{}%
\providecommand \@@endlink[0]{}%
\providecommand \url  [0]{\begingroup\@sanitize@url \@url }%
\providecommand \@url [1]{\endgroup\@href {#1}{\urlprefix }}%
\providecommand \urlprefix  [0]{URL }%
\providecommand \Eprint [0]{\href }%
\providecommand \doibase [0]{http://dx.doi.org/}%
\providecommand \selectlanguage [0]{\@gobble}%
\providecommand \bibinfo  [0]{\@secondoftwo}%
\providecommand \bibfield  [0]{\@secondoftwo}%
\providecommand \translation [1]{[#1]}%
\providecommand \BibitemOpen [0]{}%
\providecommand \bibitemStop [0]{}%
\providecommand \bibitemNoStop [0]{.\EOS\space}%
\providecommand \EOS [0]{\spacefactor3000\relax}%
\providecommand \BibitemShut  [1]{\csname bibitem#1\endcsname}%
\let\auto@bib@innerbib\@empty
\bibitem [{\citenamefont {Aspuru-Guzik}\ \emph {et~al.}(2005)\citenamefont
  {Aspuru-Guzik}, \citenamefont {Dutoi}, \citenamefont {Love},\ and\
  \citenamefont {Head-Gordon}}]{Aspuru-Guzik2005}%
  \BibitemOpen
  \bibfield  {author} {\bibinfo {author} {\bibfnamefont {A.}~\bibnamefont
  {Aspuru-Guzik}}, \bibinfo {author} {\bibfnamefont {A.~D.}\ \bibnamefont
  {Dutoi}}, \bibinfo {author} {\bibfnamefont {P.~J.}\ \bibnamefont {Love}}, \
  and\ \bibinfo {author} {\bibfnamefont {M.}~\bibnamefont {Head-Gordon}},\
  }\href {\doibase 10.1126/science.1113479} {\bibfield  {journal} {\bibinfo
  {journal} {Science}\ }\textbf {\bibinfo {volume} {309}},\ \bibinfo {pages}
  {1704} (\bibinfo {year} {2005})}\BibitemShut {NoStop}%
\bibitem [{\citenamefont {Lloyd}(1996)}]{Lloyd1996}%
  \BibitemOpen
  \bibfield  {author} {\bibinfo {author} {\bibfnamefont {S.}~\bibnamefont
  {Lloyd}},\ }\href {http://dx.doi.org/10.1126/science.273.5278.1073}
  {\bibfield  {journal} {\bibinfo  {journal} {Science}\ }\textbf {\bibinfo
  {volume} {273}},\ \bibinfo {pages} {1073} (\bibinfo {year}
  {1996})}\BibitemShut {NoStop}%
\bibitem [{\citenamefont {Abrams}\ and\ \citenamefont
  {Lloyd}(1997)}]{Abrams1997}%
  \BibitemOpen
  \bibfield  {author} {\bibinfo {author} {\bibfnamefont {D.~S.}\ \bibnamefont
  {Abrams}}\ and\ \bibinfo {author} {\bibfnamefont {S.}~\bibnamefont {Lloyd}},\
  }\href {https://doi.org/10.1103/PhysRevLett.79.2586} {\bibfield  {journal}
  {\bibinfo  {journal} {Physical Review Letters}\ }\textbf {\bibinfo {volume}
  {79}},\ \bibinfo {pages} {4} (\bibinfo {year} {1997})}\BibitemShut {NoStop}%
\bibitem [{\citenamefont {{Cody Jones}}\ \emph {et~al.}(2012)\citenamefont
  {{Cody Jones}}, \citenamefont {Whitfield}, \citenamefont {McMahon},
  \citenamefont {Yung}, \citenamefont {Meter}, \citenamefont {Aspuru-Guzik},\
  and\ \citenamefont {Yamamoto}}]{Jones2012}%
  \BibitemOpen
  \bibfield  {author} {\bibinfo {author} {\bibfnamefont {N.}~\bibnamefont
  {{Cody Jones}}}, \bibinfo {author} {\bibfnamefont {J.~D.}\ \bibnamefont
  {Whitfield}}, \bibinfo {author} {\bibfnamefont {P.~L.}\ \bibnamefont
  {McMahon}}, \bibinfo {author} {\bibfnamefont {M.-H.}\ \bibnamefont {Yung}},
  \bibinfo {author} {\bibfnamefont {R.~V.}\ \bibnamefont {Meter}}, \bibinfo
  {author} {\bibfnamefont {A.}~\bibnamefont {Aspuru-Guzik}}, \ and\ \bibinfo
  {author} {\bibfnamefont {Y.}~\bibnamefont {Yamamoto}},\ }\href {\doibase
  10.1088/1367-2630/14/11/115023} {\bibfield  {journal} {\bibinfo  {journal}
  {New Journal of Physics}\ }\textbf {\bibinfo {volume} {14}},\ \bibinfo
  {pages} {115023} (\bibinfo {year} {2012})}\BibitemShut {NoStop}%
\bibitem [{\citenamefont {Veis}\ and\ \citenamefont
  {Pittner}(2010)}]{Veis2010}%
  \BibitemOpen
  \bibfield  {author} {\bibinfo {author} {\bibfnamefont {L.}~\bibnamefont
  {Veis}}\ and\ \bibinfo {author} {\bibfnamefont {J.}~\bibnamefont {Pittner}},\
  }\href {\doibase 10.1063/1.3503767} {\bibfield  {journal} {\bibinfo
  {journal} {The Journal of Chemical Physics}\ }\textbf {\bibinfo {volume}
  {133}},\ \bibinfo {pages} {194106} (\bibinfo {year} {2010})}\BibitemShut
  {NoStop}%
\bibitem [{\citenamefont {Wang}\ \emph {et~al.}(2015)\citenamefont {Wang},
  \citenamefont {Dolde}, \citenamefont {Biamonte}, \citenamefont {Babbush},
  \citenamefont {Bergholm}, \citenamefont {Yang}, \citenamefont {Jakobi},
  \citenamefont {Neumann}, \citenamefont {Aspuru-Guzik}, \citenamefont
  {Whitfield},\ and\ \citenamefont {Wrachtrup}}]{Wang2014}%
  \BibitemOpen
  \bibfield  {author} {\bibinfo {author} {\bibfnamefont {Y.}~\bibnamefont
  {Wang}}, \bibinfo {author} {\bibfnamefont {F.}~\bibnamefont {Dolde}},
  \bibinfo {author} {\bibfnamefont {J.}~\bibnamefont {Biamonte}}, \bibinfo
  {author} {\bibfnamefont {R.}~\bibnamefont {Babbush}}, \bibinfo {author}
  {\bibfnamefont {V.}~\bibnamefont {Bergholm}}, \bibinfo {author}
  {\bibfnamefont {S.}~\bibnamefont {Yang}}, \bibinfo {author} {\bibfnamefont
  {I.}~\bibnamefont {Jakobi}}, \bibinfo {author} {\bibfnamefont
  {P.}~\bibnamefont {Neumann}}, \bibinfo {author} {\bibfnamefont
  {A.}~\bibnamefont {Aspuru-Guzik}}, \bibinfo {author} {\bibfnamefont {J.~D.}\
  \bibnamefont {Whitfield}}, \ and\ \bibinfo {author} {\bibfnamefont
  {J.}~\bibnamefont {Wrachtrup}},\ }\href {\doibase 10.1021/acsnano.5b01651}
  {\bibfield  {journal} {\bibinfo  {journal} {ACS Nano}\ }\textbf {\bibinfo
  {volume} {9}},\ \bibinfo {pages} {7769} (\bibinfo {year} {2015})}\BibitemShut
  {NoStop}%
\bibitem [{\citenamefont {Li}\ \emph {et~al.}(2011)\citenamefont {Li},
  \citenamefont {Yung}, \citenamefont {Chen}, \citenamefont {Lu}, \citenamefont
  {Whitfield}, \citenamefont {Peng}, \citenamefont {Aspuru-Guzik},\ and\
  \citenamefont {Du}}]{Li2011}%
  \BibitemOpen
  \bibfield  {author} {\bibinfo {author} {\bibfnamefont {Z.}~\bibnamefont
  {Li}}, \bibinfo {author} {\bibfnamefont {M.-H.}\ \bibnamefont {Yung}},
  \bibinfo {author} {\bibfnamefont {H.}~\bibnamefont {Chen}}, \bibinfo {author}
  {\bibfnamefont {D.}~\bibnamefont {Lu}}, \bibinfo {author} {\bibfnamefont
  {J.~D.}\ \bibnamefont {Whitfield}}, \bibinfo {author} {\bibfnamefont
  {X.}~\bibnamefont {Peng}}, \bibinfo {author} {\bibfnamefont {A.}~\bibnamefont
  {Aspuru-Guzik}}, \ and\ \bibinfo {author} {\bibfnamefont {J.}~\bibnamefont
  {Du}},\ }\href {\doibase doi:10.1038/srep00088} {\bibfield  {journal}
  {\bibinfo  {journal} {Scientific Reports}\ }\textbf {\bibinfo {volume} {1}},\
  \bibinfo {pages} {88} (\bibinfo {year} {2011})}\BibitemShut {NoStop}%
\bibitem [{\citenamefont {Yung}\ \emph {et~al.}(2014)\citenamefont {Yung},
  \citenamefont {Casanova}, \citenamefont {Mezzacapo}, \citenamefont {McClean},
  \citenamefont {Lamata}, \citenamefont {Aspuru-Guzik},\ and\ \citenamefont
  {Solano}}]{Yung2013}%
  \BibitemOpen
  \bibfield  {author} {\bibinfo {author} {\bibfnamefont {M.-H.}\ \bibnamefont
  {Yung}}, \bibinfo {author} {\bibfnamefont {J.}~\bibnamefont {Casanova}},
  \bibinfo {author} {\bibfnamefont {A.}~\bibnamefont {Mezzacapo}}, \bibinfo
  {author} {\bibfnamefont {J.}~\bibnamefont {McClean}}, \bibinfo {author}
  {\bibfnamefont {L.}~\bibnamefont {Lamata}}, \bibinfo {author} {\bibfnamefont
  {A.}~\bibnamefont {Aspuru-Guzik}}, \ and\ \bibinfo {author} {\bibfnamefont
  {E.}~\bibnamefont {Solano}},\ }\href {\doibase 10.1038/srep03589} {\bibfield
  {journal} {\bibinfo  {journal} {Scientific Reports}\ }\textbf {\bibinfo
  {volume} {4}},\ \bibinfo {pages} {9} (\bibinfo {year} {2014})}\BibitemShut
  {NoStop}%
\bibitem [{\citenamefont {Kassal}\ \emph {et~al.}(2008)\citenamefont {Kassal},
  \citenamefont {Jordan}, \citenamefont {Love}, \citenamefont {Mohseni},\ and\
  \citenamefont {Aspuru-Guzik}}]{Kassal2008}%
  \BibitemOpen
  \bibfield  {author} {\bibinfo {author} {\bibfnamefont {I.}~\bibnamefont
  {Kassal}}, \bibinfo {author} {\bibfnamefont {S.~P.}\ \bibnamefont {Jordan}},
  \bibinfo {author} {\bibfnamefont {P.~J.}\ \bibnamefont {Love}}, \bibinfo
  {author} {\bibfnamefont {M.}~\bibnamefont {Mohseni}}, \ and\ \bibinfo
  {author} {\bibfnamefont {A.}~\bibnamefont {Aspuru-Guzik}},\ }\href
  {http://dx.doi.org/10.1073/pnas.0808245105} {\bibfield  {journal} {\bibinfo
  {journal} {Proceedings of the National Academy of Sciences}\ }\textbf
  {\bibinfo {volume} {105}},\ \bibinfo {pages} {18681} (\bibinfo {year}
  {2008})}\BibitemShut {NoStop}%
\bibitem [{\citenamefont {Toloui}\ and\ \citenamefont
  {Love}(2013)}]{Toloui2013}%
  \BibitemOpen
  \bibfield  {author} {\bibinfo {author} {\bibfnamefont {B.}~\bibnamefont
  {Toloui}}\ and\ \bibinfo {author} {\bibfnamefont {P.~J.}\ \bibnamefont
  {Love}},\ }\href {http://arxiv.org/abs/1312.2579} {\bibfield  {journal}
  {\bibinfo  {journal} {e-print arXiv:1312.2579}\ } (\bibinfo {year}
  {2013})}\BibitemShut {NoStop}%
\bibitem [{\citenamefont {Whitfield}(2013)}]{Whitfield2013b}%
  \BibitemOpen
  \bibfield  {author} {\bibinfo {author} {\bibfnamefont {J.~D.}\ \bibnamefont
  {Whitfield}},\ }\href {\doibase 10.1063/1.4812566} {\bibfield  {journal}
  {\bibinfo  {journal} {Journal of Chemical Physics}\ }\textbf {\bibinfo
  {volume} {139}},\ \bibinfo {pages} {021105} (\bibinfo {year}
  {2013})}\BibitemShut {NoStop}%
\bibitem [{\citenamefont {Whitfield}(2015)}]{Whitfield2015}%
  \BibitemOpen
  \bibfield  {author} {\bibinfo {author} {\bibfnamefont {J.~D.}\ \bibnamefont
  {Whitfield}},\ }\href {http://arxiv.org/abs/1502.03771} {\bibfield  {journal}
  {\bibinfo  {journal} {e-print arXiv:1502.03771}\ } (\bibinfo {year}
  {2015})}\BibitemShut {NoStop}%
\bibitem [{\citenamefont {Wecker}\ \emph {et~al.}(2014)\citenamefont {Wecker},
  \citenamefont {Bauer}, \citenamefont {Clark}, \citenamefont {Hastings},\ and\
  \citenamefont {Troyer}}]{Wecker2014}%
  \BibitemOpen
  \bibfield  {author} {\bibinfo {author} {\bibfnamefont {D.}~\bibnamefont
  {Wecker}}, \bibinfo {author} {\bibfnamefont {B.}~\bibnamefont {Bauer}},
  \bibinfo {author} {\bibfnamefont {B.~K.}\ \bibnamefont {Clark}}, \bibinfo
  {author} {\bibfnamefont {M.~B.}\ \bibnamefont {Hastings}}, \ and\ \bibinfo
  {author} {\bibfnamefont {M.}~\bibnamefont {Troyer}},\ }\href {\doibase
  10.1103/PhysRevA.90.022305} {\bibfield  {journal} {\bibinfo  {journal}
  {Physical Review A}\ }\textbf {\bibinfo {volume} {90}},\ \bibinfo {pages} {1}
  (\bibinfo {year} {2014})}\BibitemShut {NoStop}%
\bibitem [{\citenamefont {Hastings}\ \emph {et~al.}(2015)\citenamefont
  {Hastings}, \citenamefont {Wecker}, \citenamefont {Bauer},\ and\
  \citenamefont {Troyer}}]{Hastings2015}%
  \BibitemOpen
  \bibfield  {author} {\bibinfo {author} {\bibfnamefont {M.~B.}\ \bibnamefont
  {Hastings}}, \bibinfo {author} {\bibfnamefont {D.}~\bibnamefont {Wecker}},
  \bibinfo {author} {\bibfnamefont {B.}~\bibnamefont {Bauer}}, \ and\ \bibinfo
  {author} {\bibfnamefont {M.}~\bibnamefont {Troyer}},\ }\href
  {http://www.rintonpress.com/xxqic15/qic-15-12/0001-0021.pdf} {\bibfield
  {journal} {\bibinfo  {journal} {Quantum Information \& Computation}\ }\textbf
  {\bibinfo {volume} {15}},\ \bibinfo {pages} {1} (\bibinfo {year}
  {2015})}\BibitemShut {NoStop}%
\bibitem [{\citenamefont {Poulin}\ \emph {et~al.}(2015)\citenamefont {Poulin},
  \citenamefont {Hastings}, \citenamefont {Wecker}, \citenamefont {Wiebe},
  \citenamefont {Doherty},\ and\ \citenamefont {Troyer}}]{Poulin2014}%
  \BibitemOpen
  \bibfield  {author} {\bibinfo {author} {\bibfnamefont {D.}~\bibnamefont
  {Poulin}}, \bibinfo {author} {\bibfnamefont {M.~B.}\ \bibnamefont
  {Hastings}}, \bibinfo {author} {\bibfnamefont {D.}~\bibnamefont {Wecker}},
  \bibinfo {author} {\bibfnamefont {N.}~\bibnamefont {Wiebe}}, \bibinfo
  {author} {\bibfnamefont {A.~C.}\ \bibnamefont {Doherty}}, \ and\ \bibinfo
  {author} {\bibfnamefont {M.}~\bibnamefont {Troyer}},\ }\href
  {http://www.rintonpress.com/xxqic15/qic-15-56/0361-0384.pdf} {\bibfield
  {journal} {\bibinfo  {journal} {Quantum Information \& Computation}\ }\textbf
  {\bibinfo {volume} {15}},\ \bibinfo {pages} {361} (\bibinfo {year}
  {2015})}\BibitemShut {NoStop}%
\bibitem [{\citenamefont {McClean}\ \emph {et~al.}(2014)\citenamefont
  {McClean}, \citenamefont {Babbush}, \citenamefont {Love},\ and\ \citenamefont
  {Aspuru-Guzik}}]{McClean2014}%
  \BibitemOpen
  \bibfield  {author} {\bibinfo {author} {\bibfnamefont {J.~R.}\ \bibnamefont
  {McClean}}, \bibinfo {author} {\bibfnamefont {R.}~\bibnamefont {Babbush}},
  \bibinfo {author} {\bibfnamefont {P.~J.}\ \bibnamefont {Love}}, \ and\
  \bibinfo {author} {\bibfnamefont {A.}~\bibnamefont {Aspuru-Guzik}},\ }\href
  {\doibase 10.1021/jz501649m} {\bibfield  {journal} {\bibinfo  {journal} {The
  Journal of Physical Chemistry Letters}\ }\textbf {\bibinfo {volume} {5}},\
  \bibinfo {pages} {4368} (\bibinfo {year} {2014})}\BibitemShut {NoStop}%
\bibitem [{\citenamefont {Babbush}\ \emph {et~al.}(2015)\citenamefont
  {Babbush}, \citenamefont {McClean}, \citenamefont {Wecker}, \citenamefont
  {Aspuru-Guzik},\ and\ \citenamefont {Wiebe}}]{BabbushTrotter}%
  \BibitemOpen
  \bibfield  {author} {\bibinfo {author} {\bibfnamefont {R.}~\bibnamefont
  {Babbush}}, \bibinfo {author} {\bibfnamefont {J.}~\bibnamefont {McClean}},
  \bibinfo {author} {\bibfnamefont {D.}~\bibnamefont {Wecker}}, \bibinfo
  {author} {\bibfnamefont {A.}~\bibnamefont {Aspuru-Guzik}}, \ and\ \bibinfo
  {author} {\bibfnamefont {N.}~\bibnamefont {Wiebe}},\ }\href
  {https://doi.org/10.1103/PhysRevA.91.022311} {\bibfield  {journal} {\bibinfo
  {journal} {Physical Review A}\ }\textbf {\bibinfo {volume} {91}},\ \bibinfo
  {pages} {022311} (\bibinfo {year} {2015})}\BibitemShut {NoStop}%
\bibitem [{\citenamefont {Babbush}\ \emph {et~al.}(2014)\citenamefont
  {Babbush}, \citenamefont {Love},\ and\ \citenamefont
  {Aspuru-Guzik}}]{BabbushAQChem}%
  \BibitemOpen
  \bibfield  {author} {\bibinfo {author} {\bibfnamefont {R.}~\bibnamefont
  {Babbush}}, \bibinfo {author} {\bibfnamefont {P.~J.}\ \bibnamefont {Love}}, \
  and\ \bibinfo {author} {\bibfnamefont {A.}~\bibnamefont {Aspuru-Guzik}},\
  }\href {\doibase 10.1038/srep06603} {\bibfield  {journal} {\bibinfo
  {journal} {Scientific Reports}\ }\textbf {\bibinfo {volume} {4}},\ \bibinfo
  {pages} {6603} (\bibinfo {year} {2014})}\BibitemShut {NoStop}%
\bibitem [{\citenamefont {Peruzzo}\ \emph {et~al.}(2014)\citenamefont
  {Peruzzo}, \citenamefont {McClean}, \citenamefont {Shadbolt}, \citenamefont
  {Yung}, \citenamefont {Zhou}, \citenamefont {Love}, \citenamefont
  {Aspuru-Guzik},\ and\ \citenamefont {O'Brien}}]{Peruzzo2013}%
  \BibitemOpen
  \bibfield  {author} {\bibinfo {author} {\bibfnamefont {A.}~\bibnamefont
  {Peruzzo}}, \bibinfo {author} {\bibfnamefont {J.}~\bibnamefont {McClean}},
  \bibinfo {author} {\bibfnamefont {P.}~\bibnamefont {Shadbolt}}, \bibinfo
  {author} {\bibfnamefont {M.-H.}\ \bibnamefont {Yung}}, \bibinfo {author}
  {\bibfnamefont {X.-Q.}\ \bibnamefont {Zhou}}, \bibinfo {author}
  {\bibfnamefont {P.~J.}\ \bibnamefont {Love}}, \bibinfo {author}
  {\bibfnamefont {A.}~\bibnamefont {Aspuru-Guzik}}, \ and\ \bibinfo {author}
  {\bibfnamefont {J.~L.}\ \bibnamefont {O'Brien}},\ }\href {\doibase
  10.1038/ncomms5213} {\bibfield  {journal} {\bibinfo  {journal} {Nature
  Communications}\ }\textbf {\bibinfo {volume} {5}},\ \bibinfo {pages} {4213}
  (\bibinfo {year} {2014})}\BibitemShut {NoStop}%
\bibitem [{\citenamefont {McClean}\ \emph {et~al.}(2016)\citenamefont
  {McClean}, \citenamefont {Romero}, \citenamefont {Babbush},\ and\
  \citenamefont {Aspuru-Guzik}}]{McClean2015}%
  \BibitemOpen
  \bibfield  {author} {\bibinfo {author} {\bibfnamefont {J.~R.}\ \bibnamefont
  {McClean}}, \bibinfo {author} {\bibfnamefont {J.}~\bibnamefont {Romero}},
  \bibinfo {author} {\bibfnamefont {R.}~\bibnamefont {Babbush}}, \ and\
  \bibinfo {author} {\bibfnamefont {A.}~\bibnamefont {Aspuru-Guzik}},\ }\href
  {http://iopscience.iop.org/article/10.1088/1367-2630/18/2/023023/} {\bibfield
   {journal} {\bibinfo  {journal} {New Journal of Physics}\ }\textbf {\bibinfo
  {volume} {18}},\ \bibinfo {pages} {23023} (\bibinfo {year}
  {2016})}\BibitemShut {NoStop}%
\bibitem [{\citenamefont {O'Malley}\ \emph {et~al.}(2016)\citenamefont
  {O'Malley}, \citenamefont {Babbush}, \citenamefont {Kivlichan}, \citenamefont
  {Romero}, \citenamefont {McClean}, \citenamefont {Barends}, \citenamefont
  {Kelly}, \citenamefont {Roushan}, \citenamefont {Tranter}, \citenamefont
  {Ding}, \citenamefont {Campbell}, \citenamefont {Chen}, \citenamefont {Chen},
  \citenamefont {Chiaro}, \citenamefont {Dunsworth}, \citenamefont {Fowler},
  \citenamefont {Jeffrey}, \citenamefont {Megrant}, \citenamefont {Mutus},
  \citenamefont {Neill}, \citenamefont {Quintana}, \citenamefont {Sank},
  \citenamefont {Vainsencher}, \citenamefont {Wenner}, \citenamefont {White},
  \citenamefont {Coveney}, \citenamefont {Love}, \citenamefont {Neven},
  \citenamefont {Aspuru-Guzik},\ and\ \citenamefont {Martinis}}]{OMalley2016}%
  \BibitemOpen
  \bibfield  {author} {\bibinfo {author} {\bibfnamefont {P.~J.~J.}\
  \bibnamefont {O'Malley}}, \bibinfo {author} {\bibfnamefont {R.}~\bibnamefont
  {Babbush}}, \bibinfo {author} {\bibfnamefont {I.~D.}\ \bibnamefont
  {Kivlichan}}, \bibinfo {author} {\bibfnamefont {J.}~\bibnamefont {Romero}},
  \bibinfo {author} {\bibfnamefont {J.~R.}\ \bibnamefont {McClean}}, \bibinfo
  {author} {\bibfnamefont {R.}~\bibnamefont {Barends}}, \bibinfo {author}
  {\bibfnamefont {J.}~\bibnamefont {Kelly}}, \bibinfo {author} {\bibfnamefont
  {P.}~\bibnamefont {Roushan}}, \bibinfo {author} {\bibfnamefont
  {A.}~\bibnamefont {Tranter}}, \bibinfo {author} {\bibfnamefont
  {N.}~\bibnamefont {Ding}}, \bibinfo {author} {\bibfnamefont {B.}~\bibnamefont
  {Campbell}}, \bibinfo {author} {\bibfnamefont {Y.}~\bibnamefont {Chen}},
  \bibinfo {author} {\bibfnamefont {Z.}~\bibnamefont {Chen}}, \bibinfo {author}
  {\bibfnamefont {B.}~\bibnamefont {Chiaro}}, \bibinfo {author} {\bibfnamefont
  {A.}~\bibnamefont {Dunsworth}}, \bibinfo {author} {\bibfnamefont {A.~G.}\
  \bibnamefont {Fowler}}, \bibinfo {author} {\bibfnamefont {E.}~\bibnamefont
  {Jeffrey}}, \bibinfo {author} {\bibfnamefont {A.}~\bibnamefont {Megrant}},
  \bibinfo {author} {\bibfnamefont {J.~Y.}\ \bibnamefont {Mutus}}, \bibinfo
  {author} {\bibfnamefont {C.}~\bibnamefont {Neill}}, \bibinfo {author}
  {\bibfnamefont {C.}~\bibnamefont {Quintana}}, \bibinfo {author}
  {\bibfnamefont {D.}~\bibnamefont {Sank}}, \bibinfo {author} {\bibfnamefont
  {A.}~\bibnamefont {Vainsencher}}, \bibinfo {author} {\bibfnamefont
  {J.}~\bibnamefont {Wenner}}, \bibinfo {author} {\bibfnamefont {T.~C.}\
  \bibnamefont {White}}, \bibinfo {author} {\bibfnamefont {P.~V.}\ \bibnamefont
  {Coveney}}, \bibinfo {author} {\bibfnamefont {P.~J.}\ \bibnamefont {Love}},
  \bibinfo {author} {\bibfnamefont {H.}~\bibnamefont {Neven}}, \bibinfo
  {author} {\bibfnamefont {A.}~\bibnamefont {Aspuru-Guzik}}, \ and\ \bibinfo
  {author} {\bibfnamefont {J.~M.}\ \bibnamefont {Martinis}},\ }\href {\doibase
  10.1103/PhysRevX.6.031007} {\bibfield  {journal} {\bibinfo  {journal}
  {Physical Review X}\ }\textbf {\bibinfo {volume} {6}},\ \bibinfo {pages}
  {031007} (\bibinfo {year} {2016})}\BibitemShut {NoStop}%
\bibitem [{\citenamefont {Babbush}\ \emph {et~al.}(2016)\citenamefont
  {Babbush}, \citenamefont {Berry}, \citenamefont {Kivlichan}, \citenamefont
  {Wei}, \citenamefont {Love},\ and\ \citenamefont
  {Aspuru-Guzik}}]{BabbushSparse1}%
  \BibitemOpen
  \bibfield  {author} {\bibinfo {author} {\bibfnamefont {R.}~\bibnamefont
  {Babbush}}, \bibinfo {author} {\bibfnamefont {D.~W.}\ \bibnamefont {Berry}},
  \bibinfo {author} {\bibfnamefont {I.~D.}\ \bibnamefont {Kivlichan}}, \bibinfo
  {author} {\bibfnamefont {A.~Y.}\ \bibnamefont {Wei}}, \bibinfo {author}
  {\bibfnamefont {P.~J.}\ \bibnamefont {Love}}, \ and\ \bibinfo {author}
  {\bibfnamefont {A.}~\bibnamefont {Aspuru-Guzik}},\ }\href
  {https://dx.doi.org/10.1088/1367-2630/18/3/033032} {\bibfield  {journal}
  {\bibinfo  {journal} {New Journal of Physics}\ }\textbf {\bibinfo {volume}
  {18}},\ \bibinfo {pages} {033032} (\bibinfo {year} {2016})}\BibitemShut
  {NoStop}%
\bibitem [{\citenamefont {Berry}\ \emph {et~al.}(2006)\citenamefont {Berry},
  \citenamefont {Ahokas}, \citenamefont {Cleve},\ and\ \citenamefont
  {Sanders}}]{Berry2006}%
  \BibitemOpen
  \bibfield  {author} {\bibinfo {author} {\bibfnamefont {D.~W.}\ \bibnamefont
  {Berry}}, \bibinfo {author} {\bibfnamefont {G.}~\bibnamefont {Ahokas}},
  \bibinfo {author} {\bibfnamefont {R.}~\bibnamefont {Cleve}}, \ and\ \bibinfo
  {author} {\bibfnamefont {B.~C.}\ \bibnamefont {Sanders}},\ }\href {\doibase
  10.1007/s00220-006-0150-x} {\bibfield  {journal} {\bibinfo  {journal}
  {Communications in Mathematical Physics}\ }\textbf {\bibinfo {volume}
  {270}},\ \bibinfo {pages} {359} (\bibinfo {year} {2006})}\BibitemShut
  {NoStop}%
\bibitem [{\citenamefont {Wiebe}\ \emph {et~al.}(2011)\citenamefont {Wiebe},
  \citenamefont {Berry}, \citenamefont {Hoyer},\ and\ \citenamefont
  {Sanders}}]{Wiebe2011}%
  \BibitemOpen
  \bibfield  {author} {\bibinfo {author} {\bibfnamefont {N.}~\bibnamefont
  {Wiebe}}, \bibinfo {author} {\bibfnamefont {D.~W.}\ \bibnamefont {Berry}},
  \bibinfo {author} {\bibfnamefont {P.}~\bibnamefont {Hoyer}}, \ and\ \bibinfo
  {author} {\bibfnamefont {B.~C.}\ \bibnamefont {Sanders}},\ }\href {\doibase
  10.1088/1751-8113/44/44/445308} {\bibfield  {journal} {\bibinfo  {journal}
  {Journal of Physics A: Mathematical and Theoretical}\ }\textbf {\bibinfo
  {volume} {44}},\ \bibinfo {pages} {445308} (\bibinfo {year}
  {2011})}\BibitemShut {NoStop}%
\bibitem [{\citenamefont {Gibney}(2014)}]{Gibney2014}%
  \BibitemOpen
  \bibfield  {author} {\bibinfo {author} {\bibfnamefont {E.}~\bibnamefont
  {Gibney}},\ }\href {\doibase 10.1038/516024a} {\bibfield  {journal} {\bibinfo
   {journal} {Nature}\ }\textbf {\bibinfo {volume} {516}},\ \bibinfo {pages}
  {24} (\bibinfo {year} {2014})}\BibitemShut {NoStop}%
\bibitem [{\citenamefont {Mueck}(2015)}]{Mueck2015}%
  \BibitemOpen
  \bibfield  {author} {\bibinfo {author} {\bibfnamefont {L.}~\bibnamefont
  {Mueck}},\ }\href {\doibase 10.1038/nchem.2248} {\bibfield  {journal}
  {\bibinfo  {journal} {Nature Chemistry}\ }\textbf {\bibinfo {volume} {7}},\
  \bibinfo {pages} {361} (\bibinfo {year} {2015})}\BibitemShut {NoStop}%
\bibitem [{\citenamefont {Berry}\ \emph {et~al.}(2015)\citenamefont {Berry},
  \citenamefont {Childs}, \citenamefont {Cleve}, \citenamefont {Kothari},\ and\
  \citenamefont {Somma}}]{Berry2015}%
  \BibitemOpen
  \bibfield  {author} {\bibinfo {author} {\bibfnamefont {D.~W.}\ \bibnamefont
  {Berry}}, \bibinfo {author} {\bibfnamefont {A.~M.}\ \bibnamefont {Childs}},
  \bibinfo {author} {\bibfnamefont {R.}~\bibnamefont {Cleve}}, \bibinfo
  {author} {\bibfnamefont {R.}~\bibnamefont {Kothari}}, \ and\ \bibinfo
  {author} {\bibfnamefont {R.~D.}\ \bibnamefont {Somma}},\ }\href {\doibase
  10.1103/PhysRevLett.114.090502} {\bibfield  {journal} {\bibinfo  {journal}
  {Physical Review Letters}\ }\textbf {\bibinfo {volume} {114}},\ \bibinfo
  {pages} {090502} (\bibinfo {year} {2015})}\BibitemShut {NoStop}%
\bibitem [{\citenamefont {Berry}\ \emph {et~al.}(2014)\citenamefont {Berry},
  \citenamefont {Childs}, \citenamefont {Cleve}, \citenamefont {Kothari},\ and\
  \citenamefont {Somma}}]{Berry2013}%
  \BibitemOpen
  \bibfield  {author} {\bibinfo {author} {\bibfnamefont {D.~W.}\ \bibnamefont
  {Berry}}, \bibinfo {author} {\bibfnamefont {A.~M.}\ \bibnamefont {Childs}},
  \bibinfo {author} {\bibfnamefont {R.}~\bibnamefont {Cleve}}, \bibinfo
  {author} {\bibfnamefont {R.}~\bibnamefont {Kothari}}, \ and\ \bibinfo
  {author} {\bibfnamefont {R.~D.}\ \bibnamefont {Somma}},\ }in\ \href
  {https://doi.org/10.1145/2591796.2591854} {\emph {\bibinfo {booktitle}
  {Proceedings of the Fourty-Sixth Annual ACM Symposium on Theory of Computing
  -- STOC '14}}}\ (\bibinfo {year} {2014})\ pp.\ \bibinfo {pages}
  {283--292}\BibitemShut {NoStop}%
\bibitem [{\citenamefont {Huzinaga}(1985)}]{Huzinaga85}%
  \BibitemOpen
  \bibfield  {author} {\bibinfo {author} {\bibfnamefont {S.}~\bibnamefont
  {Huzinaga}},\ }\href {\doibase 10.1016/0167-7977(85)90003-6} {\bibfield
  {journal} {\bibinfo  {journal} {Computer Physics Reports}\ }\textbf {\bibinfo
  {volume} {2}},\ \bibinfo {pages} {281} (\bibinfo {year} {1985})}\BibitemShut
  {NoStop}%
\bibitem [{\citenamefont {Kivlichan}\ \emph {et~al.}(2016)\citenamefont
  {Kivlichan}, \citenamefont {Wiebe}, \citenamefont {Babbush},\ and\
  \citenamefont {Aspuru-Guzik}}]{Kivlichan2016}%
  \BibitemOpen
  \bibfield  {author} {\bibinfo {author} {\bibfnamefont {I.~D.}\ \bibnamefont
  {Kivlichan}}, \bibinfo {author} {\bibfnamefont {N.}~\bibnamefont {Wiebe}},
  \bibinfo {author} {\bibfnamefont {R.}~\bibnamefont {Babbush}}, \ and\
  \bibinfo {author} {\bibfnamefont {A.}~\bibnamefont {Aspuru-Guzik}},\ }\href
  {{http://arxiv.org/abs/1608.05696}} {\bibfield  {journal} {\bibinfo
  {journal} {e-print arXiv:1608.05696}\ } (\bibinfo {year} {2016})}\BibitemShut
  {NoStop}%
\bibitem [{\citenamefont {Jordan}\ and\ \citenamefont
  {Wigner}(1928)}]{Jordan1928}%
  \BibitemOpen
  \bibfield  {author} {\bibinfo {author} {\bibfnamefont {P.}~\bibnamefont
  {Jordan}}\ and\ \bibinfo {author} {\bibfnamefont {E.}~\bibnamefont
  {Wigner}},\ }\href {http://dx.doi.org/10.1007/BF01331938} {\bibfield
  {journal} {\bibinfo  {journal} {Zeitschrift f\"{u}r Physik}\ }\textbf
  {\bibinfo {volume} {47}},\ \bibinfo {pages} {631} (\bibinfo {year}
  {1928})}\BibitemShut {NoStop}%
\bibitem [{\citenamefont {Somma}\ \emph {et~al.}(2002)\citenamefont {Somma},
  \citenamefont {Ortiz}, \citenamefont {Gubernatis}, \citenamefont {Knill},\
  and\ \citenamefont {Laflamme}}]{Somma2002}%
  \BibitemOpen
  \bibfield  {author} {\bibinfo {author} {\bibfnamefont {R.~D.}\ \bibnamefont
  {Somma}}, \bibinfo {author} {\bibfnamefont {G.}~\bibnamefont {Ortiz}},
  \bibinfo {author} {\bibfnamefont {J.}~\bibnamefont {Gubernatis}}, \bibinfo
  {author} {\bibfnamefont {E.}~\bibnamefont {Knill}}, \ and\ \bibinfo {author}
  {\bibfnamefont {R.}~\bibnamefont {Laflamme}},\ }\href {\doibase
  10.1103/PhysRevA.65.042323} {\bibfield  {journal} {\bibinfo  {journal}
  {Physical Review A}\ }\textbf {\bibinfo {volume} {65}},\ \bibinfo {pages}
  {17} (\bibinfo {year} {2002})}\BibitemShut {NoStop}%
\bibitem [{\citenamefont {Bravyi}\ and\ \citenamefont
  {Kitaev}(2002)}]{Bravyi2002}%
  \BibitemOpen
  \bibfield  {author} {\bibinfo {author} {\bibfnamefont {S.}~\bibnamefont
  {Bravyi}}\ and\ \bibinfo {author} {\bibfnamefont {A.}~\bibnamefont
  {Kitaev}},\ }\href {\doibase 10.1006/aphy.2002.6254} {\bibfield  {journal}
  {\bibinfo  {journal} {Annals of Physics}\ }\textbf {\bibinfo {volume}
  {298}},\ \bibinfo {pages} {210} (\bibinfo {year} {2002})}\BibitemShut
  {NoStop}%
\bibitem [{\citenamefont {Seeley}\ \emph {et~al.}(2012)\citenamefont {Seeley},
  \citenamefont {Richard},\ and\ \citenamefont {Love}}]{Seeley2012}%
  \BibitemOpen
  \bibfield  {author} {\bibinfo {author} {\bibfnamefont {J.~T.}\ \bibnamefont
  {Seeley}}, \bibinfo {author} {\bibfnamefont {M.~J.}\ \bibnamefont {Richard}},
  \ and\ \bibinfo {author} {\bibfnamefont {P.~J.}\ \bibnamefont {Love}},\
  }\href {\doibase 10.1063/1.4768229} {\bibfield  {journal} {\bibinfo
  {journal} {Journal of Chemical Physics}\ }\textbf {\bibinfo {volume} {137}},\
  \bibinfo {pages} {224109} (\bibinfo {year} {2012})}\BibitemShut {NoStop}%
\bibitem [{\citenamefont {Tranter}\ \emph {et~al.}(2015)\citenamefont
  {Tranter}, \citenamefont {Sofia}, \citenamefont {Seeley}, \citenamefont
  {Kaicher}, \citenamefont {McClean}, \citenamefont {Babbush}, \citenamefont
  {Coveney}, \citenamefont {Mintert},\ and\ \citenamefont
  {Love}}]{Tranter2015}%
  \BibitemOpen
  \bibfield  {author} {\bibinfo {author} {\bibfnamefont {A.}~\bibnamefont
  {Tranter}}, \bibinfo {author} {\bibfnamefont {S.}~\bibnamefont {Sofia}},
  \bibinfo {author} {\bibfnamefont {J.}~\bibnamefont {Seeley}}, \bibinfo
  {author} {\bibfnamefont {M.}~\bibnamefont {Kaicher}}, \bibinfo {author}
  {\bibfnamefont {J.}~\bibnamefont {McClean}}, \bibinfo {author} {\bibfnamefont
  {R.}~\bibnamefont {Babbush}}, \bibinfo {author} {\bibfnamefont {P.~V.}\
  \bibnamefont {Coveney}}, \bibinfo {author} {\bibfnamefont {F.}~\bibnamefont
  {Mintert}}, \ and\ \bibinfo {author} {\bibfnamefont {P.~J.}\ \bibnamefont
  {Love}},\ }\href {http://dx.doi.org/10.1002/qua.24969} {\bibfield  {journal}
  {\bibinfo  {journal} {International Journal of Quantum Chemistry}\ }\textbf
  {\bibinfo {volume} {115}},\ \bibinfo {pages} {1431} (\bibinfo {year}
  {2015})}\BibitemShut {NoStop}%
\bibitem [{\citenamefont {Helgaker}\ \emph {et~al.}(2002)\citenamefont
  {Helgaker}, \citenamefont {Jorgensen},\ and\ \citenamefont
  {Olsen}}]{Helgaker2002}%
  \BibitemOpen
  \bibfield  {author} {\bibinfo {author} {\bibfnamefont {T.}~\bibnamefont
  {Helgaker}}, \bibinfo {author} {\bibfnamefont {P.}~\bibnamefont {Jorgensen}},
  \ and\ \bibinfo {author} {\bibfnamefont {J.}~\bibnamefont {Olsen}},\
  }\href@noop {} {\emph {\bibinfo {title} {{Molecular Electronic Structure
  Theory}}}}\ (\bibinfo  {publisher} {Wiley},\ \bibinfo {year}
  {2002})\BibitemShut {NoStop}%
\bibitem [{\citenamefont {Aharonov}\ and\ \citenamefont
  {Ta-Shma}(2003)}]{Aharonov2003}%
  \BibitemOpen
  \bibfield  {author} {\bibinfo {author} {\bibfnamefont {D.}~\bibnamefont
  {Aharonov}}\ and\ \bibinfo {author} {\bibfnamefont {A.}~\bibnamefont
  {Ta-Shma}},\ }in\ \href {\doibase 10.1145/780542.780546} {\emph {\bibinfo
  {booktitle} {Proceedings of the Thirty-Fifth Annual ACM Symposium on Theory
  of Computing -- STOC '03}}}\ (\bibinfo {year} {2003})\ pp.\ \bibinfo {pages}
  {20--29}\BibitemShut {NoStop}%
\bibitem [{\citenamefont {Whitfield}\ \emph {et~al.}(2011)\citenamefont
  {Whitfield}, \citenamefont {Biamonte},\ and\ \citenamefont
  {Aspuru-Guzik}}]{Whitfield2010}%
  \BibitemOpen
  \bibfield  {author} {\bibinfo {author} {\bibfnamefont {J.~D.}\ \bibnamefont
  {Whitfield}}, \bibinfo {author} {\bibfnamefont {J.}~\bibnamefont {Biamonte}},
  \ and\ \bibinfo {author} {\bibfnamefont {A.}~\bibnamefont {Aspuru-Guzik}},\
  }\href {\doibase 10.1080/00268976.2011.552441} {\bibfield  {journal}
  {\bibinfo  {journal} {Molecular Physics}\ }\textbf {\bibinfo {volume}
  {109}},\ \bibinfo {pages} {735} (\bibinfo {year} {2011})}\BibitemShut
  {NoStop}%
\bibitem [{\citenamefont {Veis}\ and\ \citenamefont
  {Pittner}(2014)}]{Veis2014}%
  \BibitemOpen
  \bibfield  {author} {\bibinfo {author} {\bibfnamefont {L.}~\bibnamefont
  {Veis}}\ and\ \bibinfo {author} {\bibfnamefont {J.}~\bibnamefont {Pittner}},\
  }\href {\doibase 10.1063/1.4880755} {\bibfield  {journal} {\bibinfo
  {journal} {The Journal of Chemical Physics}\ }\textbf {\bibinfo {volume}
  {140}},\ \bibinfo {pages} {214111} (\bibinfo {year} {2014})}\BibitemShut
  {NoStop}%
\bibitem [{{\relax DLMF}()}]{NIST:DLMF}%
  \BibitemOpen
  {\relax DLMF},\ \href {http://dlmf.nist.gov/} {\enquote {\bibinfo {title}
  {{\it NIST Digital Library of Mathematical Functions}},}\ }\bibinfo
  {howpublished} {http://dlmf.nist.gov/, Release 1.0.14 of 2016-12-21},\
  \bibinfo {note} {{F}.~W.~J. Olver, A.~B. {Olde Daalhuis}, D.~W. Lozier, B.~I.
  Schneider, R.~F. Boisvert, C.~W. Clark, B.~R. Miller and B.~V. Saunders,
  eds.}\BibitemShut {Stop}%
\end{thebibliography}%

\appendix

\section{Decomposition into 1-sparse matrices}
\label{app:1sparseproof}

In \cite{Aharonov2003}, Aharonov and Ta-Shma considered the problem of simulating an arbitrary $d$-sparse Hamiltonian using the ability to query bits of the Hamiltonian. According to their prescription, we should imagine the Hamiltonian as an undirected graph where each basis state corresponds to a node and each nonzero matrix element $H^{\alpha\beta} = H^{\beta\alpha*} \neq 0$ corresponds to an edge which connects node $\ket{\alpha}$ to $\ket{\beta}$.
Since an edge coloring of a graph using $\Gamma$ colors is equivalent to the division of that graph into $\Gamma$ sets of disjoint graphs of degree 1, this edge coloring represents a decomposition of the Hamiltonian into $\Gamma$ 1-sparse matrices. Aharonov and Ta-Shma show a procedure for accomplishing the 1-sparse decomposition of any arbitrary $d$-sparse matrix using $\Theta(d^2)$ terms by coloring an arbitrary graph of degree $d$ with $\Theta(d^2)$ colors. This result was tightened from $\Theta(d^2)$ terms to $d^2$ terms in \cite{Berry2013}. Importantly, Aharonov and Ta-Shma also showed how these Hamiltonians can be efficiently simulated using an oracular scheme based on the Trotter-Suzuki decomposition. Toloui and Love used this result to show how the CI matrix can be efficiently simulated under Trotter-Suzuki decomposition with ${\cal O}(N^4)$ colors \cite{Toloui2013}.

We provide an improved 1-sparse decomposition into ${\cal O}(\eta^2 N^2)$ terms.
For convenience of notation, we denote the occupied spin-orbitals for $\ket{\alpha}$ by $\alpha_1,\ldots,\alpha_\eta$, and the occupied spin-orbitals for $\ket{\beta}$ by $\beta_1,\ldots,\beta_\eta$. We also drop the bra-ket notation for the lists of orbitals (Slater determinants); that is, we denote the list of occupied orbitals for the left portion of the graph by $\alpha$, and the list of occupied orbitals for the right portion of the graph by $\beta$.
We require both these lists of spin-orbitals to be sorted in ascending order.
According to the Slater-Condon rules, the matrix element between two Slater determinants is zero unless the determinants differ by two spin-orbitals or less. Thus, two vertices (Slater determinants) in the Hamiltonian graph are connected if and only if they differ by a single occupied orbital or two occupied orbitals.

In order to obtain the decomposition, for each color (corresponding to one of the resulting 1-sparse matrices) we need to be able to obtain $\beta$ from $\alpha$, and vice versa.
Using the approach in \cite{Berry2013}, we take the tensor product of the Hamiltonian with a $\sigma_x$ operator.
That is, we perform the simulation under the Hamiltonian $\sigma_x\otimes H$, which is bipartite and has the same sparsity as $H$.
The $\sigma_x$ operator acts on the ancilla register that determines whether we are in the left ($\alpha$) or right ($\beta$) partition of the graph.
We do this without loss of generality as simulation under $H$ can be recovered from simulation under $\sigma_x\otimes H$ using the fact that $e^{-i(\sigma_x\otimes H)t}\ket{+}\ket{\psi}=\ket{+}e^{-iHt}\ket{\psi}$ \cite{Berry2013}.

In order for the graph coloring to be suitable for the quantum algorithm, for any given color we must have a procedure for obtaining $\beta$ given $\alpha$, and another procedure for obtaining $\alpha$ given $\beta$.
For this to be a valid graph coloring, the procedure must be reversible, and different colors must not give the same $\beta$ from $\alpha$ or vice versa.

To explain the decomposition, we will first consider how it works for $\alpha$ and $\beta$ differing by only a single spin-orbital occupation.
We are given a 4-tuple $(a,b,\ell,p)$, where $a$ and $b$ are bits, $\ell$ is a number in the sorted list of occupied orbitals, and $p$ is a number that tells us how many orbitals the starting orbital is shifted by. Our notation here differs slightly from that in \sec{decomp1}, where $i$ and $j$ were used in place of $\ell$ to represent the positions of the two orbitals which differed: here we will use $i$ and $j$ for a different purpose.
To simplify the discussion, we do not perform the addition modulo $N$, and instead achieve the same effect by allowing $p$ to take positive and negative values.
If adding $p$ takes us beyond the list of allowable orbitals, then the matrix element returned is zero, and the list of occupied orbitals is unchanged (corresponding to a diagonal element of the Hamiltonian).
We will also use the convention that $\alpha_0=\beta_0=0$ and $\alpha_{\eta+1}=\beta_{\eta+1}=N+1$.
These values are not explicitly stored, but rather are dummy values to use in the algorithm when $\ell$ goes beyond the range $1,\ldots,\eta$.

The register $a$ tells us whether the $\ell$ is for $\alpha$ or $\beta$.
To simplify the discussion, when $a=0$ we take $i=\ell$, and when $a=1$ we take $j=\ell$.
In either case, we require that $\beta_j=\alpha_i+p$, but in the case $a=0$ we are given $i$ and need to work out $j$, whereas in the case $a=1$ we are given $j$ and need to work out $i$.
In particular, for $a=0$ we just take $\alpha_i$ and add $p$ to it.
Then $j$ is the new position in the list $\beta$, so $\beta_j=\alpha_i+p$.

The general principle is that, if we are given $i$ for $\alpha$ and need to determine $j$ for $\beta$, we require that $\beta_{j+1}-\beta_{j-1} \ge \alpha_{i+1}-\alpha_{i-1}$, (i.e.\ the spacing between orbitals is larger in $\beta$ than in $\alpha$).
Alternatively, if we were given $j$ for $\beta$ and needed to determine a corresponding $i$ for $\alpha$, we would require $\beta_{j+1}-\beta_{j-1} < \alpha_{i+1}-\alpha_{i-1}$ (i.e.\ the spacing between orbitals is larger in $\alpha$ than in $\beta$).
If the inequality is not consistent with the value of $a$ (i.e.\ we are proceeding in the wrong direction), then the matrix element for this term in the decomposition is taken to be zero (in the graph there is no line of that color connecting the nodes). This procedure allows for a unique connection between nodes, without double counting.

The reason for requiring these inequalities is that the list of orbitals with a larger spacing will have less ambiguity in the order of occupied orbitals.
To reduce the number of terms in the decomposition, we are only given $i$ or $j$, but not both, so we either need to be able to determine $j$ from $i$ given $\beta$, or $i$ from $j$ given $\alpha$.
When the spacing between the occupied orbitals for $\beta$ is larger, if we are given $\beta$ and $i$ there is less ambiguity in determining $j$.
In particular, when $\beta_{j+1}-\beta_{j-1} \ge \alpha_{i+1}-\alpha_{i-1}$, there can be at most two values of $j$ that could have come from $i$, and the bit $b$ is then used to distinguish between them.

There are four different cases that we need to consider.
\begin{enumerate}
	\item We are given $\beta$ and need to determine $\alpha$; $a=0$.
	\item We are given $\alpha$ and need to determine $\beta$; $a=0$.
	\item We are given $\alpha$ and need to determine $\beta$; $a=1$.
	\item We are given $\beta$ and need to determine $\alpha$; $a=1$.
\end{enumerate}
Next we explain the procedure for each of these cases in detail.
In the following we use the terminology ``\textsc{invalid}'' to indicate that we need to return $\alpha=\beta$ and a matrix element of zero.
\\ \\
\textbf{1. Given $\beta$ and need to determine $\alpha$; $a=0$.} \\ 
We are given $\beta$, but $\ell$ is the position in the list of occupied orbitals for $\alpha$.
We do not know which is the $\beta_j$ to subtract $p$ from, so we loop through all values as follows to find a list of candidates for $\alpha$, $\tilde\alpha^{(k)}$.
We define this as a procedure so we can use it later. \\ \\
procedure FindAlphas
\begin{adjustwidth}{1em}{0em}
	$k=0$  \\
	For $j=1,\ldots,\eta$:
	\begin{adjustwidth}{1em}{0em}
		Subtract $p$ from $\beta_j$ and check that this yields a valid list of orbitals, in that $\beta_j-p$ does not yield an orbital number beyond the desired range, or duplicate another orbital.
		That is:
		\\
		If $(( \beta_j-p\in\{1,\ldots,N\}) \wedge (\forall j'\in\{1,\ldots,\eta\} : \beta_j-p \ne \beta_{j'})) {\,\vee\, (p=0) }$ then
		\begin{adjustwidth}{1em}{0em}
			Sort the list of orbitals to obtain $\tilde\alpha^{(k)}$, and denote by $i$ the new position of $\beta_j-p$ in this list of occupied orbitals.
			Check that the new value of $i$ corresponds to $\ell$, and that the spacing condition for $a=0$ is satisfied, as follows.  \\
			If $(i=\ell)\wedge (\beta_{j+1}-\beta_{j-1} \ge \tilde\alpha^{(k)}_{i+1}-\tilde\alpha^{(k)}_{i-1})$ then
			\begin{adjustwidth}{1em}{0em}
				$k=k+1$
			\end{adjustwidth}
			end if
		\end{adjustwidth}
		end if
	\end{adjustwidth}
	end for
\end{adjustwidth}
end procedure\\ \\
After this procedure there is a list of at most two candidates for $\alpha$, and $k$ will correspond to how many have been found.
Depending on the value of $k$ we perform the following: \\
$\mathbf{k=0}$ We return \textsc{invalid}. \\
$\mathbf{k=1}$ If $b=0$ then return $\alpha=\tilde\alpha^{(0)}$, else return \textsc{invalid}. \\
$\mathbf{k=2}$ Return $\alpha=\tilde\alpha^{(b)}$. \\
That is, if we have two possibilities for $\alpha$, then we use $b$ to choose between them.
If there is only one, then we only return that one if $b=0$ to avoid obtaining two colors that both link $\alpha$ and $\beta$. \\ \\ 
\textbf{2. Given $\alpha$ and need to determine $\beta$; $a=0$.} \\
We are given $\alpha$, and $\ell=i$ is the position of the occupied orbital in $\alpha$ that is changed.
We therefore add $p$ to $\alpha_i$ and check that it gives a valid list of orbitals.
Not only this, we need to check that we would obtain $\alpha$ if we work backwards from the resulting $\beta$.
  \\
If $((\alpha_i+p\in\{1,\ldots,N\}) \wedge (\forall i'\in\{1,\ldots,\eta\} : \alpha_i+p\ne\alpha_{i'})) {\,\vee\, (p=0) }$ then
\begin{adjustwidth}{1em}{0em}
	We sort the new list of occupied orbitals to obtain a candidate for $\beta$, denoted $\tilde\beta$.
	We next check that the spacing condition for $a=0$ is satisfied.  \\
	If $(\tilde\beta_{j+1}-\tilde\beta_{j-1} \ge \alpha_{i+1}-\alpha_{i-1})$ then
	\begin{adjustwidth}{1em}{0em}
		Perform the procedure FindAlphas to find potential candidates for $\alpha$ that could be obtained from $\tilde\beta$.
		There can only be 1 or 2 candidates returned from this procedure. \\
		If $((k=1)\wedge (b=0)) \vee ((k=2)\wedge (\alpha=\tilde\alpha^{(b)}))$ then
		\begin{adjustwidth}{1em}{0em}
			return $\beta=\tilde\beta$
		\end{adjustwidth}
		else return \textsc{invalid}
	\end{adjustwidth}
	else return \textsc{invalid}
\end{adjustwidth}
else return \textsc{invalid} \\ \\
\textbf{3. Given $\alpha$ and need to determine $\beta$; $a=1$.} \\
This case is closely analogous to the case where we need to determine $\alpha$ from $\beta$, but $a=0$.
We are given $\alpha$, but $\ell$ is the position in the list of occupied orbitals for $\beta$.
We do not know which is the $\alpha_i$ to add $p$ to, so we loop through all values as follows to find a list of candidates for $\beta$, $\tilde\beta^{(k)}$.
We define this as a procedure so we can use it later. \\ \\
procedure FindBetas
\begin{adjustwidth}{1em}{0em}
	$k=0$  \\
	For $i=1,\ldots,\eta$:
	\begin{adjustwidth}{1em}{0em}
		Add $p$ to $\alpha_i$ and check that this yields a valid list of orbitals, in that $\alpha_i+p$ does not yield an orbital number beyond the desired range, or duplicate another orbital.
		That is:
		\\
		If $((\alpha_i+p\in\{1,\ldots,N\}) \wedge (\forall i'\in\{1,\ldots,\eta\} : \alpha_i+p\ne\alpha_{i'})) { \,\vee\, (p=0) }$ then
		\begin{adjustwidth}{1em}{0em}
			Sort the list of orbitals to obtain $\tilde\beta^{(k)}$, and denote by $j$ the new position of $\alpha_i+p$ in this list of occupied orbitals.
			Check that the new value of $j$ corresponds to $\ell$, and that the spacing condition for $a=1$ is satisfied.  \\
			If $(j=\ell)\wedge (\tilde\beta^{(k)}_{j+1}-\tilde\beta^{(k)}_{j-1} < \alpha_{i+1}-\alpha_{i-1})$ then
			\begin{adjustwidth}{1em}{0em}
				$k=k+1$
			\end{adjustwidth}
			end if
		\end{adjustwidth}
		end if
	\end{adjustwidth}
	end for
\end{adjustwidth}
end procedure\\ \\
After this procedure there is a list of at most two candidates for $\beta$, and $k$ will correspond to how many have been found.
Depending on the value of $k$ we perform the following: \\
$\mathbf{k=0}$ We return \textsc{invalid}. \\
$\mathbf{k=1}$ If $b=0$ then return $\beta=\tilde\beta^{(0)}$, else return \textsc{invalid}. \\
$\mathbf{k=2}$ Return $\beta=\tilde\beta^{(b)}$. \\
That is, if we have two possibilities for $\beta$, then we use $b$ to choose between them.
If there is only one, then we only return that one if $b=0$ to avoid obtaining two colors that both link $\alpha$ and $\beta$. \\ \\ 
\textbf{4. Given $\beta$ and need to determine $\alpha$; $a=1$.} \\
We are given $\beta$, and $\ell=j$ is the position of the occupied orbital in $\beta$ that is changed.
We therefore subtract $p$ from $\beta_j$ and check that it gives a valid list of orbitals.
Again we also need to check consistency.
That is, we work back again from the $\alpha$ to check that we correctly obtain $\beta$.
  \\
If $ ((\beta_j-p\in\{1,\ldots,N\}) \wedge (\forall j'\in\{1,\ldots,\eta\} : \beta_j-p\ne\beta_{j'})) { \,\vee\, (p=0) }$ then
\begin{adjustwidth}{1em}{0em}
	We sort the new list of occupied orbitals to obtain a candidate for $\alpha$, denoted $\tilde\alpha$.
	We next check that the spacing condition for $a=1$ is satisfied.  \\
	If $(\beta_{j+1}-\beta_{j-1} < \tilde\alpha_{i+1}-\tilde\alpha_{i-1})$ then
	\begin{adjustwidth}{1em}{0em}
		Perform the procedure FindBetas to find potential candidates for $\beta$ that could be obtained from $\tilde\alpha$.
		There can only be 1 or 2 candidates returned from this procedure. \\
		If $((k=1)\wedge (b=0)) \vee ((k=2)\wedge (\beta=\tilde\beta^{(b)}))$ then
		\begin{adjustwidth}{1em}{0em}
			return $\alpha=\tilde\alpha$
		\end{adjustwidth}
		else return \textsc{invalid}
	\end{adjustwidth}
	else return \textsc{invalid}
\end{adjustwidth}
else return \textsc{invalid}\\

To prove that this technique gives a valid coloring, we need to show that it is reversible and unique.
The most important part to show is that, provided the spacing condition holds, the ambiguity is limited to two candidates that may be resolved by the bit $b$.
We will consider the case that $p>0$; the analysis for $p<0$ is equivalent.

Consider Case 1, where we are given $\beta$ and need to determine $\alpha$, but $a=0$.
Then we take $i=\ell$, and need to determine $j$.
Let $j'$ and $j''$ be two potential values of $j$, with $j'<j''$.
For these to be potential values of $j$, they must satisfy
\begin{align}\label{con1}
\beta_{j'}-p &\in (\alpha_{i-1},\alpha_{i+1}) , \\
\label{con2}
\beta_{j'+1}-\beta_{j'-1} &\ge \alpha_{i+1}-\alpha_{i-1} , \\
\label{con3}
\beta_{j''}-p &\in (\alpha_{i-1},\alpha_{i+1}) , \\
\label{con4}
\beta_{j''+1}-\beta_{j''-1} &\ge \alpha_{i+1}-\alpha_{i-1}.
\end{align}
Condition \eqref{con1} is required because, for $j'$ to be a potential value of $j$, $\beta_{j'}-p$ must correspond to an $\alpha_i$ that is between $\alpha_{i-1}$ and $\alpha_{i+1}$ ($\alpha$ is sorted in ascending order).
Condition \eqref{con2} is the spacing condition for $a=0$.
Conditions \eqref{con3} and \eqref{con4} are simply the equivalent conditions for $j''$.

Next we consider how $\alpha$ is found from $\beta$.
In the case where $j'=i$, then we immediately know that $\alpha_{i-1}=\beta_{i-1}$ and $\alpha_{i+1}=\beta_{i+1}$.
Then the conditions \eqref{con1} and \eqref{con2} become
\begin{align}
\label{con5}
\beta_{j'}-p &\in (\beta_{i-1},\beta_{i+1}), \\
\label{con6}
\beta_{j'+1}-\beta_{j'-1} &\ge \beta_{i+1}-\beta_{i-1}.
\end{align}
In the case that $j'>i$, it is clear that $\alpha_{i-1}=\beta_{i-1}$ still holds.
Moreover, in going from the sequence of occupied orbitals for $\alpha$ to the sequence for $\beta$, we have then removed $\alpha_i$, which means that $\alpha_{i+1}$ has moved to position $i$.
That is to say, $\beta_{i}$ must be equal to $\alpha_{i+1}$.
Therefore, conditions \eqref{con1} and \eqref{con2} become
\begin{align}
\label{con9}
\beta_{j'}-p &\in (\beta_{i-1},\beta_i), \\
\label{con10}
\beta_{j'+1}-\beta_{j'-1} &\ge \beta_i-\beta_{i-1}.
\end{align}

In either case ($j'=i$ or $j'>i$), because $j''>j'$, we know that $j''>i$.
Then the same considerations as for $j'>i$ hold, and
conditions \eqref{con3} and \eqref{con4} become
\begin{align}
\label{con7}
\beta_{j''}-p &\in (\beta_{i-1},\beta_i), \\
\label{con8}
\beta_{j''+1}-\beta_{j''-1} &\ge \beta_i-\beta_{i-1}.
\end{align}
Using \eqref{con8} we have
\begin{align}
\beta_{j''+1}-p &\ge \beta_{j''-1}-p + \beta_i-\beta_{i-1} \nonumber \\
&\ge \beta_{j'}-p + \beta_i-\beta_{i-1} \nonumber \\
&> \beta_{i-1}+ \beta_i-\beta_{i-1} \nonumber \\
&= \beta_i.
\end{align}
In the second-last line we have used $\beta_{j'}-p>\beta_{i-1}$ from \eqref{con5} and \eqref{con9}, and in the second line we have used $j''>j'$.
The inequality $\beta_{j''+1}-p>\beta_i$ means that $\beta_{j''+1}-p\notin(\beta_{i-1},\beta_i)$, and therefore $\beta_{j''+1}$ could not have come from $\alpha_i$ by adding $p$.
That is because $\beta_{j''+1}$ would have to satisfy a relation similar to \eqref{con7}.
In turn, any $j>j''+1$ will satisfy $\beta_j-p>\beta_i$, because the $\beta_k$ are sorted in ascending order.

The net result of this reasoning is that, if there are two ambiguous values of $j$, then there can be no third ambiguous value.
This is because, if we call the first two ambiguous values $j'$ and $j''$, there can be no more ambiguous values for $j>j''$.
Hence, if we have a bit $b$ which tells us which of the two ambiguous values to choose, then it resolves the ambiguity and enables us to unambiguously determine $\alpha$, given $\beta$, $p$, and $i$.

Next consider Case 3, where we wish to determine $\beta$ from $\alpha$, but $a=1$.
In that case, we take $j=\ell$, and need to determine $i$.
That is, we wish to determine a value of $i$ such that adding $p$ to $\alpha_i$ gives $\beta_j$, and also require the condition $\beta_{j+1}-\beta_{j-1} < \alpha_{i+1}-\alpha_{i-1}$.
Now the situation is reversed; if we start with $\beta$, then we can immediately determine $\alpha$, but if we have $\alpha$ then we potentially need to consider multiple values of $i$ and resolve an ambiguity.
In exactly the same way as above, there are at most two possible values of $i$, and we distinguish between these using the bit $b$.

In this case, we cannot have $j=i$, because that would imply that $\alpha_k=\beta_k$ for all $k\ne j$, and the condition $\beta_{j+1}-\beta_{j-1} < \alpha_{i+1}-\alpha_{i-1}$ would be violated.
Therefore, consider two possible values of $i$, $i'$ and $i''$, with $i''<i'<j$.
The equivalents of the conditions in Eqs.~\eqref{con1} to \eqref{con4} are
\begin{align}\label{con1b}
\alpha_{i'}+p &\in (\beta_{j-1},\beta_{j+1}) , \\
\label{con2b}
\beta_{j'+1}-\beta_{j'-1} &< \alpha_{i+1}-\alpha_{i-1} , \\
\label{con3b}
\alpha_{i''}+p &\in (\beta_{j-1},\beta_{j+1}) , \\
\label{con4b}
\beta_{j''+1}-\beta_{j''-1} &< \alpha_{i+1}-\alpha_{i-1}.
\end{align}
Because $i''<i'<j$, using similar reasoning as before, we find that $\beta_{j+1}=\alpha_{j+1}$ and $\beta_{j-1}=\alpha_j$.
That means that the conditions \eqref{con1b} to \eqref{con4b} become
\begin{align}
\alpha_{i'}+p &\in (\alpha_j,\alpha_{j+1}), \\
\alpha_{j+1}-\alpha_j &< \alpha_{i'+1}-\alpha_{i'-1}, \\
\alpha_{i''}+p &\in (\alpha_j,\alpha_{j+1}), \\
\alpha_{j+1}-\alpha_j &< \alpha_{i''+1}-\alpha_{i''-1}.\label{conda}
\end{align}
Starting with Eq.~\eqref{conda} we obtain
\begin{align}
\alpha_{i''-1}+p &< \alpha_{i''+1}+p-\alpha_{j+1}+\alpha_j \nonumber \\
&\le \alpha_{i'}+p-\alpha_{j+1}+\alpha_j \nonumber \\
&< \alpha_{j+1}-\alpha_{j+1}+\alpha_j \nonumber \\
&= \alpha_j = \beta_{j-1}.
\end{align}
Hence $\alpha_{i''-1}+p$ is not in the interval $(\beta_{j-1},\beta_{j+1})$, and therefore cannot give $\beta_j$.
Therefore there can be no third ambiguous value, in the same way as above for $a=0$.
Hence the single bit $b$ is again sufficient to distinguish between any ambiguous values, and enables us to determine $\beta$ given $\alpha$, $p$, and $j$.

We now consider the requirement that the procedure is reversible.
In particular, Case 1 needs to be the reverse of Case 2, and Case 3 needs to be the reverse of Case 4.
Consider starting from a particular $\beta$ and using the method in Case 1.
We have shown that the procedure FindAlphas in Case 1 can yield at most two potential candidates for $\alpha$, and then one is chosen \textit{via} the value of $b$.
For the resulting $\alpha$, adding $p$ to $\alpha_i$ will yield the original set of occupied orbitals $\beta$.
Moreover, the inequality $\beta_{j+1}-\beta_{j-1} \ge \alpha_{i+1}-\alpha_{i-1}$ must be satisfied (otherwise Case 1 would yield \textsc{invalid}).

If Case 1 yields $\beta$ from $\alpha$, then Case 2 should yield $\beta$ given $\alpha$.
Case 2 simply adds $p$ to $\alpha_i$ (where $i$ is given), which we know should yield $\beta$.
The method in Case 2 also performs some checks, and outputs \textsc{invalid} if those fail.
These checks are:
\begin{enumerate}
\item It checks that $\beta$ is a valid list of orbitals, which must be satisfied because we started with a valid $\beta$.
\item It checks that $\beta_{j+1}-\beta_{j-1} \ge \alpha_{i+1}-\alpha_{i-1}$, which must be satisfied for Case 1 to yield $\alpha$ instead of \textsc{invalid}.
\item It checks that using Case 1 on $\beta$ would yield $\alpha$, which must be satisfied here because we considered initially using Case 1 to obtain $\alpha$ from $\beta$.
\end{enumerate}
Thus we see that, if Case 1 yields $\alpha$ from $\beta$, then Case 2 must yield $\beta$ from $\alpha$.

Going the other way, and starting with $\alpha$ and using Case 2 to find $\beta$, a result other than \textsc{invalid} will only be provided if Case 1 would yield $\alpha$ from that $\beta$.
Thus we immediately know that if Case 2 provides $\beta$ from $\alpha$, then Case 1 will provide $\alpha$ from $\beta$.
This means that the methods for Cases 1 and 2 are the inverses of each other, as required.
Via exactly the same reasoning, we can see that the methods in Cases 3 and 4 are the inverses of each other as well.

Next, consider the question of uniqueness.
The color will be unique if we can determine the color from a pair $\alpha$, $\beta$.
Given $\alpha$ and $\beta$, we will see that all the occupied orbitals are identical, except one.
Then the occupied orbitals for $\alpha$ and $\beta$ which are different will be $i$ and $j$, respectively.
We can then immediately set $p=\beta_j-\alpha_i$ for the color.
We can then compare $\beta_{j+1}-\beta_{j-1}$ and $\alpha_{i+1}-\alpha_{i-1}$.

If $\beta_{j+1}-\beta_{j-1}\ge \alpha_{i+1}-\alpha_{i-1}$ then for the color $a=0$ and $\ell=i$.
We can then find how many ambiguous values of $\alpha$ there would be if we started with $\beta$.
If $\alpha$ was obtained uniquely from $\beta$, then we would set $b=0$ for the color.
If there were two ambiguous values of $\alpha$ that could be obtained from $\beta$, then if the first was correct we would set $b=0$, and if the second were correct then we would set $b=1$.

If $\beta_{j+1}-\beta_{j-1}<\alpha_{i+1}-\alpha_{i-1}$ then for the color $a=1$ and $\ell=j$.
We can then find how many ambiguous values of $\beta$ there would be if we started with $\alpha$.
If $\beta$ was obtained uniquely from $\alpha$, then we would set $b=0$ for the color.
If there were two ambiguous values of $\beta$ that could be obtained from $\alpha$, then if the first was correct we would set $b=0$, and if the second were correct then we would set $b=1$.
In this way we can see that the pair $\alpha$, $\beta$ yields a unique color, and therefore we have a valid coloring.

So far we have considered the case where $\alpha$ and $\beta$ differ by just one orbital for simplicity.
For cases where $\alpha$ and $\beta$ differ by two orbitals, the procedure is similar.
We now need to use the above reasoning to go from $\alpha$ to $\beta$ through some intermediate list of orbitals $\chi$.
That is, we have one set of numbers $(a_1,b_1,\ell_1,p)$ that tells us how to find $\chi$ from $\alpha$, then a second set of numbers $(a_2,b_2,\ell_2,q)$ that tells us how to obtain $\beta$ from $\chi$.

First, it is easily seen that this procedure is reversible, because the steps for going from $\alpha$ to $\chi$ to $\beta$ are reversible.
Second, we need to be able to determine the color from $\alpha$ and $\beta$.
First, we find the two occupied orbitals for $\alpha$ and $\beta$ that differ.
Call the different occupied orbitals for $\alpha$ $i_1$ and $i_2$, and the different orbitals for $\beta$ $j_1$ and $j_2$ (assume in ascending order so the labels are unique).
Then there are four different ways that one could go from $\alpha$ to $\beta$, through different intermediate states $\chi$.
\begin{enumerate}
	\item $\alpha_{i_1} \mapsto \beta_{j_1}$ then $\alpha_{i_2}\mapsto\beta_{j_2}$
	\item $\alpha_{i_2}\mapsto\beta_{j_2}$ then $\alpha_{i_1} \mapsto \beta_{j_1}$
	\item $\alpha_{i_1} \mapsto \beta_{j_2}$ then $\alpha_{i_2}\mapsto\beta_{j_1}$
	\item $\alpha_{i_2}\mapsto\beta_{j_1}$ then $\alpha_{i_1} \mapsto \beta_{j_2}$
\end{enumerate}
To resolve this ambiguity we require that the color is obtained by assuming the first alternative that $\alpha_{i_1} \mapsto \beta_{j_1}$ then $\alpha_{i_2}\mapsto\beta_{j_2}$.
Then $\alpha$ and $\beta$ yield a unique color.
This also requires a slight modification of the technique for obtaining $\alpha$ from $\beta$ and vice versa.
First the color is used to obtain the pair $\alpha$, $\beta$, then it is checked whether the orbitals were mapped as in the first alternative above.
If they were not, then \textsc{invalid} is returned.

To enable us to include the matrix elements where $\alpha$ and $\beta$ differ by a single orbital or no orbitals with a coloring by an 8-tuple
$\gamma = (a_1,b_1,\ell_1,p,a_2,b_2,\ell_2,q)$, we can also allow $p=0$ (for only one differing) and $p=q=0$ (for $\alpha=\beta$).
The overall number of terms in the decomposition is then ${\cal O}(\eta^2 N^2)$.

\section{Riemann Sum Approximations of Hamiltonian Coefficients}
\label{app:integral_discretization}

The aim of this Appendix is to prove \lem{int0}, \lem{int1} and \lem{int2}.
We begin in \app{integral_discretization/preliminaries} with preliminary matters
that are integral to the proofs themselves.
We then prove the Lemmas in
\app{integral_discretization/int0_proof},
\app{integral_discretization/int1_proof} and
\app{integral_discretization/int2_proof} respectively.

Throughout this Appendix, we employ the following two notational conventions.
\begin{enumerate}
  \item   The vector symbol $\vec{\bullet}$ refers to an element of $\R^3$.
          We write $\bullet$ for the Euclidean length of $\vec{\bullet}$.
          Thus $\vec{v}$ refers to a $3$-vector of magnitude $v$.
          We denote the zero vector as $\vec{0} = (0, 0, 0)$.
          We use $\oplus$ to denote vector concatenation:
          if $\vec{v} = \left( v_1, v_2, v_3 \right)$
          and $\vec{w} = \left( w_1, w_2, w_3 \right)$, we write
          $\vec{v} \oplus \vec{w} = \left(v_1, v_2, v_3, w_1, w_2, w_3 \right)$.
          The gradient operator over $\R^6$ is then written as
          $\nabla_1 \oplus \nabla_2$.
  \item   If $x$ is a positive real number and $\vec{v}$ is a $3$-vector,
          we write $\mathcal{B}_x (\vec{v})$ for the closed ball of radius $x$
          centered at $\vec{v}$ and $\mathcal{C}_x (\vec{v})$ for the closed
          cube of side length $2x$ centered at $\vec{v}$.
          Thus $\mathcal{B}_x (\vec{v}) \subset \mathcal{C}_x (\vec{v})$
          and $\mathcal{B}_y (\vec{v}) \not\subseteq \mathcal{C}_x (\vec{v})$
          whenever $y > x$.
\end{enumerate}
This notation will be used extensively and without comment in what follows.

\subsection{Preliminaries}
\label{app:integral_discretization/preliminaries}

The purpose of this subsection is to present two key discussions
that will be needed at many points in the proofs of this Appendix.
First, in \app{integral_discretization/preliminaries/proof_structure},
we discuss the general structure of the proofs of \lem{int0}, \lem{int1}
and \lem{int2}.
Second, in \app{integral_discretization/preliminaries/exp_Coulomb_potential},
we prove an ancillary Lemma (\lem{exp_Coulomb_potential_bound})
that we use several times.

The ancillary Lemma offers bounds on the function
\begin{equation}
\label{eq:exp_Coulomb_potential}
  \Lambda_{\mu, x} (\vec{c})
  :=    \int_{\R^3 \setminus \mathcal{B}_x (\vec{0})}
        \frac{\exp(-\mu r)}{\norm{\vec{r}-\vec{c}}} \dd{\vec{r}},
\end{equation}
where $\mu$ is a positive real constant and $\vec{c}$ is a constant vector.
The Lemma is stated as follows.
\begin{lemma}
\label{lem:exp_Coulomb_potential_bound}
  Suppose $\mu$ and $x$ are positive real numbers and
  $\vec{c} \in \R^3$ is some constant vector.  Then
  \begin{equation}
  \label{eq:exp_Coulomb_potential_bound_c>x}
    \Lambda_{\mu, x} (\vec{c})
    <   \frac{16\pi}{\mu^3 c} e^{-\mu x/2},
  \end{equation}
  and for $c \leq x$,
 \begin{equation}
 \label{eq:exp_Coulomb_potential_bound_c<x}
   \Lambda_{\mu, x} (\vec{c}) < \frac{8\pi}{\mu^2} e^{-\mu x/2}.
 \end{equation}
\end{lemma}
The function $\Lambda_{\mu, x} (\vec{c})$ appears in bounds derived
in the proofs of \lem{int1} and \lem{int2}.  Although it is possible to compute
an analytic formula for the value of integral, the result is unwieldy.
The bounds of \lem{exp_Coulomb_potential_bound} are then used to ensure
meaningfully expressed bounds on the Riemann sum approximations.

\subsubsection{Structure of the Proofs}
\label{app:integral_discretization/preliminaries/proof_structure}

The proofs of \lem{int0}, \lem{int1} and \lem{int2} each roughly follow
a general structure consisting of the following three stages,
though with minor deviations.

\textbf{First stage:}
The domain of integration is truncated to a domain $D$.
The size of $D$ is specified by a positive real parameter $x$,
which the conditions of the lemmas ensure is at least $x_{\max}$.
We then bound the error due to the truncation
\begin{equation}
  \delta_\text{trunc}
  :=    \abs{
          \int_{\R^d} f\left(\vec{r}\right) \dd{\vec{r}} -
          \int_{D} f\left(\vec{r}\right) \dd{\vec{r}}
        },
\end{equation}
where $f: \R^d \rightarrow \R$ refers to the relevant integrand.

\textbf{Second stage:}
We specify a Riemann sum that is designed to approximate this truncated integral
and give a bound on the error $\delta_\text{Riemann}$ of this
Riemann sum approximation.  We specify the number of terms in the Riemann sum
in order to give the bound on $\mu$ in the lemma.  We also give a bound on
the absolute value of each term in the Riemann sum using the value of $x$
specified in the first stage.

\textbf{Third stage:}
In the final stage of each proof, we bound the total error
\begin{equation}
  \delta_\text{total} :=
  \abs{
    \int_{\R^d}  f(\vec{r}) \, \dd{\vec{r}} -
    \sum_T f(\vec{r}_T) \text{vol}(T)
  }
\end{equation}
via the triangle inequality as
\begin{equation}
\label{eq:total_error_bound_trunc_Riemann}
  \delta_\text{total}
  \leq    \delta_\text{trunc} + \delta_\text{Riemann}.
\end{equation}
Our choice of $x$ then ensures that the error is bounded by $\inter$.

To be more specific about the approach in the second stage,
we partition $D$ into regions $T$, and the Riemann sum approximates
the integral over each $T$ with the value of the integrand multiplied by
the volume of $T$.  The error due to this approximation is bounded by
observing the following.
Suppose $f: \R^d \rightarrow \R$ is once-differentiable and
$f^\prime_{\max}$ is a bound on its first derivative.
If $\vec{r}_T$ is any element of $T$, we will seek to bound the error
of the approximation
\begin{equation}
  \int_T f\left(\vec{r}\right) \dd{\vec{r}}
  \approx   f\left(\vec{r}_T\right) \text{vol}(T),
\end{equation}
where $\text{vol}(T)$ is the $d$-dimensional hypervolume of the set $T$.
The error of this approximation is
\begin{equation}
    \delta_T
    :=    \abs{
            \int_T
              \left[ f\!\left(\vec{r}\right) - f\!\left(\vec{r}_T\right) \right]
            \dd{\vec{r}}
          },
\end{equation}
which can be bounded as follows:
\begin{equation}
  \delta_T
  \leq    \int_T
          \abs{f\!\left(\vec{r}\right) - f\!\left(\vec{r}_T\right)}\dd{\vec{r}}
  \leq    \max_{\vec{r} \in T}
          \abs{f\!\left(\vec{r}\right) - f\!\left(\vec{r}_T\right)}
          \text{vol}(T)
          \leq f^\prime_{\max}\max_{\vec{r} \in T}
          \norm{\vec{r} -\vec{r}_T}
          \text{vol}(T)
\end{equation}
where
\begin{equation}
  \text{vol}(T)
  =       \int_T \dd{\vec{r}}, \qquad  f^\prime_{\max}
  =       \max_{\vec r} \norm{\nabla f(\vec r)}.
\end{equation}
We will choose the points $\vec{r}_T$ in the centers of the regions $T$, so that
\begin{equation}
  \delta_T
  \leq \frac 12 f^\prime_{\max} \text{diam}(T) \text{vol} (T),
\end{equation}
where
\begin{equation}
  \text{diam}(T) := \max_{\vec{r}_1,\vec{r}_2 \in T} \norm{\vec{r}_1-\vec{r}_2}.
\end{equation}
The Riemann sum approximations we define will then take the form
\begin{equation}
  \int_D f(\vec{r})\, \dd{\vec{r}} \approx \sum_T f(\vec{r}_T) \text{vol}(T),
\end{equation}
and the error of this approximation is
\begin{equation}
  \delta_\text{Riemann} :=
  \abs{\int_D  f(\vec{r}) \,\dd{\vec{r}}- \sum_T f(\vec{r}_T) \text{vol}(T)},
\end{equation}
which can be bounded via the triangle inequality as
\begin{equation}
\label{eq:Riemann_error}
  \delta_\text{Riemann}
  \leq    \sum_T \delta_T
  \leq  \frac 12  f^\prime_{\max} \text{vol}(D) \left(\max_T \text{diam}(T)\right).
\end{equation}

\subsubsection{Proof of \lem{exp_Coulomb_potential_bound}}
\label{app:integral_discretization/preliminaries/exp_Coulomb_potential}

We prove the Lemma by deriving exact formulae for $\Lambda_{\mu, x} (\vec{c})$
in the cases $c \leq x$ and $c > x$ and then deriving bounds on these formulae
that have simpler functional forms.

To derive exact formulae for $\Lambda_{\mu, x} (\vec{c})$,
we use the Laplace expansion
\begin{equation}
  \frac{1}{\norm{\vec{r}-\vec{c}}}
  =	\sum_{\ell = 0}^\infty \sum_{m=-\ell}^\ell (-1)^m
    I_{\ell, -m} (\vec{r}) R_{\ell, m} (\vec{c}),
\end{equation}
where $R_{\ell, m}$ and $I_{\ell, m}$ refer to the regular and irregular solid
spherical harmonic functions, respectively, and $r \geq c$.  That is to say,
\begin{equation}
  R_{\ell, m} (\vec{r})
  :=  \sqrt{\frac{4\pi}{2\ell+1}} r^\ell Y_{\ell, m} (\theta,\phi)
\end{equation}
and
\begin{equation}
  I_{\ell, m} (\vec{r})
  :=  \sqrt{\frac{4\pi}{2\ell+1}} \frac{1}{r^{\ell+1}} Y_{\ell, m} (\theta,\phi),
\end{equation}
where
\begin{equation}
  Y_{\ell, m} (\theta, \phi)
  :=  \sqrt{\frac{2\ell + 1}{4\pi} \frac{(\ell-m)!}{(\ell+m)!}}
      e^{i m \phi} P_\ell^m (\cos\theta)
\end{equation}
are the spherical harmonics (see \S 14.30(i) in \cite{NIST:DLMF}),
$P_\ell^m$ are the associated Legendre polynomials, and $\theta$
and $\phi$ are respectively the polar and azimuthal angles of $\vec{r}$.
Via Eq.~(8) of \S 14.30(ii) in \cite{NIST:DLMF}, we have
\begin{equation}
  \int_0^{2\pi} \dd{\phi} \int_0^\pi \dd{\theta}\
    I_{\ell, m} (\theta,\phi) \sin\theta
  =   \frac{4\pi}{r} \delta_{m,0} \delta_{\ell,0}
\end{equation}
and
\begin{equation}
  \int_0^{2\pi} \dd{\phi} \int_0^\pi \dd{\theta}\
    R_{\ell, m} (\theta,\phi) \sin\theta
  =   4\pi \delta_{m,0} \delta_{\ell,0},
\end{equation}
where $\delta_{a,b}$ denotes the Kronecker delta.

If $c \leq x$:
\begin{align}
\label{eq:Lamba_c<x_exact}
  \Lambda_{\mu, x} (\vec{c})
  & =   \sum_{\ell = 0}^\infty \sum_{m=-\ell}^\ell (-1)^m R_{\ell, m} (\vec{c})
        \int_{\R^3 \setminus \mathcal{B}_x (\vec{0})} \dd{\vec{r}} \,
        e^{-\mu r} I_{\ell, -m} (\vec{r}) \nn
  & =   4 \pi R_{0,0} (\vec{c}) \int_x^\infty \dd{r}\ r e^{-\mu r} \nn
  & =   4\pi \left( \frac{x}{\mu} + \frac{1}{\mu^2} \right) e^{-\mu x}.
\end{align}
\eq{exp_Coulomb_potential_bound_c<x} follows from the fact that
$(1 + z) e^{-z} < 2 e^{-z/2}$ for all $z > 0$.

If $c > x$:
\begin{align}
  \Lambda_{\mu, x} (\vec{c}) - \Lambda_{\mu, c} (\vec{c})
  & =   \int_{\mathcal{B}_c (\vec{0})\setminus\mathcal{B}_x (\vec{0})}\dd{\vec{r}}
      \ \frac{\exp(-\mu r)}{\norm{\vec{r}-\vec{c}}} \nn
  & =   \sum_{\ell = 0}^\infty \sum_{m=-\ell}^\ell (-1)^m I_{\ell,m} (\vec{c})
        \int_{\mathcal{B}_c (\vec{0}) \setminus \mathcal{B}_x (\vec{0})} \dd{\vec{r}}
        \ e^{-\mu r} R_{\ell,-m} (\vec{r}) \nn
  & =   \frac{4\pi}{c} \left[
          \left( \frac{x^2}{\mu} + \frac{2x}{\mu^2} + \frac{2}{\mu^3} \right)
          e^{-\mu x} -
          \left( \frac{c^2}{\mu} + \frac{2c}{\mu^2} + \frac{2}{\mu^3} \right)
          e^{-\mu c}
        \right].
\end{align}
Therefore,
\begin{align}
  \Lambda_{\mu, x} (\vec{c})
  & =   \Lambda_{\mu, c} (\vec{c}) + \frac{4\pi}{c} \left[
          \left( \frac{x^2}{\mu} + \frac{2x}{\mu^2} + \frac{2}{\mu^3} \right)
          e^{-\mu x} -
          \left( \frac{c^2}{\mu} + \frac{2c}{\mu^2} + \frac{2}{\mu^3} \right)
          e^{-\mu c}
        \right] \nn
  & =   \frac{4\pi}{c} \left[
          \left( \frac{x^2}{\mu} + \frac{2x}{\mu^2} + \frac{2}{\mu^3} \right)
          e^{-\mu x} -
          \left( \frac{c}{\mu^2} + \frac{2}{\mu^3} \right)
          e^{-\mu c}
        \right] \nn
  & <   \frac{4\pi}{c}
        \left( \frac{x^2}{\mu} + \frac{2x}{\mu^2} + \frac{2}{\mu^3} \right)
        e^{-\mu x} \nn
  & <   \frac{16\pi}{\mu^3 c} e^{-\mu x/2},
\end{align}
where we use the fact that $(z^2 + 2z + 2) e^{-z} < e^{-z/2}$ for any $z>0$
and the fact that
\begin{equation}
  \Lambda_{\mu, c} (\vec{c})
  =     4\pi \left( \frac{c}{\mu} + \frac{1}{\mu^2} \right) e^{-\mu c},
\end{equation}
which follows from \eq{Lamba_c<x_exact}.
This gives us \eq{exp_Coulomb_potential_bound_c>x} for $c>x$.
In the case that $c\le x$,
\begin{equation}
\frac{16\pi}{\mu^3 c} e^{-\mu x/2} \ge \frac{16\pi}{\mu^3 x} e^{-\mu x/2},
\end{equation}
and, since $(z^2 + z) e^{-z} < 4e^{-z/2}$ for all $z > 0$, we have
\begin{equation}
\frac{16\pi}{\mu^3 x} e^{-\mu x/2} >
4\pi \left( \frac{x}{\mu} + \frac{1}{\mu^2} \right) e^{-\mu x}.
\end{equation}
Therefore the bound \eq{exp_Coulomb_potential_bound_c>x} holds for $c\le x$ as well.

\subsection{Proof of \lem{int0}}
\label{app:integral_discretization/int0_proof}

Our proof for \lem{int0} roughly follows the three stages presented in
\app{integral_discretization/preliminaries/proof_structure}.
Here we give the proof in summary form and relegate some of the details
to the later subsections.

\subsubsection{First stage for \lem{int0}}

The first part of the proof corresponds to the first stage discussed in
\app{integral_discretization/preliminaries/proof_structure}.
We choose
\begin{align}
\label{eq:xvallem1}
  x_0
  &:=     \frac{2}{\alpha} x_\text{max} \log \left(
          \frac{K_0 \varphi_\text{max}^2 x_\text{max}}{\inter}
        \right), \nn
  D_0 &:= \mathcal{C}_{x_0} \left( \vec{c}_i \right).
\end{align}
The condition \eq{sensible0} ensures that $x_0 \geq x_\text{max}$.
We show in \app{integral_discretization/int0_proof/trunc} that the error due to
this truncation can be bounded as
\begin{equation}
  \delta_\text{trunc}^{(0)}
  :=    \abs{
          S_{ij}^{(0)} \!\left( \R^3 \right) -
          S_{ij}^{(0)} \!\left( D_0 \right)
        }
  <     \frac{8\pi\gamma_2}{\alpha^3} \varphi_\text{max}^2 x_\text{max}
        \exp \left( -\frac{\alpha}{2} \frac{x_0}{x_\text{max}} \right).
\end{equation}

\subsubsection{Green's identity for \lem{int0}}

The next part of the proof is specific to \lem{int0},
and is not one of the general stages outlined in
\app{integral_discretization/preliminaries/proof_structure}.
The integral is given in the form with a second derivative of an orbital,
which means that to bound the error we would need additional bounds on
the third derivatives of the orbitals.
We have not assumed such bounds, so we would like to reexpress the integral
in terms of first derivatives before approximating it as a Riemann sum.
We have already truncated the domain, though, so we will obtain terms from
the boundary of the truncated domain.

We reexpress the integral via Green's first identity, which gives
\begin{equation}\label{eq:greensfirst}
	S_{ij}^{(0)} \!\left( D_0 \right)
	=   \frac{1}{2} \int_{D_0}
        \nabla \varphi_i^* (\vec{r}) \cdot
        \nabla \varphi_j (\vec{r})\
      \dd{V} - \frac{1}{2} \oint_{\partial D_0}
        \varphi_i^* (\vec{r})
        \nabla \varphi_j (\vec{r}) \cdot
      \dd{\vec{S}},
\end{equation}
where $\dd{V}$ and $\dd{\vec{S}}$ are the volume and oriented surface elements,
respectively, and $\partial D_0$ is the boundary of $D_0$.
The reason why we do not make this change before truncating the domain is that
we have not made any assumptions on the rate of decay of the derivatives of
the orbitals.
 We define
\begin{equation}
	\widetilde{S}_{ij}^{(0)} \!\left( D_0 \right)
	:=		\frac{1}{2} \int_{D_0}
          \nabla \varphi_i^* (\vec{r}) \cdot
          \nabla \varphi_j (\vec{r})
       \  \dd{V}.
\end{equation}
We show (in \app{integral_discretization/int0_proof/Green}) that
\begin{equation}\label{eq:greenserror}
  \delta_\text{Green}^{(0)}
  :=      \abs{S_{ij}^{(0)}\! \left( D_0 \right) -
          \widetilde{S}_{ij}^{(0)}\! \left( D_0 \right)}
	<    		\frac{26 \gamma_1}{\alpha^2} \varphi_\text{max}^2 x_\text{max}
			    \exp \left( - \frac{\alpha}{2} \frac{x_0}{x_\text{max}} \right).
\end{equation}

\subsubsection{Second stage for \lem{int0}}

Next we consider the discretization into a Riemann sum for \lem{int0}.
We define
\begin{equation}
\label{eq:partition_size_int0}
  N_0
  :=  \left\lceil
        \left( \frac{x_0}{x_\text{max}} \right)^4
        \exp \left( \frac{\alpha}{2} \frac{x_0}{x_\text{max}} \right)
      \right\rceil,
\end{equation}
so that
\begin{equation}
  N_0
  =     \left\lceil
          \frac{K_0 \varphi_\text{max}^2 x_\text{max}}{\inter}
          \left[
            \frac{2}{\alpha} \log \left(
              \frac{K_0 \varphi_\text{max}^2 x_\text{max}}{\inter}
            \right)
          \right]^4
        \right\rceil.
\end{equation}
The Riemann sum is then
\begin{equation}
  \mathcal{R}_0
  :=      \sum_{\vec{k}} \frac{1}{2}
          \nabla \varphi_i^* \!\left( \vec{r}_{\vec{k}} \right) \cdot
          \nabla \varphi_j \!\left( \vec{r}_{\vec{k}} \right)
          \text{vol} \left( T_{\vec{k}}^{(0)} \right),
\end{equation}
where, for every triple of integers
$\vec{k} = \left( k_1, k_2 , k_3 \right)$
such that $0 \leq k_1, k_2, k_3 < N_0$, we define
\begin{equation}
  \vec{r}_{\vec{k}}
  =       \frac{x_0}{N_0} \left[
            2\vec{k} - \left(N_0-1,N_0-1,N_0-1\right)
          \right]
\end{equation}
and
\begin{equation}
  T_{\vec{k}}^{(0)}
  :=    \mathcal{C}_{x_0/N_0} \left( \vec{r}_{\vec{k}} \right).
\end{equation}
Thus we have partitioned $D_0$ into $\mu = N_0^3$ equal-sized cubes
$T_{\vec{k}}^{(0)}$ that overlap on sets of measure zero.
The expression in \eq{lem1mu} of \lem{int0} then follows immediately.

Each term of $\mathcal{R}_0$ satisfies
\begin{align}
\label{eq:int0_summand_bound}
  \left\|
    \frac{1}{2}
    \nabla \varphi_i^*\! \left( \vec{r}_{\vec{k}} \right) \cdot
    \nabla \varphi_j\! \left( \vec{r}_{\vec{k}} \right)
    \text{vol} \left( T_{\vec{k}}^{(0)} \right)
  \right\|
  & \leq  \frac{1}{2}
          \left\|\nabla \varphi_i^* \!\left( \vec{r}_{\vec{k}} \right)\right\|
          \left\|\nabla \varphi_j \!\left( \vec{r}_{\vec{k}} \right)\right\|
          \text{vol} \left( T_{\vec{k}}^{(0)} \right) \nn
  & \leq  \frac{1}{2} \left(
            \gamma_1 \frac{\varphi_\text{max}}{x_\text{max}}
          \right)^2 \left( \frac{2x_0}{N_0} \right)^3 \nn
  & =     \frac{1}{\mu} \times
          4 \gamma_1^2 \left( \frac{x_0}{x_\text{max}} \right)^3
          \varphi_\text{max}^2 x_\text{max},
\end{align}
where the second inequality follows from
\eq{spin-orbital_first_derivative_bound}.
Using the value of $x_0$ in \eq{xvallem1} in \eq{int0_summand_bound},
each term in the sum has the upper bound on its absolute value
(corresponding to \eq{lem1bnd} in \lem{int0})
\begin{equation}
  \frac{1}{\mu} \times32 \frac{\gamma_1^2}{\alpha^3}
  \left[
    \log \left(
      \frac{K_0 \varphi_\text{max}^2 x_\text{max}}{\inter}
    \right)
  \right]^3 \varphi_\text{max}^2 x_\text{max}.
\end{equation}

We show (in \app{integral_discretization/int0_proof/Riemann}) that
\begin{equation}
  \delta_\text{Riemann}^{(0)}
  :=      \abs{
            \widetilde{S}_{ij}^{(0)} \!\left( D_0 \right) - \mathcal{R}_0
          }
  <       16\sqrt{3} \gamma_1 \gamma_2 \varphi_\text{max}^2 x_\text{max}
          \exp \left( - \frac{\alpha}{2} \frac{x_0}{x_\text{max}} \right).
\end{equation}

\subsubsection{Third stage for \lem{int0}}
In the final part of the proof of \lem{int0} we show that the total error is
properly bounded.  By the triangle inequality, we have
\begin{equation}
  \delta_\text{total}^{(0)}
  :=      \abs{S_{ij}^{(0)} \!\left( \R^3 \right) - \mathcal{R}_0}
  \leq    \delta_\text{trunc}^{(0)} +
          \delta_\text{Green}^{(0)} +
          \delta_\text{Riemann}^{(0)}.
\end{equation}
We therefore arrive at a simple bound on the total error:
\begin{equation}
\label{eq:total_error_int0}
  \delta_\text{total}^{(0)}
  <       K_0 \varphi_\text{max}^2 x_\text{max}
          \exp \left( - \frac{\alpha}{2} \frac{x_0}{x_\text{max}} \right),
\end{equation}
where
\begin{equation}
  K_0
  :=      \frac{26 \gamma_1}{\alpha^2} +
          \frac{8\pi\gamma_2}{\alpha^3} +
          16\sqrt{3} \gamma_1 \gamma_2.
\end{equation}
To ensure that $\delta^{(0)}_\text{total} \leq \inter$, we should have
\begin{equation}
  K_0 \varphi_\text{max}^2 x_\text{max}
  \exp \left( - \frac{\alpha}{2} \frac{x_0}{x_\text{max}} \right)
  \le     \inter.
\end{equation}
We can satisfy this inequality with $x_0$ given by \eq{xvallem1}.
This last step completes our proof.  The remainder of this subsection
gives the details for some of the steps above.

\subsubsection{Bounding $\delta_\text{trunc}^{(0)}$ for \lem{int0}}
\label{app:integral_discretization/int0_proof/trunc}

Observe first that
\begin{equation}
  \delta_\text{trunc}^{(0)}
  =       \abs{
            S_{ij}^{(0)} \!\left(\R^3\right) - S_{ij}^{(0)} \!\left( D_0 \right)
          }
  \leq    \frac{1}{2} \int_{\R^3 \setminus D_0}
            \abs{\varphi_i^* \!\left(\vec{r}\right)}
            \abs{\nabla^2 \varphi_j \!\left(\vec{r}\right)}
          \dd{\vec{r}}
  \leq    \frac{1}{2} \int_{\R^3 \setminus \mathcal{B}_x \left(\vec{c}_i\right)}
            \abs{\varphi_i^* \!\left(\vec{r}\right)}
            \abs{\nabla^2 \varphi_j \!\left(\vec{r}\right)}
          \dd{\vec{r}},
\end{equation}
where the last inequality follows from the fact that
$\mathcal{B}_x \left(\vec{c}_i\right) \subset D_0$.
Using this fact together with assumptions 2 and 3 from \sec{Riemann}, we have
\begin{equation}
  \delta_\text{trunc}^{(0)}
  \leq    \frac{\gamma_2}{2} \frac{\varphi_\text{max}^2}{x_\text{max}^2}
          \int_{\R^3 \setminus \mathcal{B}_{x_0} \left(\vec{c}_i\right)}
          \exp \left(
            -\alpha \frac{\norm{\vec{r} - \vec{c}_i}}{x_\text{max}}
          \right)
          \dd{\vec{r}}.
\end{equation}
We simplify this expression by expressing $\vec{r}$ in polar coordinates
with center $\vec{c}_i$.  After integrating over the solid angles, we have
\begin{equation}
  \delta_\text{trunc}^{(0)}
  \leq    2\pi \gamma_2 \frac{\varphi_\text{max}^2}{x_\text{max}^2}
          \int_{x_0}^\infty s^2 e^{-\alpha s / x_\text{max}} \dd{s}.
\end{equation}
Noting that
\begin{equation}
  \int_{x_0}^\infty s^2 e^{-\mu s} \dd{s}
  =     \left( \frac{x_0^2}{\mu} + \frac{2x_0}{\mu^2} + \frac{2}{\mu^3} \right)
        e^{-\mu x_0}
  <     \frac{4}{\mu^3} e^{-\mu x_0/2},
\end{equation}
we have
\begin{equation}
  \delta_\text{trunc}^{(0)}
  <     \frac{8\pi\gamma_2}{\alpha^3} \varphi_\text{max}^2 x_\text{max}
        \exp \left( -\frac{\alpha}{2} \frac{x_0}{x_\text{max}} \right).
\end{equation}

\subsubsection{Bounding $\delta_\text{Green}^{(0)}$ for \lem{int0}}
\label{app:integral_discretization/int0_proof/Green}

Using \eq{greensfirst} and \eq{greenserror} we have
\begin{equation}
\delta_\text{Green}^{(0)} = \abs{\frac{1}{2} \oint_{\partial D_0}
                \varphi_i^* (\vec{r})
                \nabla \varphi_j (\vec{r}) \cdot
              \dd{\vec{S}}}.
\end{equation}
Then using \eq{spin-orbital_decay} and
\eq{spin-orbital_first_derivative_bound} gives us
\begin{equation}
	\delta_\text{Green}^{(0)}
	\leq		\frac{\gamma_1}{2} \frac{\varphi_\text{max}^2}{x_\text{max}}
			    \oint_{\partial D_0}
			    \exp \left(
            -\alpha \frac{\norm{\vec{r}-\vec{c}_i}}{x_\text{max}}
          \right) \dd{S}.
\end{equation}
We further observe that $\norm{\vec{r}-\vec{c}_i} \geq x$
for all $\vec{r} \in \partial D_0$, and the cube with side length $2x$ has
surface area $24 x^2$, giving
\begin{equation}
	\delta_\text{Green}^{(0)}
	\leq		12 \gamma_1 \frac{\varphi_\text{max}^2}{x_\text{max}}
			    x_0^2 \exp \left( -\alpha \frac{x_0}{x_\text{max}} \right)
	<    		\frac{26 \gamma_1}{\alpha^2} \varphi_\text{max}^2 x_\text{max}
			    \exp \left( - \frac{\alpha}{2} \frac{x_0}{x_\text{max}} \right),
\end{equation}
where we have noted $12 z^2 e^{-z} < 26 e^{-z/2}$ for all $z > 0$.

\subsubsection{Bounding $\delta_\text{Riemann}^{(0)}$}
\label{app:integral_discretization/int0_proof/Riemann}

First we bound the derivative of the integrand.
We use the chain rule, the triangle inequality,
\eq{spin-orbital_first_derivative_bound} and
\eq{spin-orbital_second_derivative_bound} to find
\begin{equation}
	\norm{
    \nabla \left( \nabla \varphi_i^* (\vec{r}) \cdot
    \nabla \varphi_j (\vec{r}) \right)
  }
  \leq    \abs{\nabla^2 \varphi_i^* (\vec{r})}
          \norm{\nabla \varphi_j (\vec{r})} +
          \abs{\nabla^2 \varphi_j (\vec{r})}
          \norm{\nabla \varphi_i^* (\vec{r})}
	\leq		2 \gamma_1 \gamma_2 \frac{\varphi_\text{max}^2}{x_\text{max}^3}.
\end{equation}
We have
\begin{equation}
  \text{vol} \left( D_0 \right) = 8 x_0^3
\end{equation}
and
\begin{equation}
  \text{diam} \left( T_{\vec{k}}^{(0)} \right) = 2 \sqrt{3} x_0 / N_0
\end{equation}
for all $\vec{k}$.
Using \eq{Riemann_error} and \eq{partition_size_int0}, and
noting that $1/\lceil z \rceil \leq 1/z$ for any $z\in\R$, we have
\begin{equation}
  \delta_\text{Riemann}^{(0)}
  \leq    16\sqrt{3} \gamma_1 \gamma_2
          \frac{\varphi_\text{max}^2}{x_\text{max}^3} \frac{x_0^4}{N_0}
  \leq    16\sqrt{3} \gamma_1 \gamma_2 \varphi_\text{max}^2 x_\text{max}
          \exp \left( - \frac{\alpha}{2} \frac{x_0}{x_\text{max}} \right).
\end{equation}

\subsection{Proof of \lem{int1}}
\label{app:integral_discretization/int1_proof}

For this proof, the discretization into the Riemann sum will be performed
differently depending on whether spin-orbital $i$ is considered distant from
or nearby to nucleus $q$.  If the nucleus is far from the spin-orbital,
the singularity in the integrand is not inside our truncated
domain of integration and we need not take special care with it.  Otherwise,
we can remove the singularity by defining spherical polar coordinates
centred at the nucleus.  In each case, we select different truncated integration
domains and therefore different Riemann sums.

We focus on the centre of spin-orbital $i$ for simplicity; in principle,
the centre of spin-orbital $j$ could also be taken into account.

\subsubsection{First stage for \lem{int1}}

We again start by truncating the domain of integration.  We select
\begin{equation}\label{eq:xvallem2}
x_1=\frac{2}{\alpha} x_\text{max} \log \left(
\frac{K_1 Z_q \varphi_\text{max}^2 x_\text{max}^2}{\inter}\right).
\end{equation}
The condition in \eq{sensible1} ensures that $x_1 \geq x_\text{max}$.
We regard the spin-orbital as distant from the nucleus if
\begin{equation}
  \norm{\vec{R}_q - \vec{c}_i} \geq \sqrt{3}x_1 + x_\text{max}.
\end{equation}
If so, we use the truncated domain
\begin{equation}
  D_{1, q, \text{non-singular}}
  :=    \mathcal{C}_{x_1}\! \left( \vec{c}_i \right) .
\end{equation}
Otherwise, we use
\begin{equation}
D_{1, q, \text{singular}}
:=    \mathcal{B}_{4x_1}\! \left( \vec{R}_q \right).
\end{equation}
We define
\begin{align}
\delta_\text{trunc}^{(1, q, \text{non-singular})}
& :=    \abs{
        S_{ij}^{(1,q)}\! \left( \R^3 \right) -
        S_{ij}^{(1,q)}\! \left( D_{1, q, \text{non-singular}} \right)
      },
\\
  \delta_\text{trunc}^{(1, q, \text{singular})}
&  :=    \abs{
          S_{ij}^{(1,q)}\! \left( \R^3 \right) -
          S_{ij}^{(1,q)}\! \left( D_{1, q, \text{singular}} \right)
        },
\\
  \delta_\text{trunc}^{(1, q)}
&  :=    \max \left\{
          \delta_\text{trunc}^{(1, q, \text{non-singular})},
          \delta_\text{trunc}^{(1, q, \text{singular})}
        \right\},
\end{align}
and show in \app{integral_discretization/int1_proof/trunc} that
\begin{equation}
  \delta_\text{trunc}^{(1, q)}
  <       \frac{8\pi^2}{\alpha^3}\left( \alpha+2 \right)
          Z_q \varphi_\text{max}^2 x_\text{max}^2
          \exp \left( - \frac{\alpha}{2} \frac{x_1}{x_\text{max}} \right).
\end{equation}

\subsubsection{Second stage for \lem{int1} with Cartesian coordinates}

Now we consider in the discretization of the integral for the case that
$\norm{\vec{R}_q - \vec{c}_i} \geq \sqrt{3}x_1 + x_\text{max}$, so orbital $i$
can be regarded as distant from the nucleus.  We set
\begin{equation}
\label{eq:partition_size_int1}
  N_1
  :=    \left\lceil
          \left( \frac{x_1}{x_\text{max}} \right)^4
          \exp \left( \frac{\alpha}{2} \frac{x_1}{x_\text{max}} \right)
        \right\rceil
\end{equation}
and define two different Riemann sums containing $\mu = N_1^3$ terms.
We also use this expression for $N_1$ in the case that  the spin-orbital is
near the nucleus.  Using our value of $x_1$ in \eq{xvallem2},
\begin{equation}
  N_1
  =       \left\lceil
            \frac{K_1 Z_q \varphi_\text{max}^2 x_\text{max}^2}{\inter}
            \left[
              \frac{2}{\alpha} \log \left(
                \frac{K_1 Z_q \varphi_\text{max}^2 x_\text{max}^2}{\inter}
              \right)
            \right]^4
          \right\rceil  .
\end{equation}
Then, noting that $\mu = N_1^3$ is the number of terms in either Riemann sum,
we obtain the bound on $\mu$ in \eq{lem2mu} of \lem{int1}.

We approximate
$S_{ij}^{(1,q)} \!\left( D_{1, q, \text{non-singular}} \right)$
with the sum
\begin{equation}
  \mathcal{R}_{1, q, \text{non-singular}}
  :=      \sum_{\vec{k}} -Z_q \frac{
            \varphi_i^* \!\left( \vec{r}_{\vec{k}} \right)
            \varphi_j \!\left( \vec{r}_{\vec{k}} \right)
          }{
            \|\vec{R}_q - \vec{r}_{\vec{k}}\|
          }
          \text{vol} \left(
            T_{\vec{k}}^{(1, q, \text{non-singular})}
          \right),
\end{equation}
where, for every triple of integers
$\vec{k} = \left( k_1, k_2 , k_3 \right)$
such that $0 \leq k_1, k_2, k_3 < N_1$, we define
\begin{equation}
  \vec{r}_{\vec{k}}
  =       \frac{x_1}{N_1} \left[
            2\vec{k} - \left(N_1-1, N_1-1, N_1-1\right)
          \right]
\end{equation}
and
\begin{equation}
  T_{\vec{k}}^{(1, q, \text{non-singular})}
  :=    \mathcal{C}_{x_1/N_1}\! \left( \vec{r}_{\vec{k}} \right).
\end{equation}
Thus we have partitioned $D_{1, q, \text{non-singular}}$ into $N_1^3$
equal-sized cubes $T_{\vec{k}}^{(1, q, \text{non-singular})}$
that overlap on sets of measure zero.
Each term of $\mathcal{R}_{1, q, \text{non-singular}}$ satisfies
\begin{align}
\label{eq:int1_summand_bound_non-sing}
  \abs{
    -Z_q \frac{
      \varphi_i^* \!\left( \vec{r}_{\vec{k}} \right)
      \varphi_j \!\left( \vec{r}_{\vec{k}} \right)
    }{
      \|\vec{R}_q - \vec{r}_{\vec{k}}\|
    }
    \text{vol} \left(
      T_{\vec{k}}^{(1, q, \text{non-singular})}
    \right)
  }
  & \leq  Z_q \frac{
            \abs{\varphi_i^* \!\left( \vec{r}_{\vec{k}} \right)}
            \abs{\varphi_j \!\left( \vec{r}_{\vec{k}} \right)}
          }{\|\vec{R}_q - \vec{r}_{\vec{k}}\|}
          \text{vol} \left(
            T_{\vec{k}}^{(1, q, \text{non-singular})}
          \right) \nn
  & \leq  Z_q \frac{\varphi_\text{max}^2}{x_\text{max}}
          \left( \frac{2x_1}{N_1} \right)^3 \nn
  & =     \frac{1}{\mu} \times 8 x_1^3
          \frac{Z_q \varphi_\text{max}^2}{x_\text{max}},
\end{align}
where we have used \eq{spin-orbital_bound} and the fact that
$\|\vec{R}_q - \vec{r}\| \geq x_\text{max}$ for every
$\vec{r} \in D_{1, q, \text{non-singular}}$.
This upper bound is no greater than
\begin{equation}\label{eq:bndlem2}
  \frac{1}{\mu} \times 32 \pi^2 x_1^3
    \frac{Z_q \varphi_\text{max}^2}{x_\text{max}},
\end{equation}
Now substituting our value of $x_1$ from \eq{xvallem2} shows that
no term has absolute value greater than (corresponding to \eq{lem2bnd} in \lem{int1})
\begin{equation}
  \frac{1}{\mu} \times
  \frac{256\pi^2}{\alpha^3}
  Z_q \varphi_\text{max}^2 x_\text{max}^2 \left[
    \log \left(
      \frac{K_1 Z_q \varphi_\text{max}^2 x_\text{max}^2}{\inter}
    \right)
  \right]^3.
\end{equation}

We show in \app{integral_discretization/int1_proof/Riemann_non-sing}
that
\begin{align}
  \delta_\text{Riemann}^{(1, q, \text{non-singular})}
  & :=    \abs{
            S_{ij}^{(1, q)}\! \left( D_{1, q, \text{non-singular}} \right)
            - \mathcal{R}_{1, q, \text{non-singular}}
          } \nn
  & \leq  8 \sqrt{3} \left( 2 \gamma_1 + 1 \right)
          Z_q \varphi_\text{max}^2 x_\text{max}^2
          \exp \left(-\frac{\alpha}{2} \frac{x_1}{x_\text{max}}\right).
\end{align}

\subsubsection{Second stage for \lem{int1} with spherical polar coordinates}

Next we consider discretization of the integral for the case where
$\norm{\vec{R}_q - \vec{c}_i} < \sqrt{3}x_1 + x_\text{max}$, so orbital
$i$ is nearby the nucleus.  We express
\begin{equation}
  S_{ij}^{(1,q)} \! \left( D_{1, q, \text{singular}} \right)
  =     - 16 x_1^2 Z_q
        \int_0^{2\pi} \dd{\phi} \int_0^\pi \dd{\theta} \int_0^1 \dd{s}\
        f_1 (s, \theta, \phi),
\end{equation}
where we define $\vec{s} := ( \vec{r} - \vec{R}_q )/(4x_1)$ and
\begin{equation}
  f_1 (s, \theta, \phi)
  :=    \varphi_i^* \! \left( 4x_1 \vec{s} + \vec{R}_q \right)
        \varphi_j \! \left( 4x_1 \vec{s} + \vec{R}_q \right) s \sin\theta.
\end{equation}
Here we use $\theta$ and $\phi$ to refer to the polar and azimuthal
angles, respectively, of the vector $\vec{s}$.  Note that
the singularity in the nuclear Coulomb potential has been absorbed into
the spherical polar volume form
$s^2 \sin\theta\ \dd{s}\ \dd{\theta}\ \dd{\phi}$.
For every triple of natural numbers
$\vec{k} = \left( k_s, k_\theta, k_\phi \right)$ such that
$0 \leq k_s, k_\theta, k_\phi < N_1$, we define
\begin{equation}
  T_{\vec{k}}^{(1, q, \text{singular})}
  :=    \left\{ \vec{s}
          \left|
            \begin{array}{c}
              k_s/N_1 \leq s \leq \left(k_s + 1\right)/N_1 \\
              k_\theta \pi/N_1 \leq
              \theta \leq \left(k_\theta + 1\right) \pi/N_1 \\
              2 k_\phi \pi/N_1 \leq
              \phi \leq 2 \left(k_\phi + 1\right) \pi/N_1 \\
            \end{array}
          \right.
        \right\}
\end{equation}
so that $\bigcup_{\vec{k}} T_{\vec{k}}^{(1, q, \text{singular})} =
D_{1, q, \text{singular}}$.  We select
\begin{equation}
  \left( s_{\vec{k}}, \theta_{\vec{k}}, \phi_{\vec{k}} \right)
  =     \left(
          \frac{1}{N_1} \left(k_s + \tfrac{1}{2}\right),
          \frac{\pi}{N_1} \left(k_\theta + \tfrac{1}{2}\right),
          \frac{2\pi}{N_1} \left(k_\phi + \tfrac{1}{2}\right)
        \right).
\end{equation}
Thus our Riemann sum approximation is
\begin{equation}
  \mathcal{R}_{1, q, \text{singular}}
  :=    \sum_{\vec{k}}
          -16x_1^2 Z_q
          f_1\! \left( s_{\vec{k}},\theta_{\vec{k}},\phi_{\vec{k}} \right)
          \text{vol} \left(
            T_{\vec{k}}^{(1, q, \text{singular})}
          \right),
\end{equation}
where we emphasize that
\begin{equation}
\label{eq:int1_nonstandard_volume}
  \text{vol} \left( T_{\vec{k}}^{(1, q, \text{singular})} \right)
  =         \int_{T_{\vec{k}}^{(1, q, \text{singular})}}
            \dd{s}\, \dd{\theta}\, \dd{\phi}
  =         \frac{2 \pi^2}{N_1^3}
\end{equation}
is \emph{not} the volume of $T_{\vec{k}}^{(1, q, \text{singular})}$
when considered as a subset of $\R^3$ under the usual Euclidean metric.
The reason for this discrepancy is that we absorbed the Jacobian
introduced by switching from Cartesian to spherical polar coordinates
into the definition of $f_1$.  Thus we are integrating $f_1$
with respect to the volume form $\dd{s}\,\dd{\theta}\,\dd{\phi}$,
not $s^2 \sin\theta\,\dd{s}\,\dd{\theta}\,\dd{\phi}$.  The terms of
$\mathcal{R}_{1, q, \text{singular}}$ are bounded by
\begin{align}
\label{eq:integrandlem22}
  \abs{
    -16x_1^2 Z_q f_1 \!\left( s_{\vec{k}},
                          \theta_{\vec{k}},
                          \phi_{\vec{k}} \right)
    \text{vol} \left(
      T_{\vec{k}}^{(1, q, \text{singular})}
    \right)
  }
  & =       16x_1^2 Z_q \abs{
              f_1 \!\left( s_{\vec{k}},
                        \theta_{\vec{k}},
                        \phi_{\vec{k}} \right)
            }
            \text{vol} \left(
              T_{\vec{k}}^{(1, q, \text{singular})}
            \right) \nn
  & \leq    \frac{1}{\mu} \times
            32 \pi^2 x_1^2 Z_q \varphi_\text{max}^2,
\end{align}
where the inequality follows from \eq{spin-orbital_bound}.
Again this expression is upper bounded by \eq{bndlem2}, so substituting
our value of $x_1$ from \eq{xvallem2} gives the upper bound in \eq{lem2bnd}.

We show in \app{integral_discretization/int1_proof/Riemann-sing} that
\begin{align}
  \delta_\text{Riemann}^{(1, q, \text{singular})}
  & :=    \abs{
            S_{ij}^{(1, q)} \left( D_{1, q, \text{singular}} \right)
            - \mathcal{R}_{1, q, \text{singular}}
          } \nn
  & <     1121 \left( 8 \gamma_1 + \sqrt{2} \right)
          Z_q \varphi_\text{max}^2 x_\text{max}^2
          \exp \left( -\frac{\alpha}{2} \frac{x_1}{x_\text{max}} \right).
\end{align}

\subsubsection{Third stage for \lem{int1}}

We again finish the proof by showing that the total error is bounded by $\inter$.
From \eq{total_error_bound_trunc_Riemann}, we have
\begin{align}
  \delta_\text{total}^{(1, q, \text{non-singular})}
 & :=      \abs{
            S_{ij}^{(1,q)} \left( \R^3 \right) -
            \mathcal{R}_{1, q, \text{non-singular}}
          }
  \leq    \delta_\text{trunc}^{(1, q, \text{non-singular})} +
          \delta_\text{Riemann}^{(1, q, \text{non-singular})},\\
  \delta_\text{total}^{(1, q, \text{singular})}
&  :=      \abs{
            S_{ij}^{(1,q)} \left( \R^3 \right) -
            \mathcal{R}_{1, q, \text{singular}}
          }
  \leq    \delta_\text{trunc}^{(1, q, \text{singular})} +
          \delta_\text{Riemann}^{(1, q, \text{singular})}.
\end{align}
We have given a bound that holds simultaneously for both
$\delta_\text{trunc}^{(1, q, \text{non-singular})}$
and $\delta_\text{trunc}^{(1, q, \text{singular})}$,
and we have given a bound for
$\delta_\text{Riemann}^{(1, q, \text{singular})}$
that is larger (as a function of $x$) than our bound for
$\delta_\text{Riemann}^{(1, q, \text{non-singular})}$.
We are therefore able to assert that the error of our Riemann sum
approximation, no matter which we choose, is always bounded above by
\begin{equation}
\label{lem2errbnd}
  K_1 Z_q \varphi_\text{max}^2 x_\text{max}^2
  \exp \left( -\frac{\alpha}{2} \frac{x_1}{x_\text{max}} \right),
\end{equation}
where
\begin{equation}
  K_1
  :=    \frac{2\pi^2}{\alpha^3}\left( 5\alpha+1 \right) +
        1121 \left( 8 \gamma_1 + \sqrt{2} \right).
\end{equation}
We have found two different upper bounds on the magnitudes of the terms in the
Riemann sums given in Eqs.~\eqref{eq:int1_summand_bound_non-sing}
and \eqref{eq:integrandlem22}.  Finally, we note that by substituting our value
of $x_1$ from \eq{xvallem2}, this expression is upper bounded by $\inter$.
This last step completes our proof of \lem{int1}.  The remainder of this
subsection gives the details for some of the steps above.

\subsubsection{Bounding $\delta_\text{trunc}^{(1,q)}$ for \lem{int1}}
\label{app:integral_discretization/int1_proof/trunc}

Note first that
$\mathcal{B}_{x_1}\! \left( \vec{c}_i \right) \subset D_{1, q, \text{non-singular}}$
and
$\mathcal{B}_{x_1}\! \left( \vec{c}_i \right) \subset D_{1, q, \text{singular}}$.
To see the latter, note that we only consider $D_{1, q, \text{singular}}$
in the case that $\|\vec{R}_q - \vec{c}_i \| \leq \sqrt{3}x_1 + x_\text{max}$,
which implies that
\begin{equation}
  \max_{\vec{r} \in \mathcal{B}_{x_1} \left( \vec{c}_i \right)}
  \|\vec{R}_q - \vec{r}\|
  \leq    x_1 + \|\vec{R}_q - \vec{c}_i\|
  \leq    (\sqrt{3} + 1)x_1  + x_\text{max}
  <       4x_1 .
\end{equation}
As we have
\begin{equation}
  \abs{S^{(1,q)}\! \left(\R^3\right) - S^{(1,q)} (D)}
  \leq  Z_q \int_{\R^3 \setminus D} \dd{\vec{r}}\
        \frac{
          \abs{\varphi_i^* (\vec{r}) \varphi_j (\vec{r})}
        }{
          \|\vec{R}_q - \vec{r}\|
        }
  \leq  Z_q \int_{\R^3 \setminus \mathcal{B}_{x_1} \left( \vec{c}_i \right)}
        \dd{\vec{r}}\
        \frac{
          \abs{\varphi_i^* (\vec{r}) \varphi_j (\vec{r})}
        }{
          \|\vec{R}_q - \vec{r}\|
        }
\end{equation}
for any $D$ such that $\mathcal{B}_x \left( \vec{c}_i \right) \subset D$,
we may compute
\begin{equation}
	\delta_\text{trunc}^{(1,q)}
	\leq  Z_q \varphi_\text{max}^2
        \int_{\R^3 \setminus \mathcal{B}_{x_1} \left( \vec{c}_i \right)}
        \frac{
          \exp \left(
            - \alpha \frac{\norm{\vec{r} - \vec{c}_i}}{x_\text{max}}
          \right)
        }{\|\vec{R}_q - \vec{r}\|} \dd{\vec{r}} \\
	=     Z_q \varphi_\text{max}^2
        \Lambda_{\alpha/x_\text{max}, x_1}\!
        \left( \vec{R}_q - \vec{c}_i \right),
\end{equation}
where the function $\Lambda$ is as defined in \eq{exp_Coulomb_potential}
and the inequality follows from \eq{spin-orbital_decay}.
By \lem{exp_Coulomb_potential_bound}, in the case $\|\vec{R}_q - \vec{c}_i\|>x_1$ we have
\begin{align}
  \delta_\text{trunc}^{(1,q)}
   &<    \frac{16 \pi^2}{\alpha^3} Z_q \varphi_\text{max}^2
        \frac{x_\text{max}^3}{\|\vec{R}_q - \vec{c}_i\|}
        \exp \left( -\frac{\alpha}{2} \frac{x_1}{x_\text{max}} \right)\nn
   &<    \frac{16 \pi^2}{\alpha^3} Z_q \varphi_\text{max}^2 x_\text{max}^2
        \exp \left( -\frac{\alpha}{2} \frac{x_1}{x_\text{max}} \right),
\end{align}
where the second inequality follows from $x_1\ge x_\text{max}$.
In the case $\|\vec{R}_q - \vec{c}_i\|\le x_1$ we can use
\begin{equation}
  \delta_\text{trunc}^{(1,q)}
   <    \frac{8 \pi^2}{\alpha^2} Z_q \varphi_\text{max}^2
        x_\text{max}^2
        \exp \left( -\frac{\alpha}{2} \frac{x_1}{x_\text{max}} \right).
\end{equation}
We can add the bounds to find, in general, that
\begin{equation}
  \delta_\text{trunc}^{(1,q)}
   <    \frac{8 \pi^2}{\alpha^3} (\alpha+2) Z_q \varphi_\text{max}^2
        x_\text{max}^2
        \exp \left( -\frac{\alpha}{2} \frac{x_1}{x_\text{max}} \right).
\end{equation}

\subsubsection{Bounding $\delta_\text{Riemann}^{(1, q, \text{non-singular})}$ for \lem{int1}}
\label{app:integral_discretization/int1_proof/Riemann_non-sing}

Following \app{integral_discretization/preliminaries/proof_structure},
we note that
\begin{equation}
  \text{vol} \left( D_{1, q, \text{non-singular}} \right)
  =     8x_1^3
\end{equation}
and
\begin{equation}
  \text{diam} \left( T_{\vec{k}}^{(1, q, \text{non-singular})} \right)
  =     2 \sqrt{3} x_1 / N_1
\end{equation}
for each $\vec{k}$.
We can bound the derivative of the integrand using
the product rule and the triangle inequality as follows:
\begin{equation}
  \norm{
    \nabla \frac{
      \varphi_i^* (\vec{r}) \varphi_j (\vec{r})
    }{
      \|\vec{R}_q - \vec{r}\|
    }
  }
  \leq    \frac{
            \norm{\nabla \varphi_i^* (\vec{r})} \abs{\varphi_j (\vec{r})}
          }{
            \|\vec{R}_q - \vec{r}\|
          } +
           \frac{
             \abs{\varphi_i^* (\vec{r})} \norm{\nabla \varphi_j (\vec{r})}
           }{
             \|\vec{R}_q - \vec{r}\|
           } +
          \frac{
            \abs{\varphi_i^* (\vec{r}) \varphi_j (\vec{r})}
          }{
            \|\vec{R}_q - \vec{r}\|^2
          }
  \leq    \left( 2\gamma_1 + 1 \right)
          \frac{\varphi_\text{max}^2}{x_\text{max}^2},
\end{equation}
where the last inequaity follows from \eq{spin-orbital_bound} and
\eq{spin-orbital_first_derivative_bound}, as well as the fact that
$\|\vec{R}_q - \vec{r}\| \geq x_\text{max}$ for any
$\vec{r} \in D_{1, q, \text{non-singular}}$.
From \eq{Riemann_error} and \eq{partition_size_int1},
and noting $1/\lceil z \rceil \leq 1/z$, we have
\begin{equation}
  \delta_\text{Riemann}^{(1, q, \text{non-singular})}
  \leq    8 \sqrt{3} Z_q \left( 2 \gamma_1 + 1 \right)
          \frac{\varphi_\text{max}^2}{x_\text{max}^2} \frac{x_1^4}{N_1}
  \leq    8 \sqrt{3} \left( 2 \gamma_1 + 1 \right)
          Z_q \varphi_\text{max}^2 x_\text{max}^2
          \exp \left(-\frac{\alpha}{2} \frac{x_1}{x_\text{max}}\right).
\end{equation}

\subsubsection{Bounding $\delta_\text{Riemann}^{(1, q, \text{singular})}$ for \lem{int1}}
\label{app:integral_discretization/int1_proof/Riemann-sing}

Recalling that we are using a non-standard metric to evaluate the volumes
and diameters of sets, we find
\begin{equation}
  \text{vol} \left( D_{1, q, \text{singular}} \right) = 2\pi^2
\end{equation}
and
\begin{equation}
  \text{diam} \left( T^{(1, q, \text{singular})}_{\vec{k}} \right)
  = \frac{1}{N_1}\sqrt{5\pi^2 + 1}.
\end{equation}
By \eq{Riemann_error}, it remains to find a bound on the derivative of $f_1$.
Throughout this subsection, we write $f^\prime_\text{max}$ for this bound.

To bound this derivative, we consider the gradient in three different ways.
First there is $\nabla$, which is the gradient with respect to the unscaled position coordinates.
Second there is $\nabla_s$, which is the gradient with respect to the spherical polar coordinates,
but just taking the derivatives with respect to each coordinate.
That is,
\begin{equation}
\nabla_s := \left( \frac{\partial}{\partial s} , \frac{\partial}{\partial \theta}, \frac{\partial}{\partial \phi} \right)
\end{equation}
We use this because we are treating the coordinates like they were Euclidean for the discretized integral.
Third, there is the usual gradient in spherical polar coordinates,
\begin{equation}
\nabla'_s := \left( \frac{\partial}{\partial s} , \frac 1s \frac{\partial}{\partial \theta}, \frac 1{s\sin\theta} \frac{\partial}{\partial \phi} \right)
\end{equation}
Because we consider $s\in[0,1]$, the components of the gradient $\nabla_s$ are upper bounded by the components of the gradient $\nabla'_s$.
Therefore
\begin{equation}
\norm{\nabla_s \left[\varphi_j \! \left( 4x_1 \vec{s} + \vec{R}_q \right)\right]} \le \norm{\nabla'_s \left[\varphi_j \! \left( 4x_1 \vec{s} + \vec{R}_q \right)\right]}
\end{equation}
The restriction on the magnitude of the gradient in \eq{spin-orbital_first_derivative_bound} holds on the usual gradient in spherical polar coordinates.
This means that
\begin{equation}
\norm{\nabla'_s \left[\varphi_j \! \left( 4x_1 \vec{s} + \vec{R}_q \right)\right]} = 4x_1 \norm{\nabla \left[\varphi_j \! \left( 4x_1 \vec{s} + \vec{R}_q \right)\right]} \le 4x_1 \gamma_1 \frac{\varphi_{\max}}{x_{\max}}.
\end{equation}

Using these results, we have
\begin{align}
f^\prime_\text{max} &= \norm{\nabla_s \left[\varphi_i^* \! \left( 4x_1 \vec{s} + \vec{R}_q \right)
                \varphi_j \! \left( 4x_1 \vec{s} + \vec{R}_q \right) s \sin\theta\right]}\nn
                &\le  \left|\varphi_j \! \left( 4x_1 \vec{s} + \vec{R}_q \right) s \sin\theta\right| \norm{\nabla_s \left[\varphi_i^* \! \left( 4x_1 \vec{s} + \vec{R}_q \right)\right]
               }+
                \left|\varphi_i^* \! \left( 4x_1 \vec{s} + \vec{R}_q \right) s \sin\theta\right| \norm{ \nabla_s \left[\varphi_j \! \left( 4x_1 \vec{s} + \vec{R}_q \right)\right] }\nn
& \quad                +\left| \varphi_i^* \! \left( 4x_1 \vec{s} + \vec{R}_q \right) \varphi_j \! \left( 4x_1 \vec{s} + \vec{R}_q \right) \right| \norm{ \nabla_s \left[s \sin\theta\right]}\nn
&\le 4x_1\varphi_{\max}\norm{\nabla \left[\varphi_i^* \! \left( 4x_1 \vec{s} + \vec{R}_q \right)\right]}
+ 4x_1\varphi_{\max} \norm{ \nabla \left[\varphi_j \! \left( 4x_1 \vec{s} + \vec{R}_q \right)\right] } + \varphi_{\max} \sqrt{2} \nn
&\le 8x_1\gamma_1 \frac{\varphi_{\max}^2}{x_{\max}} + \sqrt{2}  \varphi_{\max}.
\end{align}
Thus we have the bound
\begin{equation}
  f^\prime_\text{max}
  \leq   \left( 8 \gamma_1 \frac{x_1}{x_\text{max}} + \sqrt{2} \right)
          \varphi_\text{max}^2.
\end{equation}

We now can give a bound for our approximation to
$S^{(1,q)}_{ij} \!\left( D_{1, \text{singular}} \right)$.
Using the above definitions of
$f^\prime_\text{max}$, $\text{vol} \left( D_{1, \text{singular}} \right)$
and $\text{diam} \left( T^{(1, q, \text{singular})}_{\vec{k}} \right)$, we have
\begin{equation}
  \delta_\text{Riemann}^{(1, q, \text{singular})}
  \leq   \frac 12 16x_1^2 Z_q f^\prime_\text{max} \text{diam} \left( T_{\vec{k}} \right)
          \text{vol} \left( D_{1, \text{singular}} \right)
  \leq    16 \pi^2 \sqrt{5\pi^2 + 1}\sqrt{3}
          \left( 8 \gamma_1 \frac{x_1}{x_\text{max}} + 1 \right)
          \frac{Z_q x_1^2 \varphi_\text{max}^2}{N_1}.
\end{equation}
Using \eq{partition_size_int1} and noting $1/\lceil z \rceil \leq 1/z$, we have
\begin{equation}
\begin{split}
  \delta_\text{Riemann}^{(1, q, \text{singular})}
  & <     16 \pi^2 \sqrt{5\pi^2 + 1}
          \left( 8 \gamma_1 + \frac{x_\text{max}}{x_1} \right)
          Z_q \varphi_\text{max}^2 x_\text{max}^2 \frac{x_\text{max}}{x_1}
          \exp \left( -\frac{\alpha}{2} \frac{x_1}{x_\text{max}} \right) \\
  & \leq  1121 \left( 8 \gamma_1 + \sqrt{2} \right)
          Z_q \varphi_\text{max}^2 x_\text{max}^2
          \exp \left( -\frac{\alpha}{2} \frac{x_1}{x_\text{max}} \right),
\end{split}
\end{equation}
where we have used $x_1 \geq x_\text{max}$.

\subsection{Proof of \lem{int2}}
\label{app:integral_discretization/int2_proof}

As in \app{integral_discretization/int1_proof},
we separate our proof into two cases, depending on whether
the singularity of the integrand is relevant or not.
If the orbitals $i$ and $j$ are distant, then the singularity is unimportant
and we can use rectangular coordinates.
If these orbitals are nearby, then we use spherical polar coordinates
to eliminate the singularity from the integrand.
We do not consider the distance between the orbitals $k$ and $\ell$
in order to simplify the analysis.

\subsubsection{First stage for \lem{int2}}

Again the first stage is to truncate the domain of integration.
We take
\begin{equation}\label{xlem3}
 x_2 :=   \frac{x_\text{max}}{\alpha} \log \left(
                \frac{K_2 \varphi_\text{max}^4 x_\text{max}^5}{\inter}
              \right)
        \end{equation}
to be the size of the truncation region.
The condition in \eq{sensible2} ensures that $x \geq x_\text{max}$.
We regard the orbitals as distant if
$\norm{\vec{c}_i - \vec{c}_j} \geq 2\sqrt{3}x_2 + x_\text{max}$.
Then we take the truncation region
\begin{equation}
  D_{2, \text{non-singular}}
  :=    \mathcal{C}_{x_2} \!\left( \vec{c}_i \right) \times
        \mathcal{C}_{x_2} \! \left( \vec{c}_j \right).
\end{equation}
Otherwise, if the orbitals are nearby we take the truncation region
\begin{equation}
  D_{2, \text{singular}}
  :=      \left\{
            \vec{r}_1 \oplus \vec{r}_2
            \left|
              \vec{r}_1 \in \mathcal{C}_{x_2} \!\left( \vec{c}_i \right),
              \vec{r}_1 - \vec{r}_2 \in
                \mathcal{B}_{\mlfac x_2} ( \vec{0} )
            \right.
          \right\},
\end{equation}
Where $\mlfac:=2\sqrt{3}+3$.
The error in the first case is
\begin{equation}
  \delta_\text{trunc}^{(2, \text{non-singular})}
  :=      \abs{
            S_{ijk\ell}^{(2)}\! \left( \R^6 \right) -
            S_{ijk\ell}^{(2)}\! \left( D_{2, \text{non-singular}} \right)
          }
\end{equation}
and the error in the second case is
\begin{equation}
  \delta_\text{trunc}^{(2, \text{singular})}
  :=      \abs{
            S_{ijk\ell}^{(2)} \!\left( \R^6 \right) -
            S_{ijk\ell}^{(2)} \!\left( D_{2, \text{singular}} \right)
          }.
\end{equation}
The maximum error for either case is denoted
\begin{equation}
  \delta_\text{trunc}^{(2)}
  :=      \max \left\{
            \delta_\text{trunc}^{(2, \text{non-singular})},
            \delta_\text{trunc}^{(2, \text{singular})}
          \right\} .
\end{equation}
We upper bound this error in \app{integral_discretization/int2_proof/trunc} as
\begin{equation}
  \delta_\text{trunc}^{(2)}
  <         \frac{128\pi}{\alpha^6} (\alpha+2)
            \varphi_\text{max}^4 x_\text{max}^5
            \exp\left( -\alpha \frac{x_2}{x_\text{max}} \right).
\end{equation}

\subsubsection{Second stage for \lem{int2} with Cartesian coordinates}

The second stage for the proof of \lem{int2} is to discretize the integrals into Riemann sums.
In this subsection we consider the case that orbitals $i$ and $j$ are distant,
so we wish to approximate the truncated integral
$S_{ijk\ell}^{(2)} \!\left( D_{2, \text{non-singular}} \right)$.
In the next subsection we consider discretization in the case where orbitals
$i$ and $j$ are nearby, and we wish to approximate
$S_{ijk\ell}^{(2)} \!\left( D_{2, \text{singular}} \right)$.
Each sum contains $\mu = N_2^6$ terms, where
\begin{equation}
\label{eq:partition_size_int2}
  N_2
  :=        \left\lceil
              \left(\frac{x_2}{x_\text{max}}\right)^7
              \exp \left( \alpha \frac{x_2}{x_\text{max}} \right)
            \right\rceil.
\end{equation}
The same value of $N_2$ will be used for spherical polar coordinates.
Using the value of $x_2$ from Eq.~\eqref{xlem3} gives
\begin{equation}
  N_2
  =         \left\lceil
              \frac{
                K_2 \varphi_\text{max}^4 x_\text{max}^5
              }{
                \inter
              }
              \left[
                \frac{1}{\alpha} \log \left(
                  \frac{
                    K_2 \varphi_\text{max}^4 x_\text{max}^5
                  }{
                    \inter
                  }
                \right)
              \right]^7
            \right\rceil .
\end{equation}
Since $\mu = N_2^6$ is the number of terms in either Riemann sum, we obtain the
lower bound on $\mu$ in \eq{lem3mu} of \lem{int2}.

We approximate $S_{ijk\ell}^{(2)} \!\left(D_{2, \text{non-singular}}\right)$
with the sum
\begin{equation}
  \mathcal{R}_{2, \text{non-singular}}
  :=      \sum_{\vec{k}_1, \vec{k}_2} \frac{
            \varphi_i^* ( \vec{r}_{\vec{k}_1} )\,
            \varphi_j^* ( \vec{r}_{\vec{k}_2} )\,
            \varphi_k ( \vec{r}_{\vec{k}_2} )\,
            \varphi_\ell ( \vec{r}_{\vec{k}_1} )
          }{
            \|\vec{r}_{\vec{k}_1} - \vec{r}_{\vec{k}_2}\|
          }
          \text{vol} \left(
            T_{\vec{k}_1, \vec{k}_2}^{(2, \text{non-singular})}
          \right),
\end{equation}
where, for every triple of integers
$\vec{k} = \left( k_1, k_2 , k_3 \right)$
such that $0 \leq k_1, k_2, k_3 < N_2$, we define
\begin{equation}
  \vec{r}_{\vec{k}}
  =       \frac{x_2}{N_2} \left[
            2\vec{k} - \left(N_2-1, N_2-1, N_2-1\right]
          \right)
\end{equation}
and
\begin{equation}
  T_{\vec{k}_1, \vec{k}_2}^{(2, \text{non-singular})}
  :=    \mathcal{C}_{x_2/N_2} ( \vec{r}_{\vec{k}_1} ) \times
        \mathcal{C}_{x_2/N_2} ( \vec{r}_{\vec{k}_2} ).
\end{equation}
Thus we have partitioned $D_{2, \text{non-singular}}$ into $\mu$
equal-sized regions that overlap on sets of measure zero.
Each term of $\mathcal{R}_{2, \text{non-singular}}$ has absolute value
no greater than
\begin{equation}
\label{eq:lem3val1}
  \frac{
    \abs{
      \varphi_i^* ( \vec{r}_{\vec{k}_1} )\,
      \varphi_j^* ( \vec{r}_{\vec{k}_2} )\,
      \varphi_k ( \vec{r}_{\vec{k}_2} )\,
      \varphi_\ell ( \vec{r}_{\vec{k}_1} )
    }
  }{
    \|\vec{r}_{\vec{k}_1} - \vec{r}_{\vec{k}_2}\|
  }
  \text{vol} \left(
    T_{\vec{k}_1, \vec{k}_2}^{(2, \text{non-singular})}
  \right)
  \leq    \frac{\varphi_\text{max}^4}{x_\text{max}}
          \left( \frac{2x}{N_2} \right)^6
  =       \frac{1}{\mu} \times
          64 \frac{\varphi_\text{max}^4}{x_\text{max}} x_2^6,
\end{equation}
where the inequality follows from \eq{spin-orbital_bound} and the fact
that the distance between $\mathcal{C}_x \!\left( \vec{c}_i \right)$ and
$\mathcal{C}_x \!\left( \vec{c}_j \right)$ is no smaller than
$x_\text{max}$ if $\norm{\vec{c}_i - \vec{c}_j} \geq
2\sqrt{3}x_2 + x_\text{max}$.
This expression is upper bounded by
       \begin{equation}\label{bndlem3}
  \frac{1}{\mu} \times
  672 \pi^2 \frac{\varphi_\text{max}^4}{x_\text{max}} x_2^6.
\end{equation}
 Substituting our value of $x_2$ from Eq.~\eqref{xlem3} shows that
no term has absolute value greater than (corresponding to \eq{lem3bnd} in \lem{int2})
\begin{equation}
  \frac{1}{\mu} \times
  \frac{672\pi^2}{\alpha^6} \left[
    \log \left(
      \frac{K_2 \varphi_\text{max}^4 x_\text{max}^5}{\inter}
    \right)
  \right]^6 \varphi_\text{max}^4 x_\text{max}^5.
\end{equation}

 We show in \app{integral_discretization/int2_proof/Riemann_non-sing} that
 the error may be bounded as
\begin{align}
  \delta_\text{Riemann}^{(2, \text{non-singular})}
  & :=    \abs{
            S_{ijk\ell}^{(2)} \!\left( D_{2, \text{non-singular}} \right)
            - \mathcal{R}_{2, \text{non-singular}}
          } \nn
  & \leq  256 \sqrt{3} \left( 4\gamma_1 + \sqrt{2} \right)
          \varphi_\text{max}^4 x_\text{max}^5
          \exp \left( -\alpha \frac{x_2}{x_\text{max}} \right).
\end{align}

\subsubsection{Second stage for \lem{int2} with spherical polar coordinates}

In this subsection we discretize the integral
$S_{ijk\ell}^{(2)} \!\left( D_{2, \text{singular}} \right)$ for the case of
nearby orbitals.
We introduce the following definition for convenience in what follows:
\begin{equation}
  \eta_{\ell \ell^\prime} (\vec{r})
  :=      \varphi_\ell^* (\vec{r}) \, \varphi_{\ell^\prime} (\vec{r}).
\end{equation}
We define $\vec{s} := \left( \vec{r}_1 - \vec{c}_i \right)/x_2$ and
$\vec{t} := \left( \vec{r}_1 - \vec{r}_2 \right)/(\mlfac x_2)$.
We write $\vec{s} = \left( s_1, s_2, s_3 \right)$ and
$\theta$ and $\phi$ for the polar and azimuthal angles of $\vec{t}$.
Next we define
\begin{equation}
  f_2 \left( s_1, s_2, s_3, t, \theta, \phi \right)
  :=     \eta_{i\ell} \!\left( x_2\vec{s} + \vec{c}_i \right)
          \eta_{jk} \!\left( x_2\vec{s} - \mlfac x_2\vec{t} + \vec{c}_i \right)
          t \sin\theta
  =    \frac 1{\mlfac x_2} \eta_{i\ell}\! \left( \vec{r}_1 \right)
          \eta_{jk} \!\left( \vec{r}_2 \right)
          \|\vec{r}_1 - \vec{r}_2\| \sin\theta .
\end{equation}
Then we can write
\begin{equation}
  S^{(2)}_{ijk\ell} \!\left( D_{2, \text{singular}} \right)
  =         \zeta^2 x^5
            \int_{-1}^1\dd{s_1}\
            \int_{-1}^1\dd{s_2}\
            \int_{-1}^1\dd{s_3}\
            \int_0^1 \dd{t}\
            \int_0^\pi \dd{\theta}\
            \int_0^{2\pi} \dd{\phi}\
            f_2 \left( s_1, s_2, s_3, t, \theta, \phi \right).
\end{equation}
Let $\vec{k}_{\vec{s}} = \left( k_1, k_2, k_3 \right)$,
where $0 \leq k_1, k_2, k_3 < N_2$,
and let $\vec{k}_{\vec{t}} = \left( k_t, k_\theta, k_\phi \right)$,
where $0 \leq k_t, k_\theta, k_\phi < N_2$.  Define
$s_{k_\ell} = \left( 2k_\ell +1-N_2 \right)/N_2$
for each $\ell = 1, 2, 3$ so that
$\vec{s}_{\vec{k}_{\vec{s}}} = \left( s_{k_1}, s_{k_2}, s_{k_3} \right)$
and define
\begin{equation}
  \left(
    t_{\vec{k}_{\vec{t}}},
    \theta_{\vec{k}_{\vec{t}}},
    \phi_{\vec{k}_{\vec{t}}}
  \right)
  =     \left(
          \frac{1}{N_2} \left(k_t + \tfrac{1}{2}\right),
          \frac{\pi}{N_2} \left(k_\theta + \tfrac{1}{2}\right),
          \frac{2\pi}{N_2} \left(k_\phi + \tfrac{1}{2}\right)
        \right)
\end{equation}
for each $\vec{k}_{\vec{t}}$.  We then define
\begin{equation}
  T_{\vec{k}_{\vec{s}} \oplus \vec{k}_{\vec{t}}}^{(2, \text{singular})}
  :=    \left\{ \left( s_1, s_2, s_3, t, \theta, \phi \right)
          \left|
            \begin{array}{c}
              s_{k_1} - \frac{1}{N_2} \leq s_1
                \leq s_{k_1} + \frac{1}{N_2} \\
              s_{k_2} - \frac{1}{N_2} \leq s_2
                \leq s_{k_2} + \frac{1}{N_2} \\
              s_{k_3} - \frac{1}{N_2} \leq s_3
                \leq s_{k_3} + \frac{1}{N_2} \\
              t_{k_t} - \frac{1}{2 N_2} \leq t
                \leq t_{k_t} + \frac{1}{2 N_2} \\
              \theta_{k_\theta} - \frac{\pi}{2 N_2} \leq \theta
                \leq \theta_{k_\theta} + \frac{\pi}{2 N_2} \\
              \phi_{k_\phi} - \frac{\pi}{N_2} \leq \phi
                \leq \phi_{k_\phi} + \frac{\pi}{N_2} \\
            \end{array}
          \right.
        \right\}.
\end{equation}
Now we define our Riemann sum:
\begin{equation}
  \mathcal{R}_{2,\text{singular}}
  :=        \sum_{\vec{k}_{\vec{s}}, \vec{k}_{\vec{t}}}
            \zeta^2 x_2^5 f_2 \left(
              s_{k_1}, s_{k_2}, s_{k_3},
              t_{k_t}, \theta_{k_\theta}, \phi_{k_\phi}
            \right)
            \text{vol} \left(
              T_{\vec{k}_{\vec{s}} \oplus \vec{k}_{\vec{t}}}
               ^{(2, \text{singular})}
            \right),
\end{equation}
where
\begin{equation}
  \text{vol} \left(
    T_{\vec{k}_{\vec{s}} \oplus \vec{k}_{\vec{t}}}
     ^{(2, \text{singular})}
  \right)
    =       \int_{T_{\vec{k}_{\vec{s}} \oplus \vec{k}_{\vec{t}}}
                            ^{(2, \text{singular})}}
            \dd{s_1}\ \dd{s_2}\ \dd{s_3}\
            \dd{t}\ \dd{\theta}\ \dd{\phi}
    =       \frac{
              \text{vol} \left( D_{2, \text{singular}} \right)
            }{
              N_2^6
            }
    =       \frac{1}{\mu} \times 16\pi^2.
\end{equation}
Here
\begin{equation}
  \text{vol} \left( D_{2, \text{singular}} \right)
  =         \int_{-1}^1\dd{s_1}\
            \int_{-1}^1\dd{s_2}\
            \int_{-1}^1\dd{s_3}\
            \int_0^1 \dd{t}\
            \int_0^\pi \dd{\theta}\
            \int_0^{2\pi} \dd{\phi}
  =         16\pi^2
\end{equation}
is \emph{not} the volume of $D_{2, \text{singular}}$ considered
under the usual Euclidean metric, as in \eq{int1_nonstandard_volume}.
We need to use this non-standard volume because the Jacobian introduced
by changing from Cartesian to spherical polar coordinates was absorbed
into the definition of our integrand $f_2$.  Therefore, each term in
the Riemann sum has absolute value no greater than
\begin{equation}
  \label{eq:lem3val2}
  \abs{ \zeta^2 x_2^5
    f_2 \left(
      s_{k_1}, s_{k_2}, s_{k_3},
      t_{k_t}, \theta_{k_\theta}, \phi_{k_\phi}
    \right)
    \text{vol} \left(
      T_{\vec{k}_{\vec{s}} \oplus \vec{k}_{\vec{t}}}
       ^{(2, \text{singular})}
    \right)
  }
  \leq      \frac{1}{\mu} \times 672 \pi^2 x_2^5 \varphi_\text{max}^4,
\end{equation}
where the inequality follows from \eq{spin-orbital_bound} applied to
the definition of $f_2$ in terms of $\eta_{i\ell}$ and $\eta_{jk}$.
Again this expression is upper bounded by Eq.~\eqref{bndlem3}
and substituting our value of $x_2$ yields the upper bound in \eq{lem3bnd}.

We show in \app{integral_discretization/int2_proof/Riemann_sing} that
\begin{equation}
  \delta_\text{Riemann}^{(2, \text{singular})}
  :=        \abs{
              S_{ijk\ell}^{(2)} \left( D_{2, \text{singular}} \right) -
              \mathcal{R}_{2,\text{singular}}
            }
  <         2161 \pi^2 \left( 20 \gamma_1 + \sqrt{2} \right)
            \varphi_\text{max}^4 x_\text{max}^5
            \exp\left( -\alpha \frac{x_2}{x_\text{max}} \right).
\end{equation}

\subsubsection{Third stage for \lem{int2}}

Lastly we show that the error is properly bounded.
 From \eq{total_error_bound_trunc_Riemann}, we have
\begin{equation}
  \delta_\text{total}^{(2, \text{non-singular})}
  :=      \abs{
            S_{ijk\ell}^{(2)} \left( \R^6 \right) -
            \mathcal{R}_{2, \text{non-singular}}
          }
  \leq    \delta_\text{trunc}^{(2, \text{non-singular})} +
          \delta_\text{Riemann}^{(2, \text{non-singular})}
\end{equation}
and
\begin{equation}
  \delta_\text{total}^{(2, \text{singular})}
  :=      \abs{
            S_{ijk\ell}^{(2)} \left( \R^6 \right) -
            \mathcal{R}_{2, \text{singular}}
          }
  \leq    \delta_\text{trunc}^{(2, \text{singular})} +
          \delta_\text{Riemann}^{(2, \text{singular})}.
\end{equation}
We have given a bound that holds simultaneously for both
$\delta_\text{trunc}^{(2, \text{non-singular})}$
and $\delta_\text{trunc}^{(2, \text{singular})}$,
and we have given a bound for
$\delta_\text{Riemann}^{(2, \text{singular})}$
that is larger (as a function of $x_2$) than our bound for
$\delta_\text{Riemann}^{(2, \text{non-singular})}$.
We are therefore able to assert that the error of our Riemann sum
approximation, no matter which we choose, is always bounded above by
\begin{equation}
  K_2 \varphi_\text{max}^4 x_\text{max}^5
  \exp \left( -\alpha \frac{x_2}{x_\text{max}} \right),
\end{equation}
where
\begin{equation}
  K_2
  :=      \frac{128\pi}{\alpha^6}(\alpha+2) +
          2161 \pi^2 \left( 20 \gamma_1 + \sqrt{2} \right).
\end{equation}
We have also found that the terms in the Riemann sum are upper bounded by
Eqs.~\eqref{eq:lem3val1} and \eqref{eq:lem3val2} in the two cases.
A bound that will hold for both is given by
\begin{equation}\label{bndlem3}
  \frac{1}{\mu} \times
  672 \pi^2 \frac{\varphi_\text{max}^4}{x_\text{max}} x_2^6,
\end{equation}
Then substituting our value of $x_2$ from Eq.~\eqref{xlem3} shows that the error
is upper bounded by $\inter$.  This last step completes our proof of \lem{int2}.
The remainder of this subsection gives the details for some of the steps above.

\subsubsection{Bounding $\delta_\text{trunc}^{(2)}$ for \lem{int2}}
\label{app:integral_discretization/int2_proof/trunc}

Note that $\mathcal{B}_{x_2} \!\left( \vec{c}_i \right) \times
\mathcal{B}_{x_2} \!\left( \vec{c}_j \right)$ is a subset of both
$D_{2, \text{non-singular}}$ and $D_{2, \text{singular}}$.
The former is immediately apparent.  To see the latter, observe that
$\norm{\vec{c}_i - \vec{c}_j} < 2\sqrt{3}x_2 + x_\text{max}$ implies that
the maximum possible value of $\norm{\vec{r}_1 - \vec{r}_2}$ for any
$\vec{r}_1 \in \mathcal{B}_x \!\left( \vec{c}_i \right)$ and
$\vec{r}_2 \in \mathcal{B}_x \!\left( \vec{c}_j \right)$ is
$2\sqrt{3}x_2 + 2x_2+ x_\text{max} \le  (2\sqrt{3}+3)x_2=\mlfac x_2$.
Therefore,
\begin{align}
  \delta_\text{trunc}^{(2)}
  & \leq    \varphi_\text{max}^4
            \int_{\R^3 \setminus \mathcal{B}_{x_2} \left( \vec{c}_i \right)}
            \dd{\vec{r}_1}
            \int_{\R^3 \setminus \mathcal{B}_{x_2} \left( \vec{c}_j \right)}
            \dd{\vec{r}_2}
            \frac{
              e^{ -\alpha \norm{\vec{r}_1 - \vec{c}_i}/x_\text{max} }
              e^{ -\alpha \norm{\vec{r}_2 - \vec{c}_j}/x_\text{max} }
            }{
              \norm{\vec{r}_1 - \vec{r}_2}
            } \nn
  & =       \varphi_\text{max}^4
            \int_{\R^3 \setminus \mathcal{B}_{x_2} ( \vec{0} )}
            \dd{\vec{s}}\ e^{ -\alpha s/x_\text{max} }
            \Lambda_{\alpha/x_\text{max}, x_2} \left(
              \vec{s} + \vec{c}_i - \vec{c}_j
            \right),
\end{align}
where we have used \eq{spin-orbital_decay} and,
with the change of variables $\vec{s} = \vec{r}_1 - \vec{c}_i$,
the definition of $\Lambda$ from \eq{exp_Coulomb_potential}.
By \lem{exp_Coulomb_potential_bound}, for $\|\vec{s} + \vec{c}_i - \vec{c}_j\|\le x_2$ we get
\begin{equation}
  \Lambda_{\alpha/x_\text{max}, x_2} \left(
    \vec{s} + \vec{c}_i - \vec{c}_j
  \right)
  \leq      \frac{8\pi x_\text{max}^2}{\alpha^2}
            e^{ -\alpha s/x_\text{max} },
\end{equation}
and for $\norm{\vec{s} + \vec{c}_i - \vec{c}_j} > x_2$ we get
\begin{align}
\Lambda_{\alpha/x_\text{max}, x_2} \left(
  \vec{s} + \vec{c}_i - \vec{c}_j
\right)
& \le \frac{16\pi x_\text{max}^3}{\alpha^3}
    \frac{e^{ -\alpha s/x_\text{max} }}{\norm{\vec{s} + \vec{c}_i - \vec{c}_j}} \nn
&<\frac{16\pi x_\text{max}^3}{x_2\alpha^3} e^{ -\alpha s/x_\text{max} }\nn
&\le  \frac{16\pi x_\text{max}^2}{\alpha^3} e^{ -\alpha s/x_\text{max} }.
\end{align}
In either case we then get
\begin{equation}
  \delta_\text{trunc}^{(2)}
   \leq    \frac{8\pi}{\alpha^3}
            \varphi_\text{max}^4 x_\text{max}^2 (\alpha+2)
            \exp\left( -\frac{\alpha}{2} \frac{x_2}{x_\text{max}} \right)
            \int_{\R^3 \setminus \mathcal{B}_{x_2} ( \vec{0} )}
            \dd{\vec{s}}\ e^{ -\alpha s/x_\text{max} }.
\end{equation}
We use the fact that $(z^2 + 2z + 2) e^{-z} < 4 e^{-z/2}$ for any $z>0$ to find
\begin{equation}
  \int_{\R^3 \setminus \mathcal{B}_{x_2}\! \left( \vec{0} \right)} \dd{\vec{s}}\
  e^{-\mu s}
  =         4\pi \int_{x_2}^\infty \dd{s}\ e^{-\mu s} s^2
  =         4\pi
            \left( \frac{x_2^2}{\mu} + \frac{2x_2}{\mu^2} + \frac{2}{\mu^3} \right)
            e^{-\mu x}
  <         \frac{16\pi}{\mu^3} e^{-\mu x_2/2},
\end{equation}
which gives us
\begin{equation}
  \delta_\text{trunc}^{(2)}
  <         \frac{128\pi}{\alpha^6}(\alpha+2)
            \varphi_\text{max}^4 x_\text{max}^5
            \exp\left( -\alpha \frac{x_2}{x_\text{max}} \right).
\end{equation}

\subsubsection{Bounding $\delta_\text{Riemann}^{(2, \text{non-singular})}$ for \lem{int2}}
\label{app:integral_discretization/int2_proof/Riemann_non-sing}

Following \app{integral_discretization/preliminaries/proof_structure}, we note that
\begin{equation}
  \text{vol} \left( D_{2, \text{non-singular}} \right)
  =     64x_2^6
\end{equation}
and
\begin{equation}
  \text{diam} \left( T_{\vec{k}_1, \vec{k}_2}^{(2, \text{non-singular})} \right)
  =     \sqrt{
          \text{diam} \left(
            \mathcal{C}_{x_2/N_2} ( \vec{r}_{\vec{k}_1} )
          \right)^2 + \text{diam} \left(
            \mathcal{C}_{x_2/N_2} ( \vec{r}_{\vec{k}_2} )
          \right)^2
        }
  =     2\sqrt{6} x_2/N_2
\end{equation}
for each $\vec{k}_1$ and $\vec{k}_2$.  To find a bound on
$\delta_\text{Riemann}^{(2, \text{non-singular})}$, it only remains
to find a bound on the derivative of the integrand.

To bound the derivative of the integrand, we first find bounds on
the gradients of the numerator and the denominator separately.
The gradient of the numerator can be bounded using
the product rule and triangle inequality, as well as
\eq{spin-orbital_bound} and \eq{spin-orbital_first_derivative_bound}:
\begin{align}
&  \norm{
    \left( \nabla_1 \oplus \nabla_2 \right) \left[
      \varphi_i^* ( \vec{r}_1 ) \, \varphi_j^* ( \vec{r}_2 )\,
      \varphi_k ( \vec{r}_2 ) \,\varphi_\ell ( \vec{r}_1 )
    \right]
  }^2 \nn
  & = \norm{
            \nabla_1 \left[
              \varphi_i^* ( \vec{r}_1 )\,
              \varphi_j^* ( \vec{r}_2 )\,
              \varphi_k ( \vec{r}_2 )\,
              \varphi_\ell ( \vec{r}_1 )
            \right]
          }^2 + \norm{
            \nabla_2 \left[
              \varphi_i^* ( \vec{r}_1 )
              \varphi_j^* ( \vec{r}_2 )
              \varphi_k ( \vec{r}_2 )
              \varphi_\ell ( \vec{r}_1 )
            \right]
          }^2 \nn
  & =   \left(  \abs{\varphi_j^* ( \vec{r}_2 )}
          \abs{\varphi_k ( \vec{r}_2 )}
          \norm{
            \nabla_1 \left[
              \varphi_i^* ( \vec{r}_1 )\,
              \varphi_\ell ( \vec{r}_1 )
            \right]
          }\right)^2 +
         \left( \abs{\varphi_i^* ( \vec{r}_1 )}
          \abs{\varphi_\ell ( \vec{r}_1 )}
          \norm{
            \nabla_2 \left[
              \varphi_j^* ( \vec{r}_2 )\,
              \varphi_k ( \vec{r}_2 )
            \right]
          }\right)^2 \nn
  & \leq \left( \varphi_\text{max}^2
          \norm{
            \nabla_1 \left[
              \varphi_i^* ( \vec{r}_1 )\,
              \varphi_\ell ( \vec{r}_1 )
            \right]
          }\right)^2 + \left(
          \varphi_\text{max}^2
          \norm{
            \nabla_2 \left[
              \varphi_j^* ( \vec{r}_2 )\,
              \varphi_k ( \vec{r}_2 )
            \right]
          }\right)^2 \nn
  & \leq \left( \varphi_\text{max}^2
          \abs{\varphi_\ell ( \vec{r}_1 )}
          \norm{
            \nabla_1 \left[
              \varphi_i^* ( \vec{r}_1 )
            \right]
          } +
          \varphi_\text{max}^2
          \abs{\varphi_i^* ( \vec{r}_1 )}
          \norm{
            \nabla_1 \left[
              \varphi_\ell ( \vec{r}_1 )
            \right]
          }\right)^2  \nn &\qquad +\left(
          \varphi_\text{max}^2
          \abs{\varphi_k ( \vec{r}_2 )}
          \norm{
            \nabla_2 \left[
              \varphi_j^* ( \vec{r}_2 )
            \right]
          } +
          \varphi_\text{max}^2
          \abs{\varphi_j^* ( \vec{r}_2 )}
          \norm{
            \nabla_2 \left[
              \varphi_k ( \vec{r}_2 )
            \right]
          }\right)^2 \nn
  & \leq 2 \left(2\gamma_1 \frac{\varphi_\text{max}^4}{x_\text{max}}\right)^2.
\end{align}
The gradient of the denominator can be computed directly:
\begin{equation}
  \norm{\left( \nabla_1 \oplus \nabla_2 \right) \norm{\vec{r}_1 - \vec{r}_2}}
  =       \norm{
            \left( \nabla_1 \norm{\vec{r}_1 - \vec{r}_2} \right) \oplus
            \left( \nabla_2 \norm{\vec{r}_1 - \vec{r}_2} \right)
          }
  =       \norm{
            \left(
              \frac{\vec{r}_1 - \vec{r}_2}{\norm{\vec{r}_1 - \vec{r}_2}}
            \right) \oplus
            \left(
              \frac{\vec{r}_2 - \vec{r}_1}{\norm{\vec{r}_2 - \vec{r}_1}}
            \right)
          }
  =       \sqrt{2}
\end{equation}
Again by the product rule and the triangle inequality,
\begin{equation}
  \norm{
    (\nabla_1 \oplus \nabla_2) \frac{
      \varphi_i^* ( \vec{r}_1 )\, \varphi_j^* ( \vec{r}_2 )\,
      \varphi_k ( \vec{r}_2 )\, \varphi_\ell ( \vec{r}_1 )
    }{
      \norm{\vec{r}_1 - \vec{r}_2}
    }
  }
  \leq      \frac{2\sqrt{2}\gamma_1}{\norm{\vec{r}_1 - \vec{r}_2}}
            \frac{\varphi_\text{max}^4}{x_\text{max}} +
            \frac{\sqrt{2}}{\norm{\vec{r}_1 - \vec{r}_2}^2}
            \varphi_\text{max}^4
  \leq      \sqrt{2}\left( 2\gamma_1 + 1 \right)
            \frac{\varphi_\text{max}^4}{x_\text{max}^2}.
\end{equation}
The last inequality follows from our assumption that
$\norm{\vec{c}_i - \vec{c}_j} > 2\sqrt{3}x_2 + x_\text{max}$, which implies that
the distance between $\mathcal{C}_{x_2} \!\left( \vec{c}_i \right)$ and
$\mathcal{C}_{x_2} \!\left( \vec{c}_j \right)$ is greater than $x_\text{max}$.
Therefore,
\begin{equation}
  \delta_\text{Riemann}^{(2)}
  \leq    \frac 12  \left( \sqrt{2} \left( 2\gamma_1 + 1 \right)
            \frac{\varphi_\text{max}^4}{x_\text{max}^2}\right)\left( 64 x_2^6 \right)\left(2\sqrt{6} \frac {x_2}{N_2}\right)
  \leq      128 \sqrt{3} \left( 2\gamma_1 + 1 \right)
            \varphi_\text{max}^4 x_\text{max}^5
            \exp \left( -\alpha \frac{x_2}{x_\text{max}} \right).
\end{equation}

\subsubsection{Bounding $\delta_\text{Riemann}^{(2, \text{singular})}$ for \lem{int2}}
\label{app:integral_discretization/int2_proof/Riemann_sing}

Following \app{integral_discretization/preliminaries/proof_structure},
we again note that $\text{vol} \left( D_{2, \text{singular}} \right) = 16\pi^2$.
We also observe that
\begin{equation}
  \text{diam} \left(
    T_{\vec{k}_{\vec{s}} \oplus \vec{k}_{\vec{t}}}
     ^{(2, \text{singular})}
  \right)
  =         \frac{1}{N_2} \sqrt{2^2 + 2^2 + 2^2 + 1^2 + \pi^2 + 4\pi^2}
  =         \frac{1}{N_2} \sqrt{13 + 5\pi^2}
  <         \frac{8}{N_2},
\end{equation}
where we are again treating the variables $s$, $\theta$ and $\phi$ formally
as Euclidean coordinates instead of spherical polar.  It then remains to
bound the derivative of the integrand
\begin{equation}
  f_2 \left( s_1, s_2, s_3, t, \theta, \phi \right)
  =         \eta_{i\ell} \!\left( x\vec{s} + \vec{c}_{i\ell} \right)
            \eta_{jk} \!\left( x\vec{s} - \mlfac x\vec{t} + \vec{c}_i \right)
            t \sin\theta,
\end{equation}
where $\vec{s} = \left( s_1, s_2, s_3 \right)$.

The derivative of the integral is bounded as follows.  Define
$\nabla_s = \left(
  \frac{\partial}{\partial s_1},
  \frac{\partial}{\partial s_2},
  \frac{\partial}{\partial s_3}
\right)$ and
$\nabla_t = \left(
  \frac{\partial}{\partial t},
  \frac{\partial}{\partial \phi},
  \frac{\partial}{\partial \theta}
\right)$.  By the product rule and the triangle inequality,
\begin{align}
  \norm{ \left( \nabla_s \oplus \nabla_t \right) f_2 }
  & \leq  \abs{
            \eta_{jk}\! \left( x_2\vec{s} - \mlfac x_2\vec{t} + \vec{c}_i \right)
            t \sin\theta
          }
          \norm{
            \left( \nabla_s \oplus \nabla_t \right)
            \eta_{i\ell} \!\left( x_2\vec{s} + \vec{c}_i \right)
          } \nn &\qquad +
          \abs{
            \eta_{i\ell} \left( x_2\vec{s} + \vec{c}_i \right)
          }
          \norm{
            \left( \nabla_s \oplus \nabla_t \right)
            \eta_{jk} \!\left( x_2\vec{s} - \mlfac x_2\vec{t} + \vec{c}_i \right)
            t \sin\theta
          } \nn
  & \leq  \varphi_\text{max}^2
          \norm{
            \left( \nabla_s \oplus \nabla_t \right)
            \eta_{i\ell} \!\left( x_2\vec{s} + \vec{c}_i \right)
          } +
          \varphi_\text{max}^2
          \norm{
            \left( \nabla_s \oplus \nabla_t \right)
            \eta_{jk} \!\left( x_2\vec{s} - \mlfac x_2\vec{t} + \vec{c}_i \right)
            t \sin\theta
          },
\end{align}
where the last inequality follows from \eq{spin-orbital_bound}.  We also have
\begin{align}
  \norm{
    \left( \nabla_s \oplus \nabla_t \right)
    \eta_{i\ell} \! \left( x_2\vec{s} + \vec{c}_i \right)
  }
  & =     \norm{\nabla_s \eta_{i\ell} \! \left( x_2\vec{s} + \vec{c}_i \right)} \nn
  & \leq  \abs{\varphi_\ell \!\left( x_2\vec{s} + \vec{c}_i \right)}
          \norm{\nabla_s \varphi_i^* \!\left( x_2\vec{s} + \vec{c}_i \right)} +
          \abs{\varphi_i^* \!\left( x_2\vec{s} + \vec{c}_i \right)}
          \norm{\nabla_s \varphi_\ell \!\left( x_2\vec{s} + \vec{c}_i \right)} \nn
  & \leq  x_2 \varphi_\text{max}
          \norm{\nabla \varphi_i^* \!\left( x_2\vec{s} + \vec{c}_i \right)} +
          x_2 \varphi_\text{max}
          \norm{\nabla \varphi_\ell \!\left( x_2\vec{s} + \vec{c}_i \right)} \nn
  & \leq  2 x_2 \gamma_1 \frac{\varphi_\text{max}^2}{x_\text{max}},
\end{align}
where $\nabla$ in the second-to-last inequality refers to the gradient operator
expressed in the usual basis and the final inequality follows from
\eq{spin-orbital_first_derivative_bound}.  Finally, we have
\begin{align}
 & \norm{
    \left( \nabla_s \oplus \nabla_t \right)
    \eta_{jk} \!\left( x_2\vec{s} - \mlfac x_2\vec{t} + \vec{c}_i \right)
    t \sin\theta
  }\nn
  & \leq  \norm{
            \nabla_s \!\left[
              \eta_{jk} \!\left( x_2\vec{s} - \mlfac x_2\vec{t} + \vec{c}_i \right)
              t \sin\theta
            \right]
          } +
          \norm{
            \nabla_t \!\left[
              \eta_{jk} \!\left( x_2\vec{s} - \mlfac x_2\vec{t} + \vec{c}_i \right)
              t \sin\theta
            \right]
          } \nn
  & \leq  \abs{t \sin\theta} \norm{
            \nabla_s \!\left[
              \eta_{jk} \!\left( x_2\vec{s} - \mlfac x_2\vec{t} + \vec{c}_i \right)
            \right]
          } +
          \abs{t \sin\theta} \norm{
            \nabla_t \!\left[
              \eta_{jk} \!\left( x_2\vec{s} - \mlfac x_2\vec{t} + \vec{c}_i \right)
            \right]
          }  +\abs{
            \eta_{jk} \!\left( x_2\vec{s} - \mlfac x_2\vec{t} + \vec{c}_i \right)
          } \norm{ \nabla_t (t \sin\theta) } \nn
  & \leq  \norm{
            \nabla_s \!\left[
              \eta_{jk} \!\left( x_2\vec{s} - \mlfac x_2\vec{t} + \vec{c}_i \right)
            \right]
          } + \norm{
            \nabla_t \!\left[
              \eta_{jk} \!\left( x_2\vec{s} - \mlfac x_2\vec{t} + \vec{c}_i \right)
            \right]
          }   +
          \varphi_\text{max}^2 \norm{ \nabla_t (t \sin\theta) },
\end{align}
where we have again used the product rule and the triangle inequality and,
in the last inequality, \eq{spin-orbital_bound}.
We have also used the bounds on the gradient operator $\nabla_t$ in the same was as in \app{integral_discretization/int1_proof/Riemann-sing}.
    We note that
$\norm{\nabla_t (t \sin\theta)} \leq \sqrt{2}$, $\norm{
  \nabla_s \left(
    \eta_{jk} \!\left( x_2\vec{s} - \mlfac x_2\vec{t} + \vec{c}_i \right)
  \right)
} \leq 2 x \gamma_1 \varphi_\text{max}^2 / x_\text{max}$ (as above) and
\begin{align}
&  \norm{
    \nabla_t
      \eta_{jk} \!\left( x_2\vec{s} - \mlfac x_2\vec{t} + \vec{c}_i \right)
  }\nn
  & \leq  \abs{\varphi_k \!\left( x_2\vec{s} - \mlfac x_2\vec{t} + \vec{c}_i \right)}
          \norm{
            \nabla_t \varphi_j^* \!\left( x_2\vec{s} - \mlfac x_2\vec{t} + \vec{c}_i \right)
          }
     +     \abs{\varphi_j^* \!\left( x_2\vec{s} - \mlfac x_2\vec{t} + \vec{c}_i \right)}
          \norm{
            \nabla_t \varphi_k \!\left( x_2\vec{s} - \mlfac x_2\vec{t} + \vec{c}_i \right)
          } \nn
  & \leq  \mlfac x_2 \varphi_\text{max} \norm{
            \nabla \varphi_j^* \!\left( x_2\vec{s} - \mlfac x_2\vec{t} + \vec{c}_i \right)
          } +
          \mlfac x_2 \varphi_\text{max} \norm{
            \nabla \varphi_k \!\left( x_2\vec{s} - \mlfac x_2\vec{t} + \vec{c}_i \right)
          } \nn
  & \leq  2\mlfac x_2 \gamma_1 \frac{\varphi_\text{max}^2}{x_\text{max}}.
\end{align}
In summary, we have shown
\begin{equation}
  \norm{ \left( \nabla_s \oplus \nabla_t \right) f_2 }
  \leq    \left( 20 \gamma_1 \frac{x_2}{x_\text{max}} + \sqrt{2} \right)
          \varphi_\text{max}^4.
\end{equation}

We can now compute our bound on $\delta_\text{Riemann}^{(2, \text{singular})}$.
Including the factor of $\zeta^2 x_2^5$ in the integral and using \eq{Riemann_error},
\begin{align}
  \delta_\text{Riemann}^{(2, \text{singular})}
  & \leq \frac 12 \zeta^2 x_2^5 f_\text{max}^\prime
          \text{vol} \left( D_{2, \text{singular}} \right)
          \max_{\vec{k}_{\vec{s}}, \vec{k}_{\vec{t}}} \text{diam} \left(
            T_{\vec{k}_{\vec{s}} \oplus \vec{k}_{\vec{t}}}
             ^{(2, \text{singular})}
          \right) \nn
  & <   \frac 12 \zeta^2 x_2^5 \times
          \left( 20 \gamma_1 \frac{x_2}{x_\text{max}} + \sqrt{2} \right)
          \varphi_\text{max}^4 \times
          2\mlfac \pi^2 \times 8/N_2 \nn
  & <  2161 \pi^2
          \left( 20 \gamma_1 + \sqrt{2} \frac{x_\text{max}}{x_2} \right)
          \varphi_\text{max}^4 x_\text{max}^5 \frac{x_\text{max}}{x_2}
          \exp\left( -\alpha \frac{x_2}{x_\text{max}} \right) \nn
  & \leq  2161 \pi^2 \left( 20 \gamma_1 + \sqrt{2} \right)
          \varphi_\text{max}^4 x_\text{max}^5
          \exp\left( -\alpha \frac{x_2}{x_\text{max}} \right).
\end{align}

\end{document}